\documentclass{article}
\usepackage[english]{babel}
\usepackage{cite}
\usepackage{pstool}
\usepackage{jheppub} 
\usepackage{bm}
\usepackage{gensymb}
\usepackage{makecell}

\usepackage{placeins}

\usepackage{epsf}
\usepackage{amssymb}
\usepackage{amsmath}
\usepackage{slashed}

\DeclareMathOperator{\Tr}{Tr}
\usepackage{amsfonts}
\usepackage{psfrag,epsfig,graphicx,graphics}

\usepackage{bm,slashed,multirow,
	soul,mathtools,xspace,array,comment}                
\allowdisplaybreaks
\usepackage{bbold}
\usepackage[dvipsnames]{xcolor}
\usepackage{subfigure}
\usepackage[printonlyused]{acronym}
\usepackage{hyperref}
\usepackage{tikz}
\usepackage[compat=1.1.0]{tikz-feynman}

\usepackage{siunitx}
\usepackage{booktabs}

\usepackage{float}

\newcommand{\APP}{App.~}
\newcommand{\TAB}{Tab.~}
\newcommand{\FIG}{Fig.~}

\newcommand{\SEC}{Sec.~}

\newcommand{\EQ}{Eq.~}
\newcommand{\EQs}{Eqs.~}

\newcommand{\GeV}{\,\mbox{GeV}}

\newcommand{\SgN}{S_{\gamma N}}

\newcommand{\pin}{p_{\mathrm{inc}}}
\newcommand{\pout}{p_{\mathrm{exc}}}

\title{Exclusive photoproduction of a di-meson pair with large invariant mass}

\author[1,2]{Saad Nabeebaccus,}
\author[1]{David Perez,}
\author[3]{Lech Szymanowski,}
\author[1]{Samuel Wallon}

\affiliation[1]{Universit\'e Paris-Saclay, CNRS/IN2P3, IJCLab, 91405 Orsay, France}
\affiliation[2]{Department of Physics \& Astronomy, University of Manchester, Manchester M13 9PL, United Kingdom}
\affiliation[3]{National Center for Nuclear Research (NCBJ), Warsaw, Poland}

\emailAdd{saad.nabeebaccus@manchester.ac.uk}
\emailAdd{david.perez@ijclab.in2p3.fr}
\emailAdd{Lech.Szymanowski@ncbj.gov.pl}
\emailAdd{samuel.wallon@ijclab.in2p3.fr}

\abstract{
The exclusive photoproduction of a pair of light mesons is studied within the framework of collinear factorisation. The amplitude factorises into a process-dependent perturbatively calculable hard part, a generalised parton distribution (GPD) and two distribution amplitudes (DAs). Specifically, we focus on the production of any combination of $\rho$ and $\pi$ mesons (of any charge and polarisation) that do not involve \textit{neutral} $C=+$ exchanges with the nucleon. This gives a total of 26 distinct channels, which are sensitive to \textit{quark} GPDs only.
We calculate the amplitude for this family of di-meson processes at leading order in the strong coupling constant $\alpha_s$ and at leading twist, in a fully-automated way. Depending on the choice of mesons in the final state, some of these processes are sensitive to chiral-odd (helicity-flip) GPDs. Particular attention is given to the treatment of poles in the 3-dimensional convolution integral of the momentum fractions connecting the hard part with the different non-perturbative components. These poles are regularised by the usual Feynman $i \epsilon$ factors, but lead to numerical instabilities if not dealt with properly.
We also discuss in detail the construction of the phase space. Importantly, we propose a resolution for the inconsistency of the kinematics of the hard part of the process, where hadron masses and other soft scales are neglected, with the rest of the process.
As a proof of concept, we explicitly evaluate the cross section, for a subset of processes whose amplitudes have been constructed, at energies typical of the CLAS12 experiment at JLab. Our results indicate that exclusive di-meson photoproduction processes have very good statistics, which can be a factor of up to a hundred more than the exclusive photon-meson photoproduction process. Therefore, the family of processes that we study here represents a great opportunity for GPD extraction.
}

\date{\today}
\begin{document}
	
	\maketitle

\preprint{}

\newpage

\section{Introduction}

\label{sec:introduction}

The study of exclusive processes provides important insights on hadron structure. By imposing specific kinematical constraints on the particles participating in the scattering process, it is possible to employ collinear factorisation theorems to extract universal partonic distributions from them, in particular the \textit{generalised parton distributions} (GPDs) \cite{Burkardt:2000za,Burkardt:2002hr,Diehl:2003ny,Belitsky:2005qn}. Such a collinear factorisation occurs at the amplitude level, in contrast to inclusive processes. This factorisation structure makes it particularly difficult to extract GPDs, since they enter in a convolution integral at the amplitude level, and is usually referred to in the literature as the \textit{deconvolution problem} \cite{Diehl:2003ny,Bertone:2021yyz}. For this reason, it is essential to have a wide range of experimentally accessible processes which can provide as many constraints as possible for the extraction of GPDs.

Since the 90s, GPDs have been studied in different processes, such as Deeply Virtual Compton Scattering (DVCS) \cite{Ji:1996nm,Radyushkin:1996nd,Collins:1998be}, Deeply Virtual Meson Production (DVMP) \cite{Radyushkin:1996ru,Collins:1996fb,Belitsky:2001nq}, timelike Compton scattering (TCS) \cite{Berger:2001xd,Pire:2011st} and exclusive heavy quarkonium photoproduction \cite{Ivanov:2004zv,Flett:2024htj}. Nevertheless, the previously mentioned processes, which are among the most well-studied, involve a \textit{single hard scale}, e.g.~virtuality of the incoming photon in DVCS and DVMP. This implies that only limited moment-type information on GPDs can be extracted from such processes, as pointed out recently in \cite{Qiu:2023mrm}. Unfortunately, they are also the only ones\footnote{It is worthwhile to mention that there has been some effort recently to attempt the experimental measurement \cite{Boer:2024hol} of double DVCS (where both the incoming and the outgoing photons are highly virtual) \cite{Muller:1994ses,Deja:2023ahc}.} that have been experimentally measured so far at JLab by the JLab Hall A collaboration \cite{JeffersonLabHallA:2015dwe,JeffersonLabHallA:2016wye} and the CLAS collaboration \cite{CLAS:2006krx,CLAS:2008rpm,CLAS:2014jpc,CLAS:2021lky}, at HERA by the H1 collaboration \cite{H1:1994nbr,H1:2001nez,H1:2009cml,H1:2013okq}, the HERMES collaboration \cite{HERMES:2000jnb,HERMES:2001bob} and the ZEUS collaboration \cite{ZEUS:1995kab,ZEUS:1995gig,ZEUS:2003pwh,ZEUS:2009asc}, and at CERN by the COMPASS collaboration \cite{COMPASS:2013fsk}. Furthermore, when in photoproduction mode, such processes can be measured in \textit{ultra-peripheral collisions} in hadron-hadron colliders, such as at LHC \cite{Pire:2008ea,LHCb:2014acg,ALICE:2014eof,LHCb:2015wlx,CMS:2018bbk}. Finally, the measurement of exclusive processes to probe GPDs is part of the science program at the EIC \cite{AbdulKhalek:2021gbh}.

It was first proposed in \cite{ElBeiyad:2010pji} to study the exclusive photoproduction of $\rho^0_T\pi^+$ pair with large invariant mass. The main motivation for considering such a process is that it gives access to \textit{chiral-odd} (helicity-flip) GPDs at the \textit{leading} twist, unlike the previously discussed processes. Moreover, due to the more involved kinematics, such a process actually possesses \textit{two hard scales}, which allows far more information to be probed from the GPDs.\footnote{The presence of two hard scales can be deduced from the fact that within the collinear factorisation framework, the hard part of the process effectively looks like a $2 \to 2$ hard scattering involving massless particles as indicated in \EQ\eqref{eq:2-body-subprocess}, meaning that two hard scales are given by 2 independent Mandelstam variables, see \EQs\eqref{eq:uprimed} to \eqref{eq:sprimed}.} Exclusive processes involving 3-body final states (including the outgoing nucleon) have been extensively studied since, such as the photoproduction of a  $\gamma \rho$ pair \cite{Boussarie:2016qop,Duplancic:2023kwe}, a $\gamma \pi$ pair \cite{Duplancic:2018bum,Duplancic:2022ffo,Crnkovic:2025man}, and a pair of photons \cite{Pedrak:2017cpp,Grocholski:2021man,Grocholski:2022rqj}. The crossed process of diphoton production in pion-nucleon scattering has also been studied in \cite{Qiu:2022bpq,Qiu:2024mny}. However, while these processes represent an excellent opportunity for GPD extraction, none of them have been measured to date. It is important to emphasise that the presence of a hard scale in the scattering process is a \textit{necessary}, but not \textit{sufficient} condition for collinear factorisation to be applicable. A proof of factorisation for a wide range of such $2 \to 3 $ exclusive processes was obtained in \cite{Qiu:2022bpq,Qiu:2022pla}. However, it was later discovered in channels which allow two-gluon exchanges with the nucleon, such as exclusive $\pi^0 \gamma$ photoproduction, collinear factorisation is broken by \textit{Glauber gluon exchanges} \cite{Nabeebaccus:2023rzr,Nabeebaccus:2024mia}.

The aim of this paper is to study a wide range of processes that correspond to the exclusive photoproduction of two \textit{light} mesons, focusing on the technical aspects of the calculation and setting the stage for numerics. We intend to address phenomenological studies in a subsequent article. We specifically consider any pairs of $\rho$ and $\pi$ mesons that \textit{cannot} exchange exactly two gluons with the nucleon, due to electric charge conservation or $C$-parity conservation, and hence the processes under consideration are only sensitive to \textit{quark} GPDs. It turns out that there are 26 such processes in total. Within the framework of collinear factorisation, the amplitude for such a process factorises
\begin{equation}
\label{eq:amplitude}
    i\mathcal{M}=\int_{-1}^{1}dx \int_{0}^1dz\int_0^{1}dv\;T_H(x,v,z)H(x)\phi_1(v)\phi_2(z)\,.
\end{equation}
The hard part, denoted by $T_H(x,v,z)$, is calculable perturbatively order by order in $\alpha_s$, whereas the non-perturbative part contains a GPD, $H(x)$, and two Distribution Amplitudes (DAs), $\phi_1(v)$ and $\phi_2(z)$. The GPD encodes the internal dynamics of the nucleon, while the DAs describe those of each meson. In the GPD, $x$ represents the average longitudinal momentum of the parton probed from the nucleon, while in the DAs, $v$ and $z$ represent the momentum fraction of each meson carried by the quark.

In this paper, $T_H(x,v,z)$ is calculated at leading order in $\alpha_s$, which can involve up to 120 diagrams, indicating the level of complication already at this order. In fact, as far as $2 \to 3$ exclusive processes within the framework of collinear factorisation are concerned, only diphoton photoproduction has been calculated at next-to-leading order \cite{Grocholski:2021man,Grocholski:2022rqj}, which is also one of the simplest processes to calculate at NLO. This is because the hard part of the process involves a single  outgoing quark-antiquark ($q\bar q$) pair, with the amplitude given by a 1-dimensional convolution integral of the hard part with a GPD. In comparison, photon-meson photoproduction involves 2 outgoing $q \bar q$ pairs and a 2-dimensional convolution integral of the hard part with a GPD and a DA. In terms of the hard part itself, half of the NLO diagrams are related by crossing symmetry to those in \cite{Nizic:1987sw,Duplancic:2006nv} in the context of $\gamma \gamma \to \pi \pi$ wide-angle scattering,\footnote{Only those that involve two $\gamma^5$ in the Dirac trace have been calculated in \cite{Nizic:1987sw,Duplancic:2006nv}, since they constitute the full set of relevant ones in that case. Diagrams which have a single $\gamma^5$ would still have to be calculated for exclusive photon-meson photoproduction.} where the amplitude is obtained by convoluting the hard part with 2 DAs. It is important to emphasise that performing the crossing symmetry for the exclusive photon-meson photoproduction case, and integrating over the relevant GPD and DA is highly non-trivial due to the need for the proper treatment of $i\epsilon$ factors in denominators, which can be safely put to zero in the wide-angle scattering case. Finally, for di-meson photoproduction, which is the subject of this paper, the hard part involves 3 outgoing $q \bar q$ pairs, and the amplitude is given by a 3-dimensional integral, as can be seen in \EQ\eqref{eq:amplitude}.

The paper is organised as follows. In \SEC\ref{sec:kinematics}, we give details on the kinematics for a generic process involving two final-state mesons. The non-perturbative inputs, namely the GPDs and DAs are discussed in \SEC\ref{sec:non-perturbative-inputs}. Then, in \SEC\ref{sec:construction-of-diagrams}, we explain in detail how to automate the construction of the hard part of the process from scratch. Next, the organisation of the amplitude is discussed in \SEC\ref{sec:organisation-amplitude}, including a detailed analysis on the different tensor structures that can appear. The numerical integration strategy is presented in \SEC\ref{sec:numerical-integration}, including the issue of how to carefully handle $i\epsilon$ factors in denominators prior to integration. \SEC\ref{sec:cross-section} deals with the construction of the cross section from the amplitudes which have already been convoluted with a GPD and two DAs. A careful treatment of the phase space, including subtle issues related to using kinematical approximations in the calculation of the hard part is presented in \SEC\ref{sec:phase-space}. To illustrate the results, we produce plots of the cross section for some of the processes in \SEC\ref{sec:results}. Finally, we end with conclusions and outlook in \SEC\ref{sec:conclusions}, followed by appendices.
Further details of the kinematics at different levels of approximation are given in \APP\ref{app:further-kinematics}.
\APP\ref{app:GPD-modelling} deals with the details of the GPD modelling.
In \APP\ref{app:gluon-GPDs}, we demonstrate that assuming collinear factorisation for di-meson photoproduction processes which can exchange two gluons with the nucleon (e.g.~$\rho^0\pi^0$) lead to a purely imaginary divergent amplitude.
In \APP\ref{app:absence-VTT-structure}, we explain using general arguments, why for processes involving the production of two transversely polarised $\rho$ mesons in specific gauge fixing choices, some tensor structures made up of the polarisation vectors of the incoming photon and the two outgoing mesons do not appear at the amplitude level, even at individual diagram level.
\APP\ref{app:ERBL-poles} explains why the poles of propagator denominators can only be crossed in the convolution integral in \EQ\eqref{eq:amplitude} when in the ERBL region (i.e.~when $|x|<\xi$).
Various symmetries of the amplitude are discussed in \APP\ref{app:symmetries}.
Finally, in \APP\ref{app:crosssectionproof}, we derive the expression for the cross section, showing the different approximations (implied by collinear factorisation) at each stage.

\section{Kinematics}

\label{sec:kinematics}

The process we consider in this paper is
\begin{align}
    \gamma(q) + N(p_N,\lambda) \to N'(p_{N'},\lambda')+ M_1(p_{M_1}) + M_2(p_{M_2})\,,
\end{align}
where $N$ and $N'$ represent two nucleons (which may or may not be the same), and $M_1$ and $M_2$ represent two generic light mesons. $\lambda$ ($\lambda'$) corresponds to the helicity of the incoming (outgoing) nucleon.

For convenience, each momentum is decomposed in the Sudakov basis using lightcone coordinates. We follow the same convention as in \cite{Duplancic:2022ffo,Duplancic:2023kwe}. To this end, we define two (dimensionful) lightcone vectors $p$ and $n$, which in Cartesian coordinates read
\begin{align}
\label{eq:lightcone-basis}
    p^\mu = \frac{\sqrt{s}}{2}(1,0,0,1)\,,\qquad n^\mu= \frac{\sqrt{s}}{2}(1,0,0,-1)\,,\qquad p \cdot  n = \frac{s}{2}\,,
\end{align}
where $\sqrt{s}$ has dimensions of mass, and is related to the centre of mass energy $\sqrt{\SgN}$ of the incoming photon-nucleon system. We will usually refer to ``plus'' (``minus'') momentum component as the one corresponding to the $p^\mu$ ($n^\mu$) component. For convenience, we define also the dimensionless lightlike vectors $\hat{p}$ and $\hat{n}$ such that $\hat{p}^\mu=\sqrt{\frac{2}{s}}\,p^\mu$ and $\hat{n}^\mu=\sqrt{\frac{2}{s}}\,n^\mu$ with $\hat{p}\cdot \hat{n}=1$.
Then, any generic 4-momentum $k$ can be represented as
\begin{align}
    k^\mu = k^- \hat{n}^\mu + k^+ \hat{p}^\mu + k_{\perp}^\mu\,,
\end{align}
where $k^-=k \cdot \hat{p}$ and $k^+=k \cdot \hat{n}$.
 Choosing the $z$-axis to be aligned with the collision axis, one can parametrise the kinematics as 
\begin{align}
    q^\mu &= n^\mu\,,\\
    p_{N}^\mu & = (1+\xi)p^\mu+\frac{M^2}{s(1+\xi)}n^\mu\,,\\
    p_{N'}^\mu &=(1-\xi)p^\mu+\frac{M^2-\Delta_{\perp}^2}{s(1-\xi)}n^\mu+\Delta_{\perp}^\mu\,,\\
    \label{eq:pM1}
    p_{M_1}^\mu &= \alpha_{M_1} n^\mu +\frac{m_{M_1}^2-(p_{\perp}+\Delta_{\perp}/2)^2}{\alpha_{M_1}s}p^\mu - p_{\perp}^\mu - \frac{\Delta^\mu_{\perp}}{2}\,,\\
    \label{eq:pM2}
    p_{M_2}^\mu &
   = \alpha_{M_2} n^\mu +\frac{m_{M_2}^2-(p_{\perp}-\Delta_{\perp}/2)^2}{\alpha_{M_2}s}p^\mu + p_{\perp}^\mu -\frac{\Delta^\mu_{\perp}}{2}\,.
\end{align}
In the above, $M$ represents the mass of the nucleon, assuming that it is the same for proton and neutron, while $m_{M_1}$ and $m_{M_2}$ correspond to the mass of the light mesons $M_1$ and $M_2$ respectively. We adopt the following convention for transverse component of a generic 4-momentum $k$: $k_\perp$ is the Minkowski 4-vector that has non-zero components in the transverse direction only, while $\vec{k}_t$ corresponds to the Euclidean 2D vector in the transverse plane. Furthermore, we often use the shorthand $k_t=|\vec{k}_t|$.

\begin{figure}
     \centering
\includegraphics[width=0.4\linewidth]{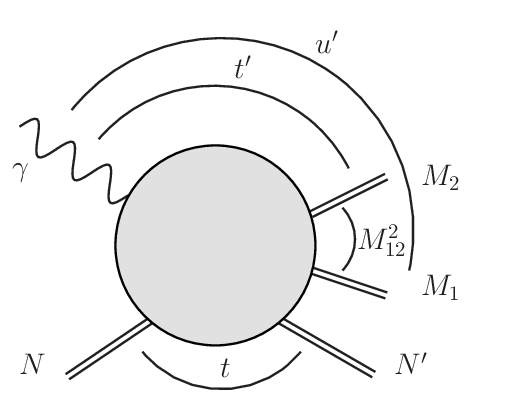}
     \caption{Definition of the kinematical variables. }
     \label{fig:schemadimeson}
 \end{figure}

We further define
\begin{align}
    P^\mu = \frac{p_N^\mu + p_{N'}^\mu}{2}\,,\qquad \Delta^\mu = p_{N'}^\mu - p_{N}^\mu\,.
\end{align}
The skewness parameter $\xi$, given by the fraction of plus momentum component transferred by the nucleon system, can be written as
\begin{align}
    \xi = -  \frac{\Delta^+}{2P^+}\,.
\end{align}
The squared centre of mass energy $\SgN$ of the incoming photon-nucleon system is given by
\begin{align}
\label{eq:Sgformula}
    \SgN = s(1+\xi)+ M^2\,,
\end{align}
while the Mandelstam variable $t$ reads
\begin{align}
    t &= (p_{N'}-p_{N})^2 = \Delta^2 = \frac{1+\xi}{1-\xi}\Delta_{\perp}^2-\frac{4\xi^2M^2}{(1-\xi^2)}\,.
    \label{eq:tformula}
\end{align}
For convenience, we also define the corresponding Mandelstam variables (denoted by primed) for the two-body subprocess
\begin{align}
\label{eq:2-body-subprocess}
    \gamma(q) + q\bar q(-\Delta) \to q \bar q(p_{M_2}) + q\bar q(p_{M_1})\,,
\end{align}
where
\begin{align}
\label{eq:uprimed}
    u' & = (q-p_{M_1})^2\,, \\
     t' &=(q-p_{M_2})^2\,,\\
    s' & =  (p_{M_1}+p_{M_2})^2 \equiv M_{12}^2 \,,
    \label{eq:sprimed}
\end{align}
 which satisfy the relation
 \begin{align}
 \label{eq:Mandelstamsum}
     s'+t'+u' = \Delta^2 + m_{M_1}^2 + m_{M_2}^2\,.
 \end{align}
The kinematical variables are illustrated in \FIG\ref{fig:schemadimeson}.
 
For collinear factorisation of the process to be valid, one requires a hard scale. This is provided by the large invariant mass $M^2_{12}$ of the di-meson pair in the final-state. In addition to this, analogous to wide-angle scattering \cite{Lepage:1979zb}, one also requires that $-u'$ and $-t'$ be large, while $-t$ should be small. It has been argued in \cite{Qiu:2022bpq,Qiu:2022pla} that a collinear factorisation of the process is applicable provided $p_t/\sqrt{-t}$ is large. It can be shown that this specific condition is parametrically equivalent to the above-mentioned set of cuts imposed on the Mandelstam variables $M^2_{12}$, $-t'$, $-u'$ and $-t$. An explicit illustration of the numerical equivalence between these two approaches is given in \APP B of Ref.~\cite{Duplancic:2022ffo}.

In practice, we use the following cuts:
\begin{equation}
    -u'>(-u')_{\mathrm{min}},\;\;\;-t'>(-t')_{\mathrm{min}},\;\;\;-t<(-t)_{\mathrm{max}}\,,
\end{equation}
with $(-u')_{\mathrm{min}}=(-t')_{\mathrm{min}}=\SI{1}{GeV^2}$ and $(-t)_{\mathrm{max}}=\SI{0.5}{GeV^2}$.

In the limit where the deflection of the nucleon is negligible ($\Delta_{t}$ small compared to $p_t$), and if we neglect all hadron masses, one obtains, through momentum conservation,
\begin{align}
\label{eq:pperpformula}
     \alpha_{M_1} + \alpha_{M_2}&=1\,,\\
    p_t^2 &= 2\xi \alpha_{M_1}\alpha_{M_2}s\,.
\end{align}
For notational convenience, we denote $\alpha \equiv \alpha_{M_2}$, which leads to the following expressions for the remaining kinematical variables
\begin{equation}
   M_{12}^2 \approx \frac{p_t^2}{\alpha \bar\alpha}\,,\qquad t' \approx -\bar\alpha M^2_{12}\,,\qquad u' \approx -\alpha M^2_{12}\,,\qquad \xi = \frac{M^2_{12}}{2\SgN - M_{12}^2}\,,
   \label{eq:approxmandelstam}
\end{equation}
where we have used the shorthand $\bar\alpha \equiv 1-\alpha$. Since $M_{12}^2>0$ and $t',u'<0$, this  implies that $0<\alpha<1$. Imposing large $M_{12}^2$, $-u'$ and $-t'$, implies that $p_t$ is large, and that $\alpha$ and $\bar \alpha$ are both not close to zero.

 In this kinematics, which we refer to as \textit{approximated kinematics} throughout the paper, the particle momenta are
\begin{equation}
    p_N^{\mu}=(1+\xi)\,p^{\mu},\quad\; p_{N'}^{\mu}=(1-\xi)\,p^{\mu}\,,\quad\;
      p_{M_1}^{\mu}=\bar \alpha \, q^{\mu}+2\alpha \xi \,p^{\mu}-p^{\mu}_\perp\,,\quad\;
    p_{M_2}^{\mu}=\alpha\, q^{\mu} + 2\bar \alpha  \xi \,p^{\mu}+p^{\mu}_\perp\,.
    \label{eq:approxkinematics}
\end{equation}
From the above, we find that any point in the phase space can be specified through the value of $\SgN$, and the two dimensionless variables $\alpha$ and $\xi$. Indeed, using \EQ\eqref{eq:approxmandelstam} from right to left, fixing $\xi, S_{\gamma N}$ implies that $M_{12}^2$ is fixed. Then, fixing $\alpha$ automatically fixes $u', t'$ and $p_t^2$. We will exploit this feature later when the phase space sampling of the amplitude is performed.

Different levels of approximation on the kinematics is described in \APP\ref{app:further-kinematics}.

The   polarisation vectors of the outgoing mesons have to satisfy
\begin{align}
    \epsilon_{M_{i},\,\lambda}\cdot p_{M_i}&=0\,,\\
    \sum_{\lambda=L,T} \epsilon_{M_i,\,\lambda}^{\mu\,*} \epsilon_{M_i,\,\lambda}^\nu &= -g^{\mu\nu}+\frac{p_{M_i}^\mu p_{M_i}^\nu }{m_{M_i}^2}\,,\\
    \epsilon_{M_{i},\,\lambda_j}^* \cdot \epsilon_{M_{i},\,\lambda_k} &=-\delta_{\lambda_j \lambda_k}\,,
\end{align}
where $L$ and $T$ stand for  longitudinal and transverse polarisation respectively.
The polarisation vectors can be parametrised as
\begin{align}
\label{eq:longpolparam}
\epsilon^{\mu}_{M_{i}}(p_{M_{i}},L) &=\frac{1}{m_{M_{i}}}p^\mu_{M_{i}}-\frac{m_{M_{i}}}{p \cdot p_{M_{i}}}p^\mu\,,\\
    \epsilon^{\mu}_{M_{i}}(p_{M_{i}},T) &= \epsilon_{M_{i}\perp}^{\mu} - \frac{\epsilon_{M_{i}\perp}\cdot p_{M_{i}}}{p \cdot p_{M_{i}}}p^{\mu}\,.\label{eq:transpolparam}
\end{align}
The photon polarisation vector is given by
\begin{align}
\label{eq:photon-pol-vector}
    \epsilon_q^\mu=\epsilon_{q\perp}^\mu\,,
\end{align}
where we have used the axial gauge $\epsilon_q \cdot p = 0$.
As usual, the normalisation of polarisation vectors satisfy  $\epsilon_{M_i}^2=\epsilon_{M_i\perp}^2=\epsilon_{q\perp}^2=-1$. Throughout the paper, whenever we simply write $\epsilon_{M_i}^\mu$, we always imply the transverse polarisation $\epsilon^{\mu}_{M_{i}}(p_{M_{i}},T)$, although this can of course be inferred from the context.

\section{Non-perturbative inputs}

\label{sec:non-perturbative-inputs}

\subsection{Generalised parton distributions}
 The scattering amplitude involves two kinds of GPDs, called chiral-even and chiral-odd. At leading twist, there are two chiral-even vector GPDs, denoted by $H^q, E^q$, and two chiral-even axial GPDs, denoted by $\Tilde{H}^q,\Tilde{E^q}$, with $q$ being the flavour of the active parton. They appear from non-forward matrix elements of quark-anti-quark operators separated by a lightlike distance \cite{Diehl:2003ny}
 \begin{align}
&\langle N'(p_{N'},\lambda')| \bar{q}\left(-\frac{z}{2}\right)\gamma^+q\left(\frac{z}{2}\right)|N(p_N,\lambda)\rangle \Big{|}_{\substack{z_\perp=0\\z^+=0}} &\nonumber\\
&=\int_{-1}^{1}dx\,e^{-i xP^+ z^-}\bar{u}(p_{N'},\lambda')\bigg[\gamma^+\,H^q(x,\xi,t)
+\frac{i}{2M}\sigma^{+\alpha}\Delta_\alpha E^q(x,\xi,t)\bigg]u(p_N,\lambda)\,,&\\
&\langle N'(p_{N'},\lambda')| \bar{q}\left(-\frac{z}{2}\right)\gamma^+\gamma^5q\left(\frac{z}{2}\right)|N(p_N,\lambda)\rangle \Big{|}_{\substack{z_\perp=0\\z^+=0}}& \nonumber\\
&=\int_{-1}^{1}dx\,e^{-i xP^+ z^-}\bar{u}(p_{N'},\lambda')\bigg[\gamma^+\gamma^5\,\Tilde{H}^q(x,\xi,t)
+\frac{1}{2M}\gamma^5\Delta^+\Tilde{E}^q(x,\xi,t)\bigg]u(p_N,\lambda)\,.&
\end{align}

On the other hand, chiral-odd (helicity-flip) GPDs describe a transfer of one unit of helicity in the $t$-channel. At leading twist, there are four of them, denoted by $H_T^q,\,\Tilde{H}_T^q,\,E_T^q$ and $\Tilde{E}_T^q$. They are defined through the following matrix element,
\begin{align}
\label{eq:transversity-GPD}
  \langle N'(p_{N'},\lambda')| \bar{q}\left(-\frac{z}{2}\right)&i\sigma^{+j}q\left(\frac{z}{2}\right)|N(p_N,\lambda)\rangle \Big{|}_{\substack{z_\perp=0\\z^+=0}}= \int_{-1}^{1}dx\,e^{-i xP^+ z^-}\bar{u}(p_{N'},\lambda')\bigg[i\sigma^{+j}H_T^q\nonumber\\
  &+\frac{P^+\Delta^j-\Delta^+\gamma^j}{M^2}\Tilde{H}^q_T+\frac{\gamma^+\Delta^j-\Delta^+\gamma^j}{2M}E^q_T+\frac{\gamma^+P^j-P^+\gamma^j}{M}\Tilde{E}^q_T\bigg]u(p_N,\lambda)\,,
\end{align}
with $\sigma^{\mu\nu}\equiv\frac{i}{2}[\gamma^\mu,\gamma^\nu]$.

It is also convenient to use a definition involving $\sigma^{+j}\gamma^5$ \cite{Diehl:2001pm},
\begin{align}
\label{eq:transversity-GPD-alt}
  \langle N'(p_{N'},\lambda')| \bar{q}\left(-\frac{z}{2}\right)&\sigma^{+j}\gamma^5 q\left(\frac{z}{2}\right)|N(p_N,\lambda)\rangle \Big{|}_{\substack{z_\perp=0\\z^+=0}}= \int_{-1}^{1}dx\,e^{-i xP^+ z^-}\bar{u}(p_{N'},\lambda')\bigg[\sigma^{+j}\gamma^5 H_T^q\nonumber\\
  &+\frac{\epsilon^{+j\alpha\beta}\Delta_\alpha P_\beta}{M^2}\Tilde{H}^q_T+\frac{\epsilon^{+j\alpha\beta}\Delta_\alpha\gamma_\beta}{2M}E^q_T+\frac{\epsilon^{+j\alpha\beta}P_\alpha\gamma_\beta}{M}\Tilde{E}^q_T\bigg]u(p_N,\lambda)\,,
\end{align}

In our study, we focus only on the production of $\pi$ and $\rho$ mesons. Since they can be charged, it is also necessary to work with transition GPDs. By isospin symmetry, they can be related to the standard proton GPD, as follows \cite{Mankiewicz:1997aa}
\begin{align}
    \langle n | \bar d \Gamma u | p \rangle = \langle p | \bar u \Gamma d | n\rangle = \langle p | \bar u \Gamma u | p \rangle - \langle p | \bar d \Gamma d | p \rangle\,,
\end{align}
where $\Gamma$ can be $ \gamma^+,\, \gamma^+ \gamma^5$ or $i \sigma^{+j}$.

The models of GPDs used in our calculation are discussed in \APP\ref{app:GPD-modelling}.

\subsection{Distribution amplitudes}

\label{sec:DAs}

 For pions, the twist 2 Distribution Amplitude (DA) reads
\begin{equation}
    \langle\pi^i(p_\pi)|\bar{q}(-y)T^i\gamma^\mu\gamma^5 q(y)|0\rangle=i p_\pi^\mu f_\pi\int_0^1dz\, e^{-i(2z-1)p_\pi \cdot y}\phi_\pi(z)\,,
\end{equation}
where $f_{\pi}=\SI{131}{MeV}$ is the pion decay constant \cite{Ball:1998je}, $i=0,\pm$ corresponds to the electric charge of the pion, and $y^\mu$ is a lightlike direction which is conjugate to $p_\pi^\mu$, similar to how $ n^\mu$ and $ p^\mu$ are defined in \EQ\eqref{eq:lightcone-basis}.
$q=(u\; d)$ is a two-dimensional vector in flavour space  and the matrices $T^i$ are defined by
\begin{equation}
    T^0=\frac{1}{\sqrt{2}}
    \begin{pmatrix}
        1 & 0\\
        0 & -1
    \end{pmatrix},\;\;T^+=\begin{pmatrix}
        0 & 0\\
        1 & 0
    \end{pmatrix},\;\;
    T^-=\begin{pmatrix}
        0 & 1\\
        0 & 0
    \end{pmatrix}.
\end{equation}
To be explicit, as an example, consider the case of $\pi^0$. There, $| \pi^0\rangle = \frac{1}{\sqrt{2}}(|\bar u u\rangle - |\bar d d \rangle)$, so 
 \begin{align}
 \langle\pi^0(p_\pi)|\bar{q}(-y)T^0\gamma^\mu\gamma^5 q(y)|0\rangle &= \frac{1}{\sqrt{2}}\left(\langle \pi^0(p_\pi)|\bar{u}(-y)\gamma^\mu\gamma^5 u(y)|0\rangle - \langle \pi^0(p_\pi)|\bar{d}(-y)\gamma^\mu\gamma^5 d(y)|0\rangle\right)\nonumber\\
 &=\left(\frac{1}{\sqrt{2}}\right)^2\left(\langle \bar u u|\bar{u}(-y)\gamma^\mu\gamma^5 u(y)|0\rangle + \langle \bar d d|\bar{d}(-y)\gamma^\mu\gamma^5 d(y)|0\rangle\right)\nonumber\\
 &= i p_\pi^\mu f_\pi\int_0^1dz\, e^{-i(2z-1)p_\pi \cdot y}\phi_\pi(z)\,.
 \end{align}
For a longitudinally polarised rho meson, we have
\begin{equation}
    \langle\rho_L^i(p_\rho,\epsilon_\rho)|\bar{q}(-y)T^i\gamma^\mu q(y)|0\rangle= p_\rho^\mu f_\rho^{||}\int_0^1dz\, e^{-i(2z-1)p_\rho \cdot y}\phi_{\rho||}(z)\,,
\end{equation}
while for a transversely polarised one,
\begin{equation}
\label{eq:tensor-DA}
    \langle\rho_T^i(p_\rho,\epsilon_\rho)|\bar{q}(-y)T^i\sigma^{\mu\nu}q(y)|0\rangle= i (\epsilon_\rho^\mu p_\rho^\nu -\epsilon_\rho^\nu p_\rho^\mu) \, f_\rho^{\perp}\int_0^1dz\, e^{-i(2z-1)p_\rho \cdot y}\phi_{\rho\perp}(z)\,,
\end{equation}
where $f_{\rho}^\perp=\SI{160}{MeV}$ and $f_{\rho}^{||}=\SI{216}{MeV}$ are the $\rho$ meson decay constants \cite{Ball:1996tb}. For the DAs, we use the simple asymptotic form, given by
\begin{align}
\label{eq:asDA}
\phi_{\pi}(z)=\phi_{\rho ||}(z)=    \phi_{\rho \perp}(z)= 6 z (1-z)\,.
\end{align}

\section{Construction of diagrams}

\label{sec:construction-of-diagrams}

\subsection{Fierz projection within collinear factorisation}

\label{sec:fierz-projection}

\begin{figure}[t!]
    \centering
    \includegraphics[width=0.5\linewidth]{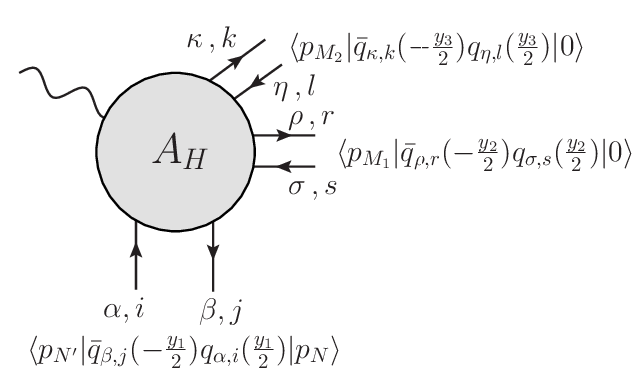}
    \caption{A schematic showing the amplitude of the process under consideration. $A_H$ corresponds to the sum of perturbatively calculable Feynman diagrams with open quark lines as external states. The external quark legs are embedded within non-perturbative correlators. The Greek (Roman) letters correspond to spinor (colour) indices.}
    \label{fig:diagbeforeFierz}
\end{figure}

The starting point towards the expression in \EQ\eqref{eq:amplitude} is diagrams of the form of those in \FIG\ref{fig:diagbeforeFierz}. $A_H$ denotes the perturbative hard part of the diagram with open quark lines as external states. The outgoing pairs of quarks represent, for now, non-perturbative correlators describing the overlap of these quark pairs with the asymptotic confined states.  

\begin{figure}[t!]
    \centering
    \includegraphics[width=0.6\linewidth]{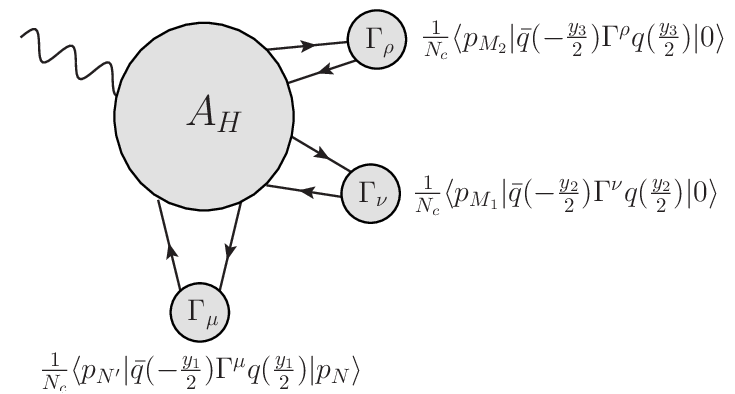}
    \caption{After performing the Fierz projections (one in colour space and the other in spinor space), the diagram becomes a product of colour and spinor traces. The $\Gamma$ matrices correspond to the relevant combination of Dirac matrices in \EQ\eqref{eq:Fierzspinor}.}
    \label{fig:diagsafterFierz}
\end{figure}

These correlators can be reexpressed through the two following Fierz identities, with $|1\rangle$ and $|2\rangle$ being generic colourless states (including nucleon, meson or vacuum),\footnote{Note that if one replaces the two $\sigma_{\mu \nu}$ in the last term by $\sigma_{\mu \nu} \gamma^5$, the prefactor stays the same (i.e. 1/8).}
\begin{align}
\nonumber
&\langle 2|\,\bar{q}_{{\alpha}}q_{{\beta}}|1\rangle
 =\frac{1}{4}{{\left(\gamma_\mu\right)}_{\beta\alpha}}\langle 2|\bar{q}\gamma^\mu q|1\rangle
 +\frac{1}{4}{(\gamma^5{\gamma_\mu})_{\beta\alpha}}\langle 2|\bar{q}\gamma^{\mu} \gamma^5 q|1\rangle\\ 
 \label{eq:Fierzspinor}
 &+\frac{1}{4}{\left(\mathbb{1}\right)_{\beta\alpha}}\langle 2|\bar{q}q|1\rangle
 +\frac{1}{4}{\left(\gamma^5\right)_{\beta\alpha}}\langle 2|\bar{q}\gamma^5 q|1\rangle
 +\frac{1}{8}{{\left(\sigma_{\mu\nu}\right)}_{\beta\alpha}}\langle 2|\bar{q}\,\sigma^{\mu\nu}q|1\rangle\,,
\end{align}
for the spinor space, where we have suppressed the spacetime arguments of the quark-antiquark fields on the RHS for conciseness. On the other hand, the Fierz decomposition in colour space reads:
 \begin{equation}
   \langle 2|\bar{q}_{\boldmath{{i}}}\Gamma^\alpha q_{\boldmath{{j}}}|1\rangle=\frac{1}{N_c}\langle 2|\bar{q}\Gamma^\alpha q|1\rangle \delta_{\boldmath{{ij}}}+2 t_{\boldmath{{ij}}}^a\overbrace{\langle 2|\bar{q}\, (t^a)^t\Gamma^\alpha q|1\rangle}^{=0}\,,
   \label{eq:Fierzcolour}
 \end{equation}
 where $\Gamma^\alpha$ is a general Dirac matrix, and the second term in \EQ\eqref{eq:Fierzcolour} is zero since it is a matrix element of a colour operator with colourless states.  This formula causes the diagrams of \FIG\ref{fig:diagbeforeFierz} to ``close'' in two traces, one over the spinor indices and the other over the colour indices. This is illustrated in \FIG\ref{fig:diagsafterFierz}. The number of traces we obtain is linked with the topology of the diagram as we will see later. 

 Not all of the Dirac matrices in \EQ\eqref{eq:Fierzspinor} will enter the diagrams since the calculation is performed at leading twist. In the case of the nucleon, only the vector, axial vector and tensor structures remain. In the case of the meson, only one of these structures contributes, depending on the quantum numbers of the meson, see \SEC\ref{sec:DAs}.

For example, the process $\gamma \,p\to p\,\rho^0_L\rho^0_L$ involves both vector and axial Dirac matrices from the nucleon side, while  $\gamma \,p\to n\,\pi^+\rho^0_T$ involves only the tensor Dirac matrix ($\sigma^{\mu\nu}$). 
 
\begin{figure}[t!]
    \centering
    \includegraphics[width=0.5\linewidth,clip=]{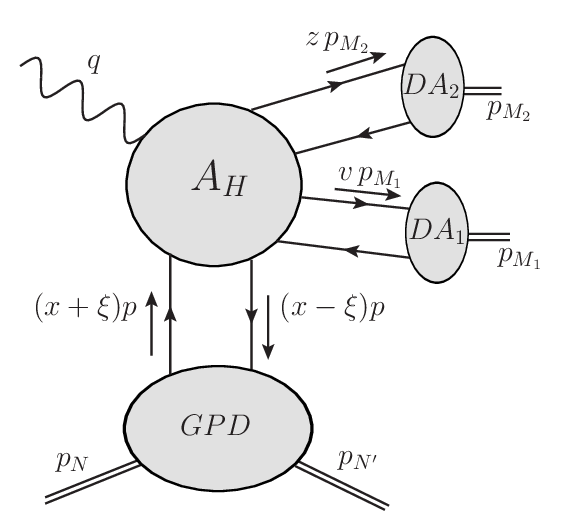}
    \caption{The structure of the amplitude, with the momenta of the external quark lines to the hard subprocess specified.}
    \label{fig:illustration}
\end{figure}

Each pair of quarks entering a meson has a momentum proportional to the momentum of the meson. We denote by $v$ and $z$ the momentum fractions carried by the quark from each pair that enters mesons $M_1$ and $M_2$ respectively, where $v,z=[0,1]$. The same is true for the quark pair attached to the nucleon, but their momenta are proportional to $\Delta$ in this case. Since $\Delta=-2\xi\,p$ in the approximated kinematics, both quarks fly along the $+$ direction. Unlike the quarks of the mesons, the quark and antiquark attached to the nucleon can be either emitted or absorbed by the nucleon. Therefore, we parametrise the momentum of the outgoing quark line from the nucleon by $(x+\xi)\,p$, and the incoming quark line to the nucleon by $(x-\xi)\,p$, with $x = [-1,1]$. This is illustrated in \FIG\ref{fig:illustration}. 

Since the quark momenta attached to a specific non-perturbative part flow in the same direction, the corresponding hadronic current has to be proportional to a momentum in that direction. For instance, for the nucleon matrix element and its vector component in the Fierz decomposition of \EQ\eqref{eq:Fierzspinor}, it means that
\begin{equation}
    \langle p_{N'}|\bar{q}\,\gamma^\mu q|p_N\rangle=\langle p_{N'}|\bar{q}\,\gamma^+ q|p_N\rangle \,\hat{p}^\mu=\langle p_{N'}|\bar{q}\,\slashed{\hat {n}} \,q|p_N\rangle \,\hat{p}^\mu.
\end{equation}
The same reasoning applies to the DAs, except that instead of the + direction, we obtain the direction of the meson. Thus, we find that collinear factorisation at leading twist further restricts which Dirac matrices enter the calculation.

Based on the above considerations, the amplitude takes the form given in \EQ\eqref{eq:amplitude} where $T_H$ includes the factor $\frac{1}{N_c^3}$ as well as the Dirac matrices arising from the Fierz projections. We stress the point that several GPDs of different chirality or flavour could contribute, depending on the valence quark decomposition of the mesons and their spins. As will be shown later, all GPDs except $H$, $\Tilde{H}$ and $H_T$ are suppressed by powers of $\xi$ in the amplitude squared. Since $\xi$ is small (typically between $10^{-5}$ and $10^{-1}$), we will keep only the GPDs $H$, $\Tilde{H}$ and $H_T$ in the computation.

\subsection{Generation and organisation of diagrams}

\begin{figure}[t!]
    \begin{minipage}{0.4\textwidth}
\includegraphics[width=\linewidth]{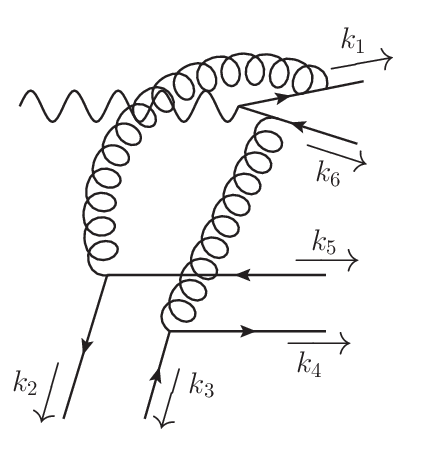}
    \end{minipage}  
    \hfill\Huge{$\Rightarrow$}\hfill
    \begin{minipage}{0.5\textwidth}
\includegraphics[width=\linewidth]{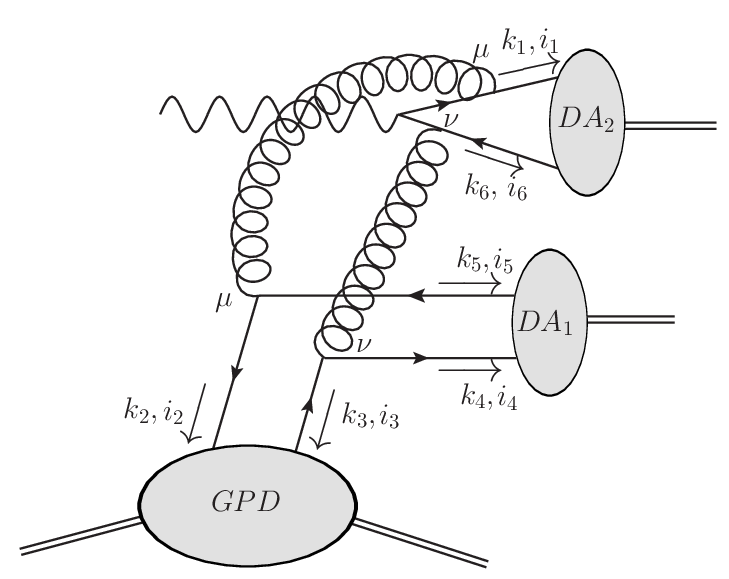}
    \end{minipage}
\caption{The projection of a diagram obtained with FeynArts and FeynCalc onto the GPD and DAs.}
    \label{fig:quark-level-process}
\end{figure}

In the collinear factorisation theorem, only connected diagrams enter the hard part. We first generate all connected $\gamma\rightarrow f\bar{f} f\bar{f}f\bar{f}$ processes, where $f$ is a quark of a specific flavour (e.g.~$f$ can be the $u$ quark), by using the FeynArts package \cite{Hahn:2000kx} in Mathematica. The diagrams are converted to expressions using FeynCalc \cite{Mertig:1990an,Shtabovenko:2016sxi,Shtabovenko:2020gxv,Shtabovenko:2023idz}.  By having only quarks $f$ of a single flavour, we are able to cover all diagrams that could enter our processes.  The correct charges of the fermionic lines are restored afterwards. There are 288 diagrams corresponding to this $1\rightarrow 6$ process. The notation of momenta follows \FIG\ref{fig:quark-level-process}.

The Fierz projection in colour space is immediately performed by setting the quark colour indices $i_2=i_3$, $i_4=i_5$, and $i_6=i_1$ and by dividing the amplitude by $N_c^3$. For the Fierz projection in spinor space, we take the chain of Dirac matrices of each fermionic line, and depending on the topology, we put them into one or two traces between the Dirac matrices associated to the GPDs and DAs, which we will denote $GPD$, $DA_1$ and $DA_2$, where the subscripts 1 and 2 correspond to the mesons $M_1$ and $M_2$. Concretely, for a vector GPD and vector DAs, $GPD=\slashed{p}$ , $DA_1=\slashed{p}_{M_1}$ and $DA_2=\slashed{p}_{M_2}$.

For instance, the diagram on the left of \FIG\ref{fig:quark-level-process} is transformed into the diagram on the right as follows
\begin{equation}
t^a_{i_2i_5}t^b_{i_4i_3}t^a_{i_1 i}t^b_{i i_6}\frac{{\bar{u}(k_1)}\boxed{\gamma_{\mu}\,(\slashed{k_1}+\slashed{k_2}+\slashed{k_5})\,\slashed{\epsilon}_q\,(\slashed{k_3}+\slashed{k_4}+\slashed{k_6})\,\gamma_\nu}{{v(k_6)\bar{u} (k_2)}} \boxed{\gamma^\mu}{v (k_5)\bar{u} (k_4)}{\boxed{{\gamma}^\nu}}{v (k_3)}}{(k_2+k_5)^2(k_3+k_4)^2(k_1+k_2+k_5)^2(k_3+k_4+k_6)^2}\nonumber
  \end{equation}
 \raisebox{-.5cm}{\makebox[\linewidth][l]{\hspace{0.5\linewidth}\scalebox{1.8}{$\Downarrow$}}}
\begin{equation}
\label{eq:hard-part}
\frac{\mathrm{Tr}(t^at^b)\mathrm{Tr}(t^at^b)}{(4N_c)^3}\frac{\mathrm{Tr}\left(\boxed{{\gamma}_{\mu}\,(\slashed{k_1}+\slashed{k_2}+\slashed{k_5})\,\slashed{\epsilon}_q\,(\slashed{k_3}+\slashed{k_4}+\slashed{k_6})\,\gamma_\nu}\,{{DA_2}}\right)\mathrm{Tr}\left(\boxed{ \gamma^\mu} \,{GPD}\,\boxed{\gamma^\nu}\,{DA_1}\right)}{(k_2+k_5)^2(k_3+k_4)^2(k_1+k_2+k_5)^2(k_3+k_4+k_6)^2}\,.\end{equation}
Note that the usual $i \epsilon$ Feynman prescription for the propagators is implicit. In the above equation, the boxes refer to the chains of Dirac matrices corresponding to each fermion line.
Each $1\to 6$ diagram given by FeynArts can be projected onto 14 different combinations of Dirac matrices. These are 
\begin{equation}
\label{eq:listcombinations}
    \text{VVV, VAV, VVA, AVV, VAA, AVA, AAV, AAA, TVT, TAT, VTT, ATT, TTV, TTA.}
\end{equation}
For example, VAV means that the GPD is Vector, the $DA_1$ is Axial and the $DA_2$ is Vector.

To obtain these 14 combinations, we used the fact that transverse structures must appear by pairs, otherwise, the Dirac traces would contain an odd number of  $\gamma$ matrices, and therefore would be zero. Let us prove this assertion.

At leading order, the diagrams contain either two (without a triple gluon vertex) or three (with a triple gluon vertex) gluons. In the case of two gluons, the diagram has two quark propagators, one quark-photon vertex and two quark-gluon vertices, that is, seven $\gamma$ matrices. In the other case, there are only five $\gamma$ matrices. Consequently, an odd number of $\gamma$ matrices (other than $\gamma^5$) must be present in the GPD and the DAs for the total number of $\gamma$ matrices to be even. One (three) transverse structure means four (six) $\gamma$ matrices coming from the soft parts, and zero (two) transverse structures lead to three (five) $\gamma$ matrices. Thus, the only way to achieve an even number of $\gamma$ matrices (different from $\gamma^5$) is to have either zero or two transverse structures. If there is an odd number of $\gamma$ matrices in the whole diagram, having more than one trace would still result in at least one trace having an odd number of $\gamma$ matrices, which would cause the whole diagram to vanish. This concludes the proof.\footnote{Note that whether the incoming photon is taken into account or not, the argument does not change. Indeed, adding the photon to any of the quark line generates two $\gamma$ matrices, one associated to the vertex itself and another related to the extra quark propagator that is created. The same argument would also hold for higher order corrections where one would add internal gluons to the basic ``skeleton'' diagrams.} This can be understood by the fact that a tensor Dirac matrix describes a chirality flip of the quarks inside the nucleon or meson, and since QCD and QED are chirality preserving in the massless limit, this chirality flip has to be compensated by another one elsewhere.

\begin{figure}[t!]
    \centering
    \includegraphics[width=\linewidth]{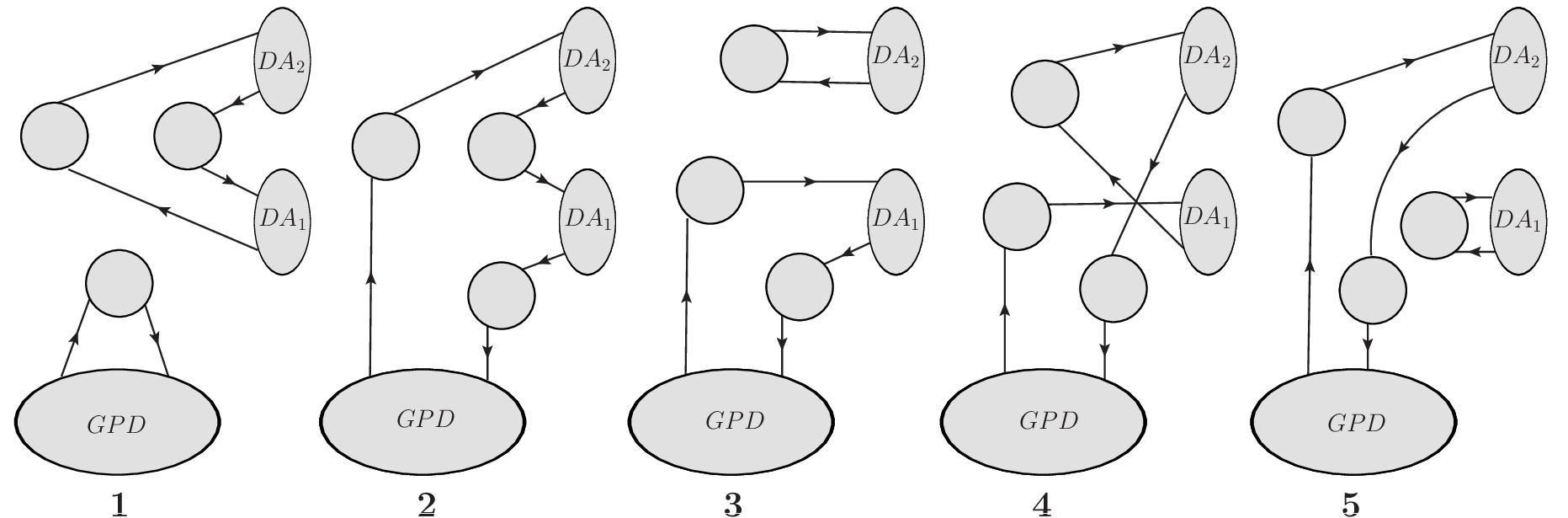}
    \caption{The five ways of connecting the fermion lines, giving five topologies identified by a number. The unmarked blobs correspond to
places where the internal gluons and/or incoming photon could attach to. Each blob may contain up to 3 attachments (at leading order).}
    \label{fig:topologies}
\end{figure}

Combining the 288 diagrams generated by FeynArts with the 14 combinations results  in 1304 \textit{non-zero} projected diagrams covering every kind of di-meson processes. They are classified according to their \textit{topology}, i.e.~the way the fermionic lines are connected. These topologies are represented on \FIG\ref{fig:topologies}. The topology corresponding to three Dirac traces does not appear because of colour conservation.

It is worth noting at this stage that many diagrams cancel for various reasons, which can be:
 \begin{itemize}
     \item Colour conservation: When a colour trace contains a single SU(3) generator in it. 
     \item $\gamma$ counting: Dirac traces with an odd number of $\gamma$ matrices vanish. This is the reason why transversity always has to appear in pairs.
     \item Antisymmetry of the flavour wavefunction $\frac{|u\bar{u}\rangle-|d\bar{d}\rangle}{\sqrt{2}}$, as illustrated in the diagrams in \FIG\ref{fig:flavourwavefunction}.
     \begin{figure}[t!]
         \centering
         \includegraphics[width=0.55\linewidth]{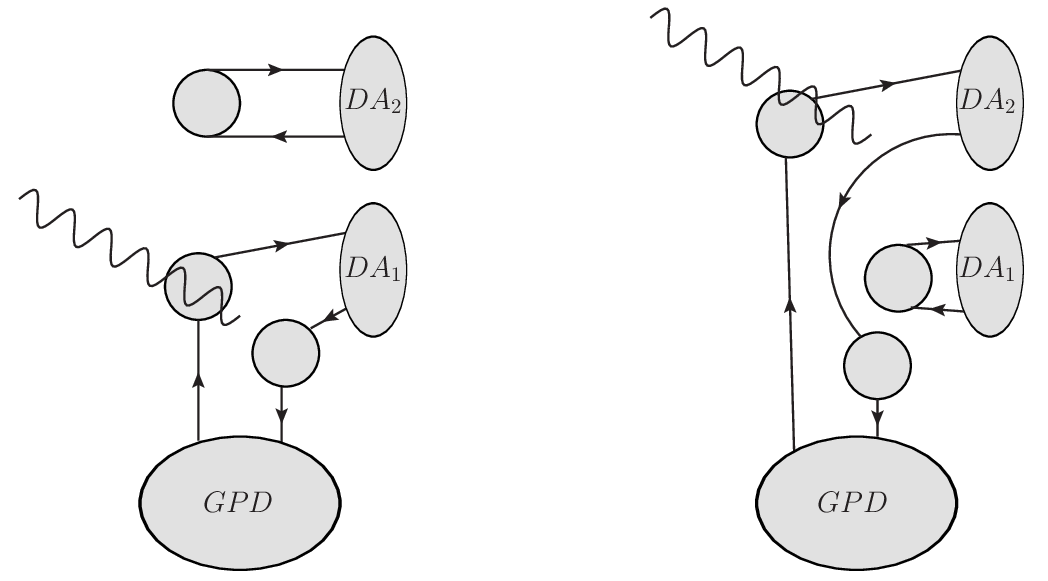}
         \caption{When a DA is traced over itself (i.e.~the same quark line leaves and enters it) and the photon is connected to the other trace, the diagram vanishes by the antisymmetry of the neutral meson wavefunction $\frac{|u\bar{u}\rangle-|d\bar{d}\rangle}{\sqrt{2}}$.  The photon should be understood as being attached to one of the two blobs connecting $GPD$ with $DA_1$ (left diagram) or $GPD$ with $DA_2$ (right diagram).} 
         \label{fig:flavourwavefunction}
     \end{figure}
     \item  Some are antisymmetric in $v\rightarrow \bar{v}$ or $z\rightarrow \bar{z}$ (when using the axial gauge $p \cdot \epsilon_q = 0 $), so they cancel after integration since the DAs $\phi_1(v)$ and $\phi_2(z)$ are symmetric. This is the case of the 4 specific diagrams represented in \FIG\ref{fig:antisymdiags}.
     \begin{figure}[t!]
         \centering
         \includegraphics[width=\linewidth]{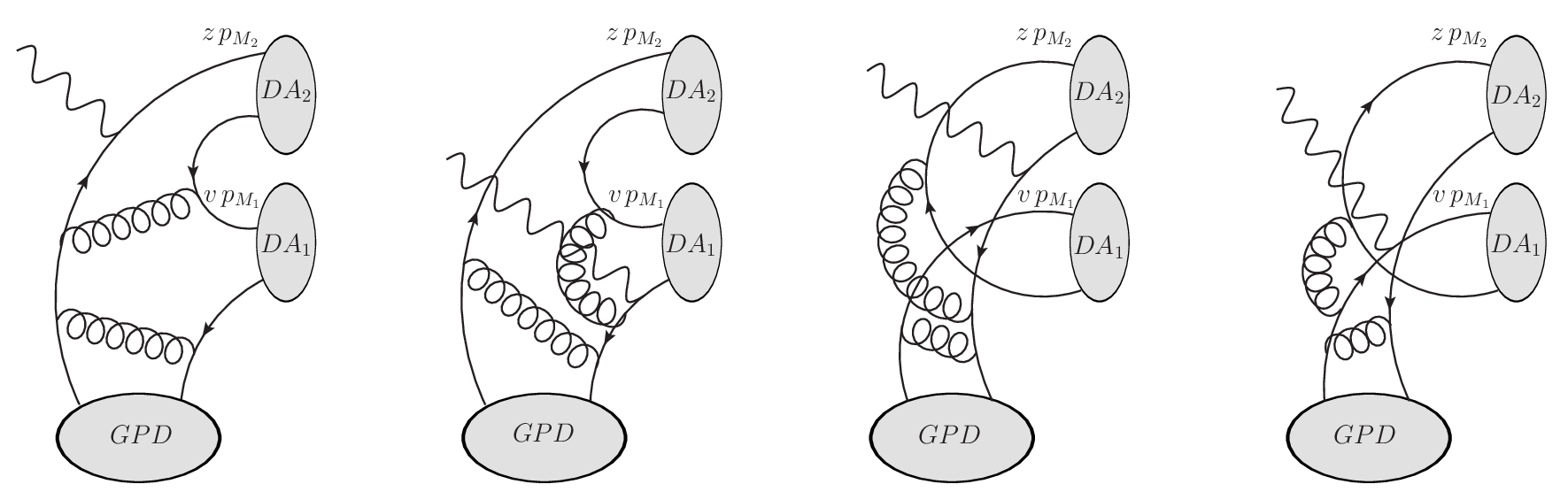}
         \caption{The four diagrams that vanish by antisymmetry in $z \to \bar z$ (first and third diagrams) and $v \to \bar v$ (second and last diagrams) whenever there are zero or two axial Dirac structures among the GPD and the DAs. This antisymmetry is a consequence of the gauge fixing choice for the incoming photon, see \EQ\eqref{eq:photon-pol-vector}. }
         \label{fig:antisymdiags}
     \end{figure}
     \item Others cancel (in the Feynman gauge) whenever the configuration of \FIG\ref{fig:bardiag} is present, where a gluon connects the quark lines coming from a transversity structure, leading to $\gamma^\alpha[\gamma^\mu,\gamma^\nu]\gamma_\alpha=0$.
     \begin{figure}[t!]
         \centering
         \includegraphics[width=0.2\linewidth]{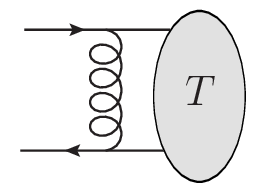}
         \caption{This configuration vanishes in the Feynman gauge, because $\gamma^\alpha[\gamma^\mu,\gamma^\nu]\gamma_\alpha=0$.}
         \label{fig:bardiag}
     \end{figure}
 \end{itemize}

Next, the momenta $k^\mu_i$ are replaced by the kinematics in the massless limit and $\Delta^\mu_\perp=0$, 
\begin{alignat}{2}
   k_1^\mu&=z\,(2\,\bar{\alpha}\,\xi\,p^\mu+\alpha\,q^\mu+p^\mu_\perp),\; \qquad &k^\mu_2&=-(x+\xi)\,p^\mu,\\
   k^\mu_3&=(x-\xi)\,p^\mu,\,& k^\mu_4&=(1-v)(2\,\alpha\,\xi\,p^\mu+\bar{\alpha}\,q^\mu-p^\mu_\perp),\\
   k^\mu_5&=v\,(2\,\alpha\,\xi\,p^\mu+\bar{\alpha}\,q^\mu-p^\mu_\perp),\qquad \;&k^\mu_6&=(1-z)(2\,\bar{\alpha}\,\xi\,p^\mu+\alpha\,q^\mu+p^\mu_\perp)\,.
\end{alignat}
If we denote the resulting term by $T(x,v,z,\alpha,\xi)$,  the contribution of the diagram of \FIG\ref{fig:quark-level-process} to the amplitude, for a vector GPD and vector DAs (VVV), then reads
\begin{equation}
\label{eq:diag255}
    i\mathcal{M}[\{255,\text{VVV}\}]= \frac{1}{p \cdot q}\int_{-1}^1 dx\int_0^1dv\int_0^1dz\,i\mathcal{T}_{j=\{255,\mathrm{VVV}\}}(x,v,z,\alpha,\xi)\bar{u}(\lambda',p_{N'}) \slashed{q} u(\lambda,p_N)\,,
\end{equation}
with\footnote{The minus sign is related to the number of anti-commutation operations that have to be performed on the quark fields. If the diagram consists of two traces, this would be $-1$, and if it has one trace, $+1$. In fact, the sign can be directly obtained by $(-1)^{\sharp\mathrm{topology}}$, where $\sharp\mathrm{topology}$ corresponds to the topology number in \FIG\ref{fig:topologies}.}
\begin{align}
\label{eq:example-amplitude}
 \mathcal{T}_{j=\{255,\mathrm{VVV}\}}(x,v,z,\alpha,\xi)=&-\,e g_s^4f(\{255,1,3,\{1,6\}\},N,N',\{M_1,P_1\},\{M_2,P_2\})\nonumber\\
&\qquad \times T_{j=\{255,\mathrm{VVV}\}}(x,v,z,\alpha,\xi)\,,
\end{align}
and   
\begin{align}
    T_{j=\{255,\mathrm{VVV}\}}(x,v,z,\alpha,\xi)&=\frac{C_F C_A}{{2}(4N_c)^3}\,{p}_{\perp}\cdot {\epsilon }_{q\perp}
    \nonumber\\
    &\times \frac{{8}
   \left(\alpha  \xi  \left(-4 v^2+8 v z-8 z^2+4 z-1\right)+2 \xi  (2 v-1)
   (v-z)+\alpha  x (2 z-1)\right)
   }{ \bar{\alpha}^2 s^2 (v-1) v
   ({{i\epsilon+x}}{{-\xi}} ) ({{-i\epsilon+x}}+{{\xi}} ) (\bar{\alpha}
   v+\alpha  z) (1-\bar{\alpha} v-\alpha  z)}\nonumber\\&
   \times\frac{1}{\left({{\frac{\xi -\xi  v (\alpha -2
   z+1)+(\alpha -2) \xi  z}{-\bar{\alpha}v-\alpha  z+1}}}+{{x+i\epsilon}}\right)\left({{-\frac{\xi  (v (\alpha +2 z-1)-\alpha  z)}{\bar{\alpha} v+\alpha
   z}}}+{{x-\text{$i\epsilon
   $}}}\right)}\,.
   \label{eq:hardpart}
\end{align}
Note that in \EQ\eqref{eq:hard-part}, the factor $(4N_c)^3$ in the denominator comes from the Fierz projections  in  both colour and spinor spaces, while the colour factor is $\frac{1}{2}C_F C_A$.

Each of the 1304 projected diagrams  is stored in a list, together with a scalar function $f$ that contains all the structural information about the diagram.
Here, the function $f(\{255,1,3,\{1,6\}\},N,N', $ $\{M_1,P_1\},\{M_2,P_2\})$ has a list of four arguments as its first entry, which respectively indicates that this amplitude
\begin{itemize}
    \item originates from diagram number 255 out of the 288 generated diagrams,
    \item was  calculated using only vector GPD and DAs (combination number 1, out of 14), see \EQ\eqref{eq:listcombinations},
    \item  belongs to the third topology  (out of 5) shown in \FIG\ref{fig:topologies}, and
    \item the photon is attached to the fermion line connecting $k_1$ to $k_6$.
\end{itemize}
Note that while the information on the topology and where the photon is attached is redundant, since the choice of a diagram fixes them, it is nevertheless useful to keep track of them when performing the calculation.
The variables $N$, $N'$, $M_1$, $M_2$, $P_1$, $P_2$ refer respectively to the incoming and outgoing nucleon species, the first and second type of meson, and their polarisations $P_1$ and $P_2$. Once these are specified, the flag becomes the prefactor of the diagram for the given process. This prefactor includes the electric charge (of the quark line which is connected to the photon), the decay constants of the mesons, and the GPD and DAs. 
For example, for the production of two longitudinal $\rho^0$ mesons from a proton, this prefactor reads
\begin{equation}
f(\{255,1,3,\{1,6\}\},p,p,\{\rho^0,L\},\{\rho^0,L\})=\frac{1}{2}f_{\rho}^2 \,\phi_1(v)\phi_2(z)(Q_u-Q_d)(H^u(x)-H^d(x))\,.
\end{equation}
One recognises the DAs $\phi_1(v)$ and $\phi_2(z)$, the square of the $\rho$ meson decay constant $f_{\rho}$, the quark electric charges $Q_u$ and $Q_d$, and the vector GPDs $H^u(x)$ and $H^d(x)$ corresponding to the quark flavours $u$ and $d$ respectively. Note that the charges factorise from the GPDs because the photon is attached to the fermion line connecting the second meson to itself, and the flavour of the line is independent of the flavour of the fermion lines to which the nucleon is attached. The factor of 1/2 comes from the two neutral meson wavefunctions.

Therefore, if one chooses a specific process, all prefactors are replaced in the list of diagrams. Often, most of them actually end up being zero, since a specific process could be completely incompatible with one or more topologies. For example, the first topology in \FIG\ref{fig:topologies} is incompatible with a transversity GPD (tensor Dirac matrix) due to the presence of an odd number of gamma matrices in one of the traces.

\begin{figure}
    \centering
    \includegraphics[width=0.9\linewidth]{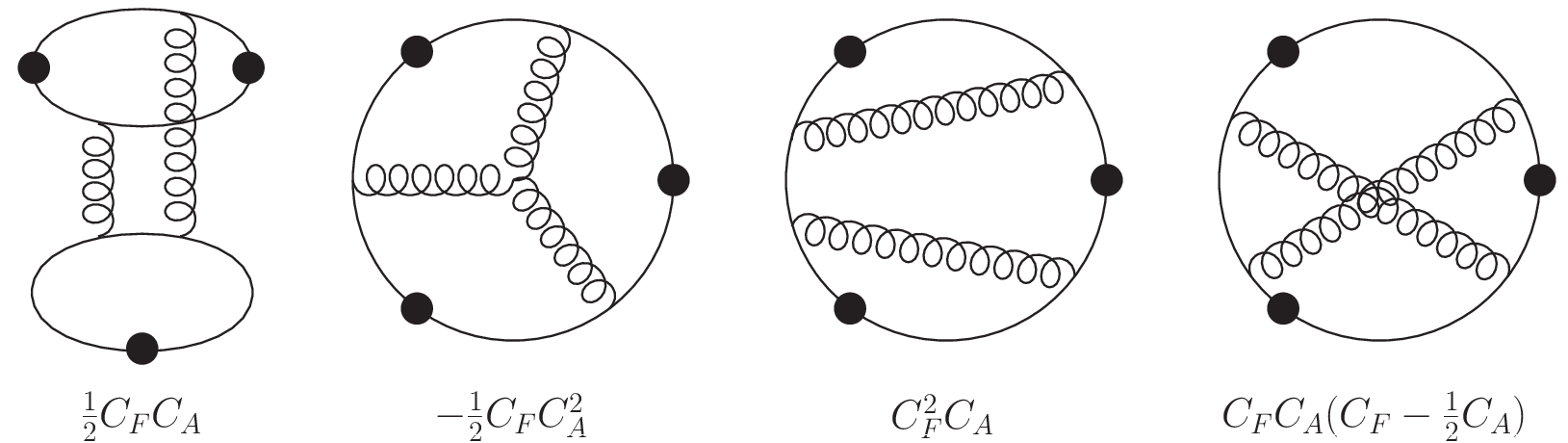}
    \caption{The four possible colour structures associated to each LO diagram, using Hermitian generators. The black dots represent the $GPD$, $DA_1$ or $DA_2$, and their exact ordering has no influence on the colour factor obtained, since all three of them simply connect the colour flow ($\delta_{ij}$ in \EQ\eqref{eq:Fierzcolour}) of the two lines entering it.}
    \label{fig:colour-structures}
\end{figure}

It is interesting to identify the different colour structures that appear in the computation. Topologies 1, 3 and 5, which have two quark traces, all correspond to the first element of  \FIG\ref{fig:colour-structures}, where the 3 black dots represent $GPD$, $DA_1$ and $DA_2$, whose exact positioning has no consequence on the colour factor. On the other hand, topologies 2 and 4, which have a single quark trace, may correspond to either of the last three elements in \FIG\ref{fig:colour-structures}, depending on how the internal gluons are arranged. The specific example considered in \EQ\eqref{eq:hardpart} (diagram number 255) corresponds to the first element of \FIG\ref{fig:colour-structures}. Note that there is a linear dependence among the colour factors in \FIG\ref{fig:colour-structures}, specifically the fourth element is the sum of the second and third elements.

Finally, it is worth noting that we do not consider processes which are sensitive to the exchange of two gluons in the $t$-channel. This is because a similar mechanism for collinear factorisation breaking, as described in \cite{Nabeebaccus:2023rzr,Nabeebaccus:2024mia} for exclusive $\pi^0 \gamma $ pair photoproduction, is expected to occur for such channels. In \APP\ref{app:gluon-GPDs}, we show that the corresponding gluon GPD contributions for such processes, when collinear factorisation is na\"ively assumed, have a divergent imaginary part, exactly analogous to the one obtained for exclusive $\pi^0 \gamma$ pair photoproduction.

\section{Organisation of the amplitude}

\label{sec:organisation-amplitude}

In a specific process, chiral odd and chiral even GPDs cannot both appear. Thus, one can rewrite the amplitude in \EQ\eqref{eq:amplitude} in terms of the generalised Compton form factors $\mathcal{H}$, $\mathcal{E}$, $\Tilde{\mathcal{H}}$, $\Tilde{\mathcal{E}}$, $\mathcal{H}_T^j$, $\tilde{\mathcal{H}}_T^j$, $\mathcal{E}_T^j$, $\Tilde{\mathcal{E}}_T^j$,
\begin{equation}
\label{eq:formfactorslong}
    \mathcal{M_{||}}=\frac{1}{p \cdot q}\bar{u}(p_{N'},\lambda')\left[ \slashed{q}\mathcal{H}(\alpha,\xi,t)+\frac{i\sigma^{q\,\beta}\Delta_\beta}{2 M}\mathcal{E}(\alpha,\xi,t)+\slashed{q}\gamma^5\mathcal{\Tilde{H}}(\alpha,\xi,t)+\frac{q \cdot \Delta}{2 M}\gamma^5\Tilde{\mathcal{E}}(\alpha,\xi,t)\right]u(p_N,\lambda)\,,
\end{equation}
for chiral even GPDs, and
\begin{align}
\nonumber
   \hspace{-0cm} \mathcal{M_{\perp}}=\frac{1}{p \cdot  q}\bar{u}(p_{N'},\lambda')\bigg[ &i\sigma^{qj}\mathcal{H}_{Tj}+\frac{P \cdot q\; \Delta^j-\Delta\cdot q\,P^j}{M^2}\Tilde{\mathcal{H}}_{Tj}\\
   &+\frac{\slashed{q}\;\Delta^j-\Delta \cdot q\; \gamma^j}{2M}\mathcal{E}_{Tj}+\frac{\slashed{q}\;P^j-P\cdot q\;\gamma^j}{M}\Tilde{\mathcal{E}}_{Tj}\bigg]u(p_N,\lambda)\,,
   \label{eq:formfactorsperp}
\end{align}
for the chiral odd case. In the above, we have used the notation $q_\mu \sigma^{\mu \nu} \equiv \sigma^{q \nu}$.

For example, the form factor ${\cal H}$ for a process with a vector GPD is
\begin{align}
    {\cal H}(\alpha, \xi, t) = \int_{-1}^1 dx\int_0^1dv\int_0^1dz \sum_{\ell}\mathcal{T}_{\ell}(x,v,z,\alpha,\xi)\,,
\end{align}
where the sum over $\ell$ is over all diagrams which involves a vector GPD, out of the whole list of 1304 projected diagrams.

\subsection{Factorisation of $s$}

\label{sec:factorisation-of-s}

The amplitude $\mathcal{M}$ expressed in \EQs\eqref{eq:formfactorslong} and \eqref{eq:formfactorsperp} has a mass dimension of $-1$. Indeed, its relation to the asymptotic Fock states is ${}_{\mathrm{out}}\langle N'M_1M_2|N\gamma\rangle_{\mathrm{in}}=i(2\pi)^4\mathcal{M}\;\delta^4(p_N+p_\gamma-p_{N'}-p_{M_1}-p_{M_2})$.  The spinors each have a mass dimension of 1/2. Therefore, we conclude that the form factors have the same mass dimension as the amplitude, namely $-1$.
GPDs are dimensionless, while DAs introduce two decay constants, each of mass dimension 1. Therefore, they contribute two units of mass dimension to the form factors. Consequently, an additional contribution of mass dimension $-3$ is required for dimensional consistency. Among the kinematic variables, only $s$, $p_\perp$, $p$ and $q$ carry mass dimension. However, since $p_\perp \cdot p=p_\perp \cdot q=p^2=q^2=0$, after squaring the amplitude and applying polarisation sum rules, these momenta can only enter through the combinations $p \cdot q$ and $p_\perp^2$. We choose $s$ as the scale we factorise from the squared amplitude,\footnote{Note that the \textit{complete} factorisation of a single scale (for which we choose $s$) uses the assumption that the squared amplitude is a rational function of polynomials of the different scales. If the functional dependence on the scales is more complicated (e.g.~logarithmic, when including the effect of GPD/DA evolution), this factorisation does not work.} keeping in mind that $p_\perp^2=-2\xi\alpha\bar{\alpha}s$ and $p \cdot q=\frac{s}{2}$. Therefore, the squared amplitude must scale as $\frac{1}{s^3}f_1^ 2f_2^2$, where $f_1$ and $f_2$ are the decay constants of the two mesons. For simplicity, we set $s=1$ inside the amplitude, and we restore the $s$-dependence after squaring the amplitude and after averaging (summing) over the polarisations of incoming (outgoing) states.

\subsection{Tensor structures}

\label{sec:tensor-structures}

The amplitude can be decomposed in terms of tensors involving the polarisation vectors of the incoming photon and/or outgoing mesons (when they are transversely polarised). The tensors that can appear also depend on the choice of spinor Fierz projection that is performed. For example, for a vector $GPD$, vector $DA_1$ and vector $DA_2$, the form factor ${\cal H}$ can be decomposed as
\begin{equation}
\label{eq:formfactoreven}
    \mathcal{H}=\mathcal{H}_A\;{p}_{\perp } \cdot {\epsilon}_{q \perp}=\mathcal{H}_A\;T_{A1}\,,
\end{equation}
while for a tensor $GPD$, an axial $DA_1$ and a tensor $DA_2$,
\begin{equation}
\label{eq:formfactorodd}
\mathcal{H}_T^j=\mathcal{H}_{TA}\,T_{A10}^j+\mathcal{H}_{TB}\,T_{B10}^j+\mathcal{H}_{TC}\,T_{C10}^j+\mathcal{H}_{TD}\,T_{D10}^j\,.
\end{equation}
All of the relevant tensors are given in \TAB\ref{tab:tensorbasis}. 

\begin{table}[t!]
\centering
\begin{tabular}{|c|c|c|}
\hline
    1&VVV & $T_{A1}={p}_{\perp }\cdot {\epsilon}_{q \perp}$\\ \hline
   2&  VAV & $T_{A2}={\epsilon }^{{\epsilon}_{q \perp}{p}\,{p}_{\perp}{q}}$ \\ \hline
  3&   VVA &  $T_{A3}={\epsilon }^{{\epsilon}_{q \perp}{p}\,{p}_{\perp}{q}}$ \\ \hline
    4& AVV &  $T_{A4}={\epsilon }^{{\epsilon}_{q \perp}{p}\,{p}_{\perp}{q}}$ \\ \hline 
    5& VAA &  $T_{A5}={p}_{\perp}\cdot {\epsilon}_{q \perp}$\\ \hline
   6&  AVA & $T_{A6}={p}_{\perp}\cdot{\epsilon}_{q \perp}$ \\ \hline 
    7& AAV & $T_{A7}={p}_{\perp}\cdot {\epsilon}_{q \perp}$ \\ \hline
    8& AAA & $T_{A8}={\epsilon }^{{\epsilon}_{q \perp}{p}\,{p}_{\perp}{q}}$ \\ \hline
    \multirow{2}{*}{9 } 
        &\multirow{2}{*}{TVT }  
        &  $T_{A9}^j=p_{\perp}^j \left({p}_{\perp}\cdot{\epsilon}_{q \perp}\right)
   \left({p}_{\perp}\cdot {\epsilon}_{M_2 \perp}\right),
   \quad
   T_{B9}^j=\epsilon_{q \perp}^j \left({p}_{\perp}\cdot {\epsilon}_{M_2 \perp}\right),$
   \\
      & &$T_{C9}^j=\epsilon_{M_2 \perp}^j \left({p}_{\perp}\cdot{\epsilon}_{q \perp}\right),
      \quad
      T_{D9}^j=p_{\perp}^j \left({\epsilon} _{q \perp}\cdot {\epsilon}_{M_2 \perp}\right)$ \\[.05cm]
        \hline
     \multirow{2}{*}{10} &\multirow{2}{*}{TAT}  &  
     $T_{A10}^j=p_{\perp}^j \left({p}_{\perp} \cdot {\epsilon}_{q \perp}\right)
   \left({p}_{\perp}\cdot {\epsilon}_{M_2 \perp}\right),\quad 
   T_{B10}^j=\epsilon_{q \perp}^j \left({p}_{\perp}\cdot {\epsilon}_{M_2 \perp}\right),$\\
   & & $
   T_{C10}^j=\epsilon_{M_2 \perp}^j \left({p}_{\perp}\cdot {\epsilon}_{q \perp}\right),\quad
   T_{D10}^j=p_{\perp}^j \left({\epsilon} _{q \perp}\cdot{\epsilon}_{M_2 \perp}\right)$\\[.05cm]
   \hline 
      \multirow{2}{*}{11}& \multirow{2}{*}{VTT} & 
     $T_{A11}=\left({p}_{\perp }\cdot {\epsilon} _{M_1 \perp}\right) \left({\epsilon}
   _{q \perp}\cdot {\epsilon} _{M_2 \perp}\right),
   \quad
   T_{B11}=\left({p}_{\perp }\cdot {\epsilon} _{M_2 \perp}\right) \left({\epsilon}
   _{q \perp}\cdot {\epsilon } _{M_1 \perp}\right),$
   \\
   & & $T_{C11}=\left({p}_{\perp}\cdot {\epsilon} _{q \perp}\right) \left({\epsilon}
   _{M_1 \perp}\cdot {\epsilon} _{M_2 \perp}\right)$ \\ \hline 
     \multirow{2}{*}{12}&\multirow{2}{*}{ATT} & 
     $T_{A12}=\left({p}_{\perp}\cdot {\epsilon} _{M_1 \perp}\right) {\epsilon
   }^{{\epsilon} _{M_2 \perp}{\epsilon}_{q \perp}{p}\,{q}},
   \quad
   T_{B12}=\left({\epsilon} _{q \perp}\cdot {\epsilon} _{M_2 \perp}\right) {\epsilon
   }^{{\epsilon }_{M_1 \perp}{p}\,{p}_{\perp }{q}},
   $\\
   & & 
   $   T_{C12}=\left({p}_{\perp}\cdot {\epsilon} _{M_2 \perp}\right) {\epsilon
   }^{{\epsilon}_{M_1 \perp}{\epsilon} _{q \perp}{p}\,{q}}$\\ 
   \hline 
    \multirow{2}{*}{13}& \multirow{2}{*}{TTV} & 
    $T_{A13}^j=p_{\perp}^j \left({p}_{\perp}\cdot {\epsilon}_{q \perp}\right)
   \left({p}_{\perp}\cdot {\epsilon}_{M_1 \perp}\right),
   \quad
   T_{B13}^j=\epsilon_{q \perp}^j \left({p}_{\perp}\cdot {\epsilon}_{M_1 \perp}\right),
   $\\
   & & 
   $ T_{C13}^j=\epsilon_{M_1 \perp}^j \left({p}_{\perp}\cdot {\epsilon}_{q \perp}\right),
   \quad
   T_{D13}^j=p_{\perp}^j \left({\epsilon}_{q \perp}\cdot{\epsilon}_{M_1 \perp}\right)$ \\[.05cm] 
   \hline 
   \multirow{2}{*}{14}&  \multirow{2}{*}{TTA} & 
   $T_{A14}^j=p_{\perp}^j \left({p}_{\perp}\cdot {\epsilon} _{q \perp}\right)
   \left({p}_{\perp}\cdot {\epsilon}_{M_1 \perp}\right),
   \quad
   T_{B14}^j=\epsilon_{q \perp}^j \left({p}_{\perp}\cdot {\epsilon}
   _{M_1 \perp}\right),
   $
\\
& & 
$   T_{C14}^j=\epsilon_{M_1 \perp}^j \left({p}_{\perp}\cdot {\epsilon}
   _{q \perp}\right),
   \quad 
   T_{D14}^j=p_{\perp}^j \left({\epsilon}_{q \perp}\cdot {\epsilon}
   _{M_1 \perp}\right)$ \\[.05cm] 
   \hline 
\end{tabular}
\caption{The tensor basis for each combination of Dirac structures. A stands for Axial, V for Vector and T for Tensor. The first letter refers to the $GPD$, the second, to $DA_1$ and the third, to $DA_2$.}
\label{tab:tensorbasis}
\end{table}
 
There are three distinct cases:
\begin{itemize}
    \item Neither the GPD nor the DAs has a transverse structure (rows 1 to 8 in \TAB\ref{tab:tensorbasis}). This implies that $\epsilon_{M_1 \perp}$ and $\epsilon_{M_2 \perp}$ are absent, and therefore, the tensor structure can only contain the vectors $\epsilon_{q\perp}$, $p_\perp$, $p$ and $q$ (with the amplitude being linear in $\epsilon_{q \perp}$).
    \begin{itemize}
                \item If only one of the 3 Dirac structures $GPD$, $DA_1$ or $DA_2$ is axial, then the only tensor structure obtained is a Levi-Civita tensor $\epsilon^{\epsilon_{q\perp}p \,p_\perp q}$.
            \item If two or none of them is axial, then the tensor structure must be simply the scalar product $p_\perp\cdot\epsilon_{q \perp}$.
    \end{itemize}
    \item The GPD has a transverse structure (rows 9, 10, 13 and 14 in \TAB\ref{tab:tensorbasis}), so only one of the DAs also has a transverse structure. Consequently, an additional polarisation vector is present. Without loss of generality, let us assume that the DA with the transverse structure is $DA_1$. The $DA_2$ can be either vector or axial. In the first case, no Levi-Civita tensor appears. In the second case, the Fierz decomposition in \EQ\eqref{eq:Fierzspinor} is modified  by replacing  $\sigma^{+j}$ with $\sigma^{+j}\gamma^5$. In this way, no Levi-Civita tensor appears either. We therefore have at our disposal the vectors $p_\perp$, $\epsilon_{q\perp}$ and $\epsilon_{M_1\perp}$. One of these vector carries a transverse index $j$, coming from $GPD$ (which will be contracted with the non-perturbative matrix element, \EQs\eqref{eq:transversity-GPD} or \eqref{eq:transversity-GPD-alt}), which leads to three possible choices. 
    If this index is carried by  a polarisation vector, the other polarisation vector must enter through a scalar product with $p_\perp$, since the amplitude is linear in them. If $j$ is carried by $p_\perp$, there are two possibilities: either  $(\epsilon_{q\perp}\cdot \epsilon_{M_1\perp})$, or $(\epsilon_{q\perp} \cdot p_\perp)(\epsilon_{M_1\perp}\cdot p_\perp)$. In total, there are four tensor structures in this case, which are listed in rows 13 and 14 of \TAB\ref{tab:tensorbasis}. Note that exchanging the roles of $DA_1$ and $DA_2$ gives the rows 9 and 10 in \TAB\ref{tab:tensorbasis} instead of rows 13 and 14 respectively.
\item Both DAs have a  transverse structure (rows 11 and 12 in \TAB\ref{tab:tensorbasis}), such that the GPD is either vector or axial.
    \begin{itemize}
        \item In the case of a vector GPD (VTT), one obtains only three tensor structures corresponding to the different ways of contracting the three polarisation vectors (either with themselves or with $p_{\perp}$) into scalar products. It should be noted that the structure $\left({\epsilon} _{ M_1\perp}\cdot{p}_{\perp } \right) \left({\epsilon}
   _{q\perp}\cdot p_\perp\right)$ $\left({\epsilon} _{ M_2\perp} \cdot p_\perp\right)$ is absent (row 11 in \TAB\ref{tab:tensorbasis}). A detailed explanation for this feature, which relies on the choice of gauge fixing and the restricted possibilities for Dirac trace structures, is presented in \APP\ref{app:absence-VTT-structure}.
   \item If the GPD is axial (ATT), each of the three polarisation vectors must be contracted either with a Levi-Civita tensor or with $p_{\perp}$. The Levi-Civita tensor  gives a  non-zero contribution only if it is contracted with exactly two transverse and two longitudinal vectors. Two cases must be distinguished:
   \begin{itemize}
       \item Only one polarisation vector is contracted with the Levi-Civita tensor (with the other contracted transverse vector being a $p_\perp$). Two polarisation vectors then remain, which can be contracted either with themselves or separately with $p_\perp$. This leads to six possible tensor structures.
       \item If two polarisation vectors enter the Levi-Civita tensor, the remaining polarisation vector must necessarily be contracted with $p_\perp$. This gives three possible tensor structures.
   \end{itemize}
However, not all of these 9 structures are independent. Indeed, the Schouten identities give 5 relations between them, which allows them to be reduced to a basis of 4 linearly independent tensors. The relations are
    \begin{align*}
(p_{\perp}\cdot\epsilon_{q\perp}){\epsilon}^{\,\epsilon_{M_1\perp}\,\epsilon_{M_2\perp}\,{p}\,{q}}
&=
(p_{\perp}\cdot\epsilon_{M_2\perp}) {\epsilon }^{\,\epsilon_{M_1\perp}\,\epsilon_{q\perp}\,{p}\,{q}}
-
(p_{\perp}\cdot\epsilon_{M_1\perp}) {\epsilon }^{\,\epsilon_{M_2\perp}\,\epsilon _{q\perp}\,{p}\,{q}}\,,\\
   (\epsilon_{M_1\perp}\cdot \epsilon _{M_2\perp} ){\epsilon}^{\,\epsilon_{q\perp}\,{p}\,p_{\perp}\,{q}}
   &=
   (\epsilon _{q\perp}\cdot\epsilon_{M_2\perp} ){\epsilon}^{\,\epsilon_{M_1\perp}\,{p}\,p_{\perp}\,{q}}
   +
  ( p_{\perp}\cdot\epsilon _{M_2\perp} ){\epsilon}^{\,\epsilon_{M_1\perp}\,\epsilon_{q\perp}\,{p}\,{q}}\,,
   \\
   (\epsilon_{q\perp}\cdot\epsilon _{M_1\perp} ){\epsilon}^{\,\epsilon _{M_2\perp}\,{p}\,p_{\perp}\,{q}}
   &=
   (\epsilon _{q\perp}\cdot\epsilon _{M_2\perp} ){\epsilon}^{\,\epsilon_{M_1\perp}\,{p}\,p_{\perp}\,{q}}
   +
   (p_{\perp}\cdot\epsilon _{M_2\perp}) {\epsilon }^{\,\epsilon_{M_1\perp}\,\epsilon_{q\perp}\,{p}\,{q}}
   \\
   &\quad
   -(p_{\perp}\cdot\epsilon_{M_1\perp}) {\epsilon }^{\,\epsilon_{M_2\perp}\,\epsilon_{q\perp}\,{p}\,{q}}\,,
   \\
   (p_{\perp}\cdot\epsilon_{q\perp}) (p_{\perp}\cdot\epsilon _{M_1\perp})
   {\epsilon }^{\,\epsilon_{M_2\perp}\,{p}\,p_{\perp}\,{q}}
   &=
 ( p_{\perp}\cdot\epsilon _{M_2\perp} )\left(-p_t^2\,{\epsilon }^{\,\epsilon _{M_1\perp}\,\epsilon_{q\perp}\,{p}\,{q}}
   +
   (p_{\perp}\cdot\epsilon_{q\perp} ){\epsilon}^{\,\epsilon_{M_1\perp}\,{p}\,p_{\perp}\,{q}}\right)
   \\
   &\quad
   +p_t^2 \,(p_{\perp}\cdot\epsilon _{M_1\perp})
   {\epsilon }^{\,\epsilon_{M_2\perp}\,\epsilon_{q\perp}\,{p}\,{q}}\,,
   \\
    (p_{\perp}\cdot\epsilon_{M_1\perp}) (p_{\perp}\cdot\epsilon_{M_2\perp})
   {\epsilon }^{\,\epsilon_{q\perp}\,{p}\,p_{\perp}\,{q}}
   &=
   (p_{\perp}\cdot\epsilon_{M_2\perp} )\left(-p_t^2\,{\epsilon }^{\,\epsilon_{M_1\perp}\,\epsilon_{q\perp}\,{p}\,{q}}
   +
   (p_{\perp}\cdot\epsilon_{q\perp} ){\epsilon}^{\,\epsilon_{M_1\perp}\,{p}\,p_{\perp}\,{q}}\right)\,.
    \end{align*}
 As in the VTT case discussed above, it is remarkable that the tensor structure with three $p_\perp$, namely $(p_{\perp}\cdot\epsilon_{M_2\perp})\,(p_{\perp}\cdot\epsilon_{q\perp}) \,{\epsilon}^{\,\epsilon_{M_1\perp}\,{p}\,p_{\perp}\,{q}}$ does not appear at all in any of our amplitudes.
    \end{itemize}
\end{itemize}

In summary, if both mesons are not transversely polarised, the amplitude involves the two independent tensor structures ${p}_{\perp }\cdot {\epsilon}_{q \perp}$ and ${\epsilon }^{{\epsilon}_{q \perp}{p}\,{p}_{\perp}{q}}$. If exactly one meson is transversely polarised, we will get four independent tensor structures, which could correspond to row 9, 10, 13 or 14. If both mesons are transversely polarised, we obtain the six independent tensor structures from rows 11 and 12. Finally, note that some of the rows are identical. They correspond to rows
\{1, 5, 6, 7\}, 
\{2, 3, 4, 8\},
\{9, 10\} and \{13, 14\}.

\section{Numerical integration}

\label{sec:numerical-integration}

Now that the analytical computation is complete, the next stage is to perform the numerical integration. For a given process, the coefficient in front of each tensor structure is extracted and integrated over $x$, $v$ and $z$.  The amplitude has contributions such as the one in \EQ\eqref{eq:example-amplitude}, which  contain  poles in the variable $x$, which are regularised using the standard Feynman $i \epsilon$ prescription. It is extremely hard to numerically integrate such a contribution by brute-force using a small and finite $i \epsilon$, since the numerical integration has to deal with large cancellations numerically, making the final result unstable. Therefore, we adopt a strategy where we instead have \textit{full analytical control} over the large cancellations, and do not rely on having a small and finite $i \epsilon$ for the numerical integration. This is achieved through a partial fraction decomposition, followed by the application of the Sokhotski-Plemelj formula. A similar strategy was also used in \cite{Crnkovic:2025man}.

\subsection{Partial fraction decomposition}
Each diagram can have between one and four poles in the variable $x$. These are split by performing a partial fraction decomposition. The resulting terms are stored in a list of 3000 elements. The final number of relevant terms is 2920 since 80 of them only contribute to $\pi^0\rho^0_L$ which is not considered in our study since it also involves gluon GPDs, and such processes were shown to break collinear factorisation \cite{Nabeebaccus:2023rzr,Nabeebaccus:2024mia}. To give an idea, the number of these terms, with only one pole in $x$ (after partial fraction in $x$),  that enter our processes varies between 142 (for $\pi^+\rho^0_T$) and 504 (for $\rho^0_L\rho^0_L$).

The partial fraction causes several spurious
divergences to appear. For example, let us consider the amplitude in \EQ\eqref{eq:example-amplitude}. Its partial fraction decomposition contains a total of four terms, among which the following two terms arise:
\begin{align}
&\frac{e g^4 N_c C_F \,{p}_{\perp}\cdot {\epsilon }_{ q\perp} {\boldsymbol{(v-z)}}
   (-\bar{\alpha} v-\alpha  z+1)^2 \left(2 \bar{\alpha}^2 v^2+\bar{\alpha} v (4
   \alpha  z-3)+\alpha  \left(\alpha +4 \alpha  z^2-3 (\alpha +1)
   z\right)+1\right)
   }{2(4N_c)^3 \bar{\alpha}^3 s^2 \bar{v}^2 v \bar{z}
   (1-\alpha (z-v)-v) {\boldsymbol{\left(
   (v-z)^2+i\epsilon \frac{(\bar{\alpha} v+\alpha  z) (\bar{\alpha}
   v+\alpha  z-1)}{\bar{\alpha} \alpha  \xi}\right)}}(\alpha  \xi  v-\xi  z (\alpha +v-1))}\nonumber\\&
   \hspace{5cm}\times\frac{f(\{255,1,3,\{1,6\}\},N,N',\{M_1,P_1\},\{M_2,P_2\})}{\left({\boldsymbol{i\epsilon+x}}{\boldsymbol{+\frac{\xi -\xi  v (\alpha -2 z+1)+(\alpha -2) \xi
   z}{(\alpha -1) v-\alpha  z+1}}}\right)}\nonumber\\
   &-\frac{e g^4 N_c C_F\, {p}_{\perp}\cdot {\epsilon }_{ q\perp} {\boldsymbol{(v-z)}}
   (\bar{\alpha} v+\alpha  z) \left(2 \bar{\alpha}^2 v^2+\bar{\alpha} v (4
   \alpha  z-1)+\alpha  z (\alpha  (4 z-1)-1)\right)
   }{2(4N_c)^3 \xi \bar{\alpha}^3 s^2 \bar{v}v^2 z
   {\boldsymbol{\left((v-z)^2+i\epsilon \frac{(\bar{\alpha}
   v+\alpha  z) (\bar{\alpha} v+\alpha  z+1)}{\bar{\alpha} \alpha  \xi }\right) }}(\xi  v (\alpha +z-1)-\alpha  \xi  z)
   }\nonumber\\
   &\hspace{5cm}
   \times \frac{f(\{255,1,3,\{1,6\}\},N,N',\{M_1,P_1\},\{M_2,P_2\})}{\left({\boldsymbol{-i\epsilon+x}}{\boldsymbol{+\frac{\xi  (v (\alpha +2 z-1)-\alpha
   z)}{(\alpha -1) v-\alpha  z}}}\right)}
   \,.\label{eq:partialfraction255}
\end{align}
Any factors of $i \epsilon$ in the numerator can be safely ignored. We identify three types of divergences in these two terms: 
\begin{itemize}
    \item The pole in the variable $x$ (in bold), which is regularised with $i \epsilon$. This is treated using the Sokhotski-Plemelj formula as described in \SEC\ref{sec:eliminatation-Ieps}.
    \item The \textit{spurious} divergence in $v=z$ (in bold), which is also regularised with $i\epsilon$. The divergences of this type are treated through a series of foldings of the $(v,z)$ plane.\footnote{We thank Goran Duplancic for his insights and suggestions regarding the use of the folding technique to eliminate spurious divergences and to improve numerical stability.} We emphasize that the amplitude \textit{before} partial fraction does not have any divergences of this type, confirming that they are indeed spurious.
    \item The divergences at $v=1$ and $v=0$ that are not cancelled by the DAs (not yet included). They are cancelled by the two other terms coming from the partial fraction not shown in \EQ\eqref{eq:partialfraction255}. More generally, any divergences on the sides of the integration domain of the $(v,z)$-plane, i.e.~in $v$ or $z$ being $0$ or $1$, are cancelled when summing over all the terms coming from the partial fraction. So, these divergences are all clearly \textit{spurious}.
    \item The divergences at the corners in the $(v,z)$-plane, for example, $v=z=0$ in the first term, originating from the denominator $\alpha  \xi  v-\xi  z (\alpha +v-1)$. These divergences are all integrable.
\end{itemize}

More generally, any divergence that has appeared only after performing the partial fraction procedure has to be spurious. This means that only the pole in $x$, which is regularised by $i \epsilon$, is a ``true'' divergence, since none of the other divergences were present before the partial fraction procedure.\footnote{In fact, one can very easily perform the integral of \EQ\eqref{eq:hardpart} numerically in the DGLAP region for $x$ (i.e.~$1>|x|>\xi$), with $i \epsilon = 0$ exactly. This confirms that the $v=z$ divergence is indeed spurious.}

\begin{figure}[t!]
    \centering
    \includegraphics[scale=.55]{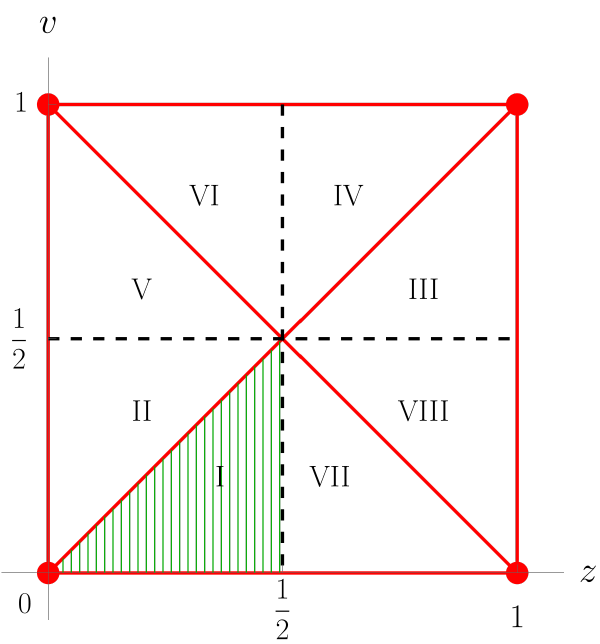}
    \caption{As a result of the series of foldings, the integration domain can be restricted to the green part. The contribution of each domain indicated by a Roman numeral corresponds to a particular term in \EQ\eqref{eq:folding}. The lines in red may exhibit divergences, but all of them are spurious and cancel after integration. The integrand is divergent at the corners, indicated by the red dots, but is integrable.}
    \label{fig:domain3}
\end{figure}

Let us examine the spurious divergences along the diagonal $v=z$, which have the form $\frac{v-z}{(v-z)^2+i \epsilon}$. One convenient way to deal with them is to perform a series of \textit{foldings} of the $(v,z)$ integration region, along the lines $z=v$, $z=1-v$, and $z=1/2$.\footnote{This last folding is not necessary to regularise the divergences along the diagonals. Nevertheless, we perform it since it helps with numerical stability, and allows the cancellation of antisymmetric terms in $z \to \bar z$.}  This essentially corresponds to  
\begin{align}
\label{eq:folding}
\int_{0}^1 dz\int_{0}^1 dv\, \mathcal{A}(z,v)=\int_0^{\frac{1}{2}}\,dz\,&\int_0^z\,dv \bigg(\underset{\rm{I}}{\mathcal{A}(z,v)}+\underset{\rm{II}}{\mathcal{A}(v,z)}+\underset{\rm{III}}{\mathcal{A}(1-v,1-z)}+\underset{\rm{IV}}{\mathcal{A}(1-z,1-v)}\nonumber\\&+\underset{\rm{V}}{\mathcal{A}(1-v, z)}+\underset{\rm{VI}}{\mathcal{A}(z,1-v)}+\underset{\rm{VII}}{\mathcal{A}(1-z,v)}+\underset{\rm{VIII}}{\mathcal{A}(v,1-z)} \bigg)\,,
\end{align}
where $\mathcal{A}$ denotes the term that contains spurious divergences in $v=z$ and/or $v=1-z$. The Roman numeral below each term in \EQ\eqref{eq:folding} corresponds to the contribution from each domain labelled in \FIG\ref{fig:domain3}.

In this way, the $i \epsilon$ factors can be safely eliminated. Indeed, if we have a generic term of the form 
\begin{equation}
\label{eq:example-term-folding}
    \frac{h(z,v)(z-v)}{(z-v)^2+i\epsilon},
\end{equation}
with $h$ being some generic function, then after performing the series of folding, one gets 
\begin{align}
\label{eq:folding-spurious-divs}
    &\left(\frac{h(z,v)(z-v)}{(z-v)^2+i\epsilon}+\frac{h(v,z)(v-z)}{(v-z)^2+i\epsilon}\right)+(z\rightarrow 1-z\;\, \text{and}\;\, v\rightarrow1-v)\nonumber\\
    &\hspace{2cm}\underset{v\rightarrow z}{\sim}
    \frac{(v-z)^2\;\left(\frac{\partial{h}(z,v)}{\partial{z}}-\frac{\partial{h(z,v)}}{\partial{v}}\right)\Big{|}_{z=v}}{(v-z)^2+i\epsilon}+(z\rightarrow 1-z\;\, \text{and}\;\, v\rightarrow1-v)\,,
\end{align}
 where $i\epsilon$ can be safely taken to 0 before integration on the right-hand side. In \EQ\eqref{eq:folding-spurious-divs}, only the four potentially singular terms resulting from the series of foldings of the term in \EQ\eqref{eq:example-term-folding} are kept. The other four terms not shown in \EQ\eqref{eq:folding-spurious-divs} potentially have a singularity on the line $v = 1-z$, but the same corresponding mechanism occurs, indicating that the singularity is also spurious in that case.

\subsection{Elimination of the $i\epsilon$ prescription}

\label{sec:eliminatation-Ieps}

In order to get completely rid of the $i\epsilon$, which is important for numerical stability, the remaining term containing $i \epsilon$ is further split using the Sokhotski-Plemelj formula, which results in
\begin{equation}
\label{eq:principalvalueformula}
    \int_{-1}^1dx\frac{g(v,z)H(x)}{x-r(v,z)\pm i\epsilon}=\int_{-1}^1dx\;g(v,z)\frac{H(x)-H(r(v,z))}{x-r(v,z)}+\left[ \ln\left(\frac{1-r(v,z)}{1+r(v,z)}\right)\mp i \pi\right] g(v,z)H(r(v,z))\,,
\end{equation}
where $H(x)$ corresponds to a generic GPD (where we have omitted the $\xi$ and $t$ arguments for conciseness),  $g(v,z)$ denotes the $x$-independent remaining part of the hard part including the DAs, and $r(v,z)$ denotes the real part of the pole in $x$. Note that after partial fraction, the only $x$ dependence in the numerator corresponds to that of the relevant GPD $H(x)$. This is always the case at leading order, since the numerator of the hard part of any diagram simply involves a polynomial in $x$. Note that the second term on the right-hand side of \EQ\eqref{eq:principalvalueformula} has already been integrated over $x$. 

\subsection{Organisation of the integration}

Now, we describe how the terms entering the amplitude are organised prior to integration. First, a function called \textit{chooseprocess} fixes the parameters $N$, $N'$, $M_1$, $M_2$, $P_1$, $P_2$. This selects only the relevant terms in the whole list of 2920 elements obtained from partial fraction, which are compatible with the process. Each term, in the notation of \EQ\eqref{eq:principalvalueformula}, is of the type 
\begin{equation*}
    \frac{g(v,z) H(x)}{\text{denLO[$k$]}}\,,
\end{equation*}
where denLO[$k$] denotes a specific $x$ dependent pole  of the form $x-r_k(v,z)+ \eta_k\,i\epsilon$, with $\eta_k=\pm1$. We have identified 14 distinct cases for $k$, given in \TAB\ref{tab:denLO-cases}.
\begin{table}[t!]
    \centering
    \begin{tabular}{|c|c|}
    \hline
       \raisebox{-0.1cm}{$\text{denLO}[1]$}&\raisebox{-0.1cm}{$ x-\xi -i \epsilon $}
       \\[0.2cm] \hline
      \raisebox{-0.1cm}{$\text{denLO}[2]$}
       &\raisebox{-0.1cm}{
       $ x+\xi+i \epsilon $}
   \\[0.2cm]   \hline
  \raisebox{-0.1cm}{$\text{denLO}[3]$}
   &\raisebox{-0.1cm}{
   $ x+\xi -i \epsilon
   $
   }\\[0.2cm]  \hline \raisebox{-0.1cm}{
   $\text{denLO}[4]$
   }&\raisebox{-0.1cm}{
   $x -\xi +i \epsilon $
   }\\[0.2cm]  \hline \raisebox{-0.1cm}{
   $\text{denLO}[5]$
   }&\raisebox{-0.1cm}{
   $x+ \frac{\xi
    (\alpha  (v-1)+v)}{\alpha  (v-1)-v}-i \epsilon
   $
   }\\[0.2cm]  \hline \raisebox{-0.1cm}{
   $\text{denLO}[6]$
   }&\raisebox{-0.1cm}{
   $x -\frac{\xi  (\alpha  v+v-1)}{(\alpha -1)
   v+1}+i \epsilon $
   }\\[0.2cm]  \hline \raisebox{-0.1cm}{
   $\text{denLO}[7]$
   }&\raisebox{-0.1cm}{
   $ x+\frac{\xi  (\alpha  (z-1)-2
   z+1)}{\alpha  (z-1)+1}-i \epsilon $
   }\\[0.2cm]  \hline \raisebox{-0.1cm}{
   $\text{denLO}[8]$
   }&\raisebox{-0.1cm}{
   $ x-\frac{\xi 
   ((\alpha -2) z+1)}{\alpha  z-1}+i \epsilon $
   }\\[0.2cm]  \hline \raisebox{-0.1cm}{ 
   $\text{denLO}[9]$
   }&\raisebox{-0.1cm}{
   $x+
   \frac{\xi -\xi  v (\alpha -2 z+1)+(\alpha -2) \xi  z}{(\alpha -1)
   v-\alpha  z+1}+i \epsilon $
   }\\[0.2cm]  \hline \raisebox{-0.1cm}{
   $\text{denLO}[10]$
   }&\raisebox{-0.1cm}{
   $x+ \frac{\xi  (v
   (\alpha +2 z-1)-\alpha  z)}{(\alpha -1) v-\alpha  z}-i \epsilon   $
   }\\[0.2cm]  \hline \raisebox{-0.1cm}{
   $\text{denLO}[11]$
   }&\raisebox{-0.1cm}{
   $x+ \frac{\xi  (-\alpha  (v+z-1)-2 v z+v+2
   z-1)}{\alpha  (v+z-1)-v+1}+i \epsilon $
   }\\[0.2cm]  \hline \raisebox{-0.1cm}{
   $\text{denLO}[12]$
   }&\raisebox{-0.1cm}{
   $x+
   \frac{\xi  (\alpha  (v+z-1)-2 v z+v)}{\alpha  (v+z-1)-v}-i
   \epsilon $
   }\\[0.2cm]  \hline \raisebox{-0.1cm}{
   $\text{denLO}[13]$
   }&\raisebox{-0.1cm}{
   $ x-\frac{\xi  (\alpha  (v+z-1)-2 v
   z+v)}{\alpha  (v+z-1)-v}+i \epsilon $
   }\\[0.2cm]  \hline \raisebox{-0.1cm}{
   $\text{denLO}[14]$
   }&\raisebox{-0.1cm}{
   $ x+\frac{\xi
    (\alpha  (v+z-1)+(v-1) (2 z-1))}{\alpha  (v+z-1)-v+1}-i \epsilon$}
    \\[0.2cm]  \hline
    \end{tabular}
    \caption{List of all possible poles in the variable $x$. Note that all of them are in the ERBL region, $-\xi \leq x \leq \xi$.}
    \label{tab:denLO-cases}
\end{table}
These terms are then grouped according to their denominator denLO[$k$], and each group is summed to yield a single term \begin{equation*}
    \frac{f(x,v,z,\alpha,\xi)}{\text{denLO[$k$]}}\,,
\end{equation*}
where  $f$ is a generic function, from which we extract the coefficients multiplying $Q_u$ and $Q_d$. Recall that the amplitude is proportional to a single charge $Q_f$ since there is a single photon attached to the hard part. Integrating the $Q_u$ and $Q_d$ contributions separately allows  us to study certain symmetries between different processes. To each coefficient, we apply the formula \EQ\eqref{eq:principalvalueformula} and we store the $x$-dependent and $x$-independent terms in two seperate lists. We subsequently extract  the coefficients multiplying the different tensor structures. 

It is interesting to note that the denLO[$k$] (as a function of $z$ and $v$ for fixed external kinematics) never vanish in the DGLAP region ($1>|x|>\xi$). All zeros are thus found in the ERBL region, where  $-\xi < x <\xi$. This feature is a consequence of the fact that at the partonic level, the DGLAP region  corresponds to a standard  $ 2 \to n$ \textit{massless} particle scattering, where the virtualities of any internal line cannot change sign for any physical point in the phase space. In contrast, the ERBL region effectively corresponds to a \textit{massless} $3 \to n$ process (since both the quark and anti-quark from the nucleon approach the hard scattering sub-process with positive energy), and the virtualities of internal lines do not necessarily have a fixed sign within the physical phase space. This discussion is further elaborated in \APP\ref{app:ERBL-poles}.

At this stage, there are two lists, one for $Q_u$ and the other for $Q_d$. Each list has two sublists of 30 elements, where each element corresponds to a tensor structure denoted by $T$ in the third column of \TAB\ref{tab:tensorbasis}. The first sublist contains the $x$-independent coefficients, while the second contains the $x$-dependent ones. In the end, the folding of the $(v,z)$-phase space is performed. The expressions to integrate are very heavy (dozens of mega-bytes), which leads to rather slow numerical integrations.

Due to the complexity of the expressions we want to integrate, a good control on the numerical precision is required. As $\xi$ decreases, the ERBL region where the GPD exhibits significant variations, becomes narrower, causing the integrand to vary abruptly in the vicinity of this region. This can introduce numerical instabilities, and significantly reduce the accuracy of our results.

To take into account the rapid variations of the integrand, we split the integration domain in $x$ in the same way as for the GPD sampling (see \APP\ref{app:GPD-sampling}), that is,
\begin{equation}
\label{eq:xinterval}
x\in[-1,1]\rightarrow x\in I_1\cup I_2\,,\quad
I_1=[-1,-\xi-\delta\,\xi]\cup[\xi+\delta\,\xi,1]\,,\quad
I_2=[-\xi-\delta\,\xi,\xi+\delta\,\xi]\,.
\end{equation}
The ERBL region, corresponding to $I_2$, is slighlty extended by a small fraction 
$\delta$, in order to ensure that the region of rapid variations of the integrand is fully covered. 
For the numerical integration, we use the \textit{LocalAdaptive} method, which adaptively subdivides the integration domain according to the variations of the integrand. 

Furthermore, we rationalise all parameters entering the integrand and the GPDs (forcing Mathematica to store them in terms of rational numbers, rather than floating numbers), and evaluate the \textit{full} integrand to very high precision. This avoids   instabilities due to imperfect cancellations between numerically large terms.

At this point, it is worth mentioning that depending on the chosen process, various symmetries may exist, such as charge conjugation symmetry, isospin symmetry and meson exchange symmetry, which either impose certain conditions on the amplitude, or relate it to another process. A detailed discussions of these symmetries is given in \APP\ref{app:symmetries}. Such symmetries also act as a check for the correct implementation of the automation of the  amplitude calculation for the family of di-meson processes considered in this article.

\section{Cross section}

\label{sec:cross-section}

Summing the amplitude squared over the nucleon helicities $\lambda$ and $\lambda'$, we have, for a process involving chiral even GPDs as in \EQ\eqref{eq:formfactorslong},
\begin{align}
\label{eq:longampsquared}
\sum_{\lambda\,\lambda'} \mathcal{M}_{||}(\lambda,\lambda')\mathcal{M}_{||}^*(\lambda,\lambda')&=8\left[(1-\xi^2)\left(|\mathcal{H}(\alpha,\xi,t)|^2+|\tilde{\mathcal{H}}(\alpha,\xi,t)|^2\right)-2\xi^2\,\Re\left(\mathcal{H}(\alpha,\xi,t)\mathcal{E}^*(\alpha,\xi,t)\right)
\right.\nonumber\\&\left.-2\xi^2\,\Re\left(\tilde{\mathcal{H}}(\alpha,\xi,t)\tilde{\mathcal{E}}^*(\alpha,\xi,t)\right)+\frac{\xi^4}{1-\xi^2}\left(|\mathcal{E}(\alpha,\xi,t)|^2+|\tilde{\mathcal{E}}(\alpha,\xi,t)|^2\right)
\right]\,,
\end{align}
in the limit $\Delta_\perp=0$.
Since $\xi$ is typically small, at most of the order of $ 10^{-1}$, one can reasonably keep the leading power of $\xi$ in \EQ\eqref{eq:longampsquared}, that is\footnote{Strictly speaking, the result should be just $8\left(|\mathcal{H}(\alpha,\xi,t)|^2+|\tilde{\mathcal{H}}(\alpha,\xi,t)|^2\right)$, but we have decided to keep the \textit{full} coefficient in front of the relevant contributing form factors, being just a multiplicative factor that does not require further numerical computations.}
\begin{equation}
\label{eq:sumevenamp}
 \sum_{\lambda\,\lambda'} \mathcal{M}_{||}(\lambda,\lambda')\mathcal{M}_{||}^*(\lambda,\lambda')\approx 8(1-\xi^2)\left(|\mathcal{H}(\alpha,\xi,t)|^2+|\tilde{\mathcal{H}}(\alpha,\xi,t)|^2\right).
\end{equation}
In the chiral odd sector, see \EQ\eqref{eq:formfactorsperp}, one instead has
\begin{align}
\sum_{\lambda\,\lambda'} \mathcal{M}_{\perp}(\lambda,\lambda')\mathcal{M}_{\perp}^*(\lambda,\lambda')=&8\left[(1-\xi^2)|\mathcal{H}_T(\alpha,\xi,t)|^2+\frac{\xi^2}{1-\xi^2}|\xi\,\mathcal{E}_T(\alpha,\xi,t)-\tilde{\mathcal{E}}_T(\alpha,\xi,t)|^2\right.\nonumber\\ 
&\qquad\left.-2\,\xi\,\Re\,\left(\mathcal{H}^*_T\cdot[\xi\,\mathcal{E}_T(\alpha,\xi,t)-\tilde{\mathcal{E}}_T(\alpha,\xi,t)]\right)\right]\,,
\end{align}
in the limit $\Delta_{\perp} =P_{\perp}=0$. 
As in the chiral-even case, we drop terms that vanish when $\xi \to 0$, and thus obtain\footnote{It can be checked that this expression is unchanged when projecting onto the $\sigma^{\mu\nu}\gamma^5$ Fierz structure instead of $\sigma^{\mu\nu}$ in \EQ\eqref{eq:Fierzspinor}.}
\begin{equation}
\label{eq:sumoddamp}
\sum_{\lambda\,\lambda'} \mathcal{M}_{\perp}(\lambda,\lambda')\mathcal{M}_{\perp}^*(\lambda,\lambda')\approx 8(1-\xi^2)|\mathcal{H}_T(\alpha,\xi,t)|^2.
\end{equation}
One must be careful about the fact that the generalised Compton form factors $\mathcal{H}_T$, $\mathcal{E}$ and $\tilde{\mathcal{E}}_T$ are 2D Euclidean vectors with two transverse components $i=1,2$ and hence, the notation $|\mathcal{H}|^2$ is to be understood as the Euclidean squared norm of $\mathcal{H}$.

Let $|\overline{\mathcal{M}}|^2$ denote the amplitude squared averaged over the polarisation of the incoming photon and nucleon and summed over the polarisation of the outgoing nucleon and transverse polarisations (if any) of the outgoing mesons. If we call $\lambda_q=1,2$ the two possible polarisations of the incoming photon, one has
\begin{equation}
\label{eq:averagesquaredamp}
    |\overline{\mathcal{M}}|^2=\frac{1}{4}\sum_{\lambda,\,\lambda',\,\lambda_q=1,2}\mathcal{M}(\lambda,\lambda',\lambda_q)\mathcal{M}^*(\lambda,\lambda',\lambda_q)\,,
\end{equation}
including an additional sum over the outgoing meson polarisations depending on the process. The $\frac{1}{4}$ factor accounts for averaging over the incoming photon polarisation and the incoming nucleon helicity.

The evaluation of $|\overline{\mathcal{M}}|^2$ is carried out using the following polarisation sum rules\footnote{It can assumed that $\epsilon_{q\perp}$ and $\epsilon_{M_{i}\perp}$ have real components, since we sum over the polarisations.}
\begin{equation}
\label{eq:polarisation-sums}
\sum_{\lambda_q=1,2}\epsilon^\mu_{q\perp,\lambda_q }\epsilon^\nu_{q\perp,\lambda_q }=\sum_{\lambda_{M_i}=1,2}\epsilon^\mu_{M_i\perp,\lambda_{M_i} }\epsilon^\nu_{M_i\perp,\lambda_{M_i} }=-g^{\mu\nu}_\perp\,,
\end{equation} 
where $\lambda_{M_i} $ corresponds to the two possible transverse polarisations of the outgoing meson $M_i$.

 As an example, consider the form factor $\mathcal{H}=\mathcal{H}_A\,T_{A1}$ as  in \EQ\eqref{eq:formfactoreven}, which corresponds to the photoproduction of two $\rho_{L}$. The averaged squared amplitude then reads  
\begin{equation}
    \sum_{\lambda_q=1,2}|T_{A1}|^2=\sum_{\lambda_q=1,2}(p_\perp\cdot\epsilon_{q\perp, \lambda_q})(p_\perp\cdot\epsilon_{q\perp, \lambda_q})=p_t^2\,.
\end{equation}
Note that $T_{A1}$ implicitly contains an index $\lambda_q$.
The sums over polarisations of all relevant tensor structures are given in \TAB\ref{tab:structureproducts}.

For a process where $M_1$ is a pion and $M_2$ is a transversely-polarised meson, one instead has the form factor decomposition $\mathcal{H}_T^j=\mathcal{H}_{TA}\,T_{A10}^j+\mathcal{H}_{TB}\,T_{B10}^j+\mathcal{H}_{TC}\,T_{C10}^j+\mathcal{H}_{TD}\,T_{D10}^j$ (see \EQ\eqref{eq:formfactorodd}). Then
\begin{align}
    &\sum_{\lambda_q,\lambda_{M_i} =1,2}|\mathcal{H}_T|^2=\sum_{\lambda_q,\lambda_{M_i} =1,2}\bigg[|\mathcal{H}_{TA}|^2|T_{A10}|^2+|\mathcal{H}_{TB}|^2|T_{B10}|^2+|\mathcal{H}_{TC}|^2|T_{C10}|^2+|\mathcal{H}_{TD}|^2|T_{D10}|^2\nonumber\\
    &+2\,\Re\,(\mathcal{H}_{TA}\mathcal{H}_{TB}^*)\,T_{A10}\cdot T_{B10}+2\,\Re\,(\mathcal{H}_{TA}\mathcal{H}_{TC}^*)\,T_{A10}\cdot T_{C10}+2\,\Re\,(\mathcal{H}_{TA}\mathcal{H}_{TD}^*)\,T_{A10}\cdot T_{D10}\nonumber\\
    &+2\,\Re\,(\mathcal{H}_{TB}\mathcal{H}_{TC}^*)\,T_{B10}\cdot T_{C10}+2\,\Re\,(\mathcal{H}_{TB}\mathcal{H}_{TD}^*)\,T_{B10}\cdot T_{D10}+2\,\Re\,(\mathcal{H}_{TC}\mathcal{H}_{TD}^*)\,T_{C10}\cdot T_{D10}\bigg],
\end{align}
which, after using the replacements in \TAB\ref{tab:structureproducts}, gives 
\begin{align}
&\sum_{\lambda_q,\lambda_{M_i} =1,2}|\mathcal{H}_T|^2=  \;p_t^6 |\mathcal{H}_{TA}|^2
+2 p_t^2|\mathcal{H}_{TB}|^2+2 p_t^2|\mathcal{H}_{TC}|^2+2 p_t^2|\mathcal{H}_{TD}|^2+2 p_t^2\,\Re\,(\mathcal{H}_{TB}\mathcal{H}_{TC}^*)\nonumber\\
&+2 p_t^2\,\Re\,(\mathcal{H}_{TB}\mathcal{H}_{TD}^*)+2 p_t^2\,\Re\,(\mathcal{H}_{TC}\mathcal{H}_{TD}^*)- 2 p_t^4\,\Re\,(\mathcal{H}_{TA}\mathcal{H}_{TB}^*)-2 p_t^4\,\Re\,(\mathcal{H}_{TA}\mathcal{H}_{TC}^*)-2 p_t^4\,\Re\,(\mathcal{H}_{TA}\mathcal{H}_{TD}^*).
\label{eq:HTCFFsumed}
\end{align}

\begin{table}[t!]
\begin{tabular}{cccc}
\begin{tabular}{|c|c|}
\hline
\raisebox{-0.1cm}{$|{T_{A1}}|^2$} &\raisebox{-0.1cm}{$p_t^2$} \\[.2cm]
\hline
\raisebox{-0.1cm}{$|{T_{A2}}|^2$} &\raisebox{-0.1cm}{$\frac{p_t^2 s^2}{4}$}\\[.2cm]
\hline
\raisebox{-0.1cm}{${T_{A1}\,T_{A2}}$} &\raisebox{-0.1cm}{$0$}\\[.2cm]
\hline
\end{tabular}
\begin{tabular}{|c|c|}
\hline
 \raisebox{-0.1cm}{ $|{T_{A10}}|^2$}&\raisebox{-0.1cm}{$p_t^6$}\\[.2cm]
  \hline
  \raisebox{-0.1cm}{$|{T_{B10}}|^2$}&\raisebox{-0.1cm}{$2p_t^2$}\\[.2cm]
  \hline
  \raisebox{-0.1cm}{$|{T_{C10}}|^2$}&\raisebox{-0.1cm}{$2p_t^2$}\\[.2cm]
  \hline
  \raisebox{-0.1cm}{$|{T_{D10}}|^2$}&\raisebox{-0.1cm}{$2p_t^2$}\\[.2cm]
  \hline
  \raisebox{-0.1cm}{${T_{A10}\,T_{B10}}$}&\raisebox{-0.1cm}{$-p_t^4$}\\[.2cm]
  \hline
  \raisebox{-0.1cm}{${T_{A10}\,T_{C10}}$}&\raisebox{-0.1cm}{$-p_t^4$}\\[.2cm]
  \hline
  \raisebox{-0.1cm}{${T_{A10}\,T_{D10}}$}&\raisebox{-0.1cm}{$-p_t^4$}\\[.2cm]
  \hline
  \raisebox{-0.1cm}{${T_{B10}\,T_{C10}}$}&\raisebox{-0.1cm}{$p_t^2$}\\[.2cm]
  \hline
  \raisebox{-0.1cm}{${T_{B10}\,T_{D10}}$}&\raisebox{-0.1cm}{$p_ t^2$}\\[.2cm]
  \hline
  \raisebox{-0.1cm}{${T_{C10}\,T_{D10}}$}&\raisebox{-0.1cm}{$p_t^2$}\\[.2cm]
  \hline
\end{tabular}
&
\begin{tabular}{|c|c|}
\hline
  \raisebox{-0.1cm}{$|{T_{A11}}|^2$}&\raisebox{-0.1cm}{$2 p_t^2$}\\[.2cm]
  \hline
  \raisebox{-0.1cm}{$|{T_{B11}}|^2$}&\raisebox{-0.1cm}{$2p_t^2$}\\[.2cm]
  \hline
  \raisebox{-0.1cm}{$|{T_{C11}}|^2$}&\raisebox{-0.1cm}{$2p_t^2$}\\[.2cm]
  \hline
  \raisebox{-0.1cm}{${T_{A11}\,T_{B11}}$}&\raisebox{-0.1cm}{$p_t^2$}\\[.2cm]
  \hline
  \raisebox{-0.1cm}{${T_{A11}\,T_{C11}}$}&\raisebox{-0.1cm}{$p_t^2$}\\[.2cm]
  \hline
  \raisebox{-0.1cm}{${T_{B11}\,T_{C11}}$}&\raisebox{-0.1cm}{$p_t^2$}\\[.2cm]
  \hline
\end{tabular}
&
\begin{tabular}{|c|c|}
\hline 
  \raisebox{-0.1cm}{$|{T_{A12}}|^2$}&\raisebox{-0.1cm}{$\frac{p_t^2s^2}{2}$}\\[.2cm]
  \hline
  \raisebox{-0.1cm}{$|{T_{B12}}|^2$}&\raisebox{-0.1cm}{$\frac{p_t^2s^2}{2}$}\\[.2cm]
  \hline
  \raisebox{-0.1cm}{$|{T_{C12}}|^2$}&\raisebox{-0.1cm}{$\frac{p_t^2s^2}{2}$}\\[.2cm]
  \hline
  \raisebox{-0.1cm}{${T_{A12}\,T_{B12}}$}&\raisebox{-0.1cm}{$0$}\\[.2cm]
  \hline
  \raisebox{-0.1cm}{${T_{A12}\,T_{C12}}$}&\raisebox{-0.1cm}{$\frac{p_t^2s^2}{4}$}\\[.2cm]
  \hline
  \raisebox{-0.1cm}{${T_{B12}\,T_{C12}}$}&\raisebox{-0.1cm}{$\frac{-p_t^2s^2}{4}$}\\[.2cm]
  \hline
\end{tabular}
\end{tabular}
\caption{The results of the relevant tensor-structure products, after applying the sum rules for the meson and photon polarisations are listed. Products involving tensor structures from rows 13 and 14 of \TAB\ref{tab:tensorbasis} can be directly inferred from the second table by substituting $\epsilon_{M_1}$ with $\epsilon_{M_2}$. }
\label{tab:structureproducts}
\end{table}

Once the integration over $x$, $v$ and $z$ entering the form factors $\mathcal{H}_{TA}$, $\mathcal{H}_{TB}$, $\mathcal{H}_{TC}$ and $\mathcal{H}_{TD}$ has been performed, \EQs\eqref{eq:HTCFFsumed}, \eqref{eq:sumoddamp} and  \eqref{eq:averagesquaredamp} are then used to obtain $|\overline{\mathcal{M}}|^2$.

The fully differential cross section, expressed in terms of the kinematical variables $-u'$, $-t$ and $M^2_{12}$, is 
\begin{equation}
    \frac{d\sigma}{d(-t)d(-u')dM_{12}^2}=\frac{|\overline{\mathcal{M}}|^2}{32(2\pi)^3S_{\gamma N}^2M_{12}^2}\,.
    \label{eq:crosssectionformula}
\end{equation}
A derivation of this formula can be found in \APP\ref{app:crosssectionproof}.

\section{Phase space}

\label{sec:phase-space}

\subsection{Exact kinematics with $\Delta_{\perp}=0$}

\label{sec:exact-kinematics-phase-space}

In studies of exclusive photon-meson photoproduction, the exact kinematics was used to determine the physical range of the variables $u'$, $M_{\gamma M}^2$ and $t$. By exact kinematics, we mean that the masses of the hadrons and $t$ are not neglected in front of the hard scales. Let us recall the main ideas, adapting them for exclusive di-meson photoproduction, with further details found in \cite{Boussarie:2016qop}. 

\begin{figure}[t!]
    \centering
    \includegraphics[width=0.48\linewidth]{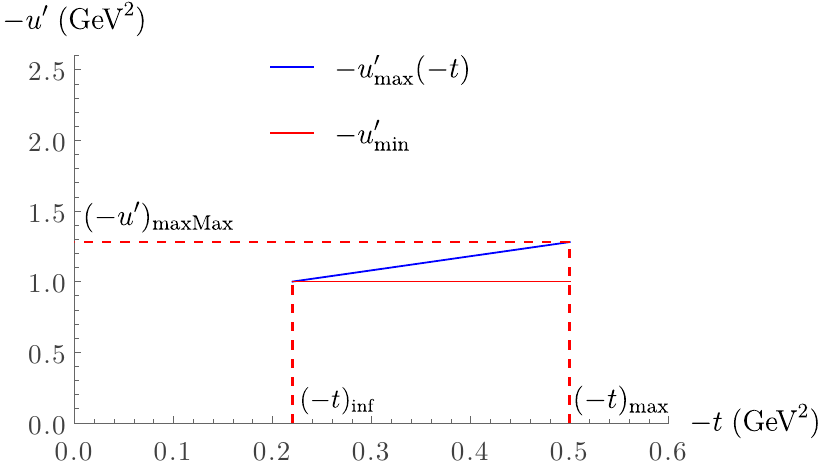}
    \includegraphics[width=0.48\linewidth]{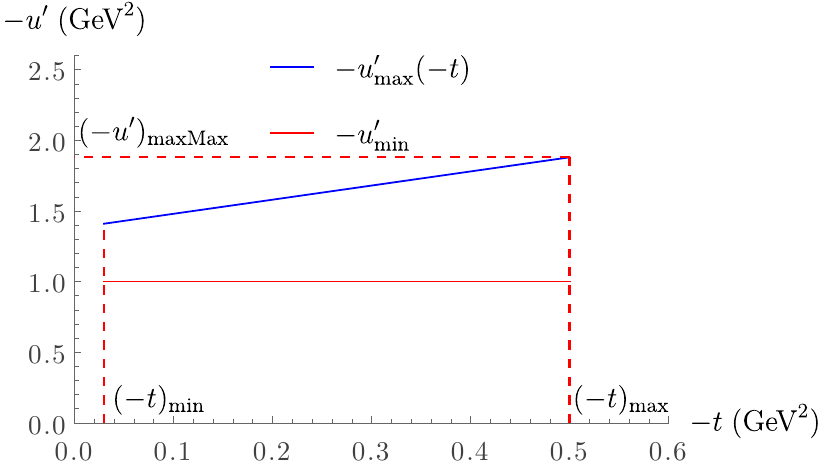}
    \caption{Phase space in the $(-t,-u')$ plane in the exact kinematics for $\Delta_\perp=0$, corresponding to the photoproduction of a $\pi\rho$ pair, with $M_{12}^2=\SI{2.4}{GeV^2}$(left) and $M_{12}^2=\SI{3}{GeV^2}$ (right), and $S_{\gamma N}=\SI{20}{GeV^2}$. The blue line corresponds to  the upper limit on $(-u')$ due to the restriction $(-t')\geq 1 \GeV^2$.}
    \label{fig:oldphasespace}
\end{figure}

First, three cutoffs are chosen in order to stay in the collinear factorisation regime, namely 
\begin{equation}
\label{eq:cut-values}
    (-u')_{\textrm{min}}=(-t')_{\textrm{min}}=\SI{1}{GeV^2}\,,\qquad    (-t)_{\textrm{max}}=\SI{0.5}{GeV^2}.
\end{equation}
We have the following relation between the Mandelstam variables,
\begin{equation}
    \label{eq:mandelstamrelation}
    M^2_{12}+t'+u'=t+m_{M_1}^2+m_{M_2}^2\,.
\end{equation}
The Mandelstam variable $u'$ thus varies in the range $[(-u')_{\textrm{min}},(-u')_{\textrm{max}}(-t)]$  at fixed $-t$ and $M_{12}^2$, with
\begin{equation}
\label{eq:uprimemaxformula}
    (-u')_{\textrm{max}}(-t)=-t-m_{M_1}^2-m_{M_2}^2+M_{12}^2-(-t')_{\textrm{min}}\,,
\end{equation}
while $(-t)\in[(-t)_{\textrm{inf}},(-t)_{\textrm{max}}]$,  where $(-t)_{\textrm{inf}}$ is the value for which $(-u')_{\textrm{max}} = (-u')_{\textrm{min}}$, so 
\begin{equation}
\label{eq:tinfformula}
    (-t)_{\textrm{inf}}=m_{M_1}^2+m_{M_2}^2-M_{12}^2+(-t')_{\textrm{min}}+(-u')_{\textrm{min}}.
\end{equation}
\EQ\eqref{eq:uprimemaxformula} represents the equation of a straight line in the plane $(-t,-u')$ with slope 1. The allowed region in $(-u')$ is thus bounded by this straight line and the line $-u'=(-u')_{\textrm{min}}$ as \FIG\ref{fig:oldphasespace} shows.

If $(-t)_{\textrm{inf}}$ is greater than $(-t)_{\textrm{min}}$, the value of $-t$ obtained from the exact kinematics by setting $\Delta_\perp$ to zero , then the physical region is a triangle, otherwise, it becomes a trapezoid, as can be seen in \FIG\ref{fig:oldphasespace}.
The value of $(-t)_{\textrm{min}}$, derived in \APP\ref{app:kinematics-Deltaperp-zero}, is given by 
\begin{equation}
    (-t)_{\textrm{min}}=\frac{1-\bar{M}_{12}^2(1+2\bar{M}^2)-\sqrt{\displaystyle 1+\bar{M}_{12}^2(\bar{M}_{12}^2-2-4\bar{M}^2)}}{2(1+\bar{M}^2)}(S_{\gamma N}-M^2)\,,
    \label{eq:tmin}
\end{equation}
where the notation with a bar implies rescaling by a factor $\frac{1}{S_{\gamma N}-M^2}$, such that the barred variables are all dimensionless. Note that in the approximation that the hadron masses are zero, $(-t)_{\mathrm{min}} = 0$.

The transition between the triangle and the trapezoid occurs when $(-t)_{\text{inf}}=(-t)_{\text{min}}$, that is, when $M_{12}^2$ equals
\begin{equation}
    M_{12\text{trans}}^2=(S_{\gamma N}-M^2)\bar{m}^2\frac{1-\bar{m}^2(1+\bar{M}^2)}{1-\bar{m}^2}\,,
\end{equation}
with 
\begin{align}
\bar{m}^2=\frac{(-u')_{\text{min}}+(-t')_{\text{min}}+m_{M_1}^2+m_{M_2}^2}{S_{\gamma N}-M^2}\,.    
\end{align}

The bounds for the variables $M_{12}^2$ and $S_{\gamma N}$ now remain to be determined. First, note that the minimal value of $M_{12}^2$ is given by the fact that the phase space becomes empty when $(-t)_{\textrm{inf}}=(-t)_{\textrm{max}}$, i.e.~when
\begin{equation}
\label{eq:Mcritformula}
    M_{12\text{crit}}^2=(-u')_{\textrm{min}}+(-t')_{\textrm{min}}+m_{M_1}^2+m_{M_2}^2-(-t)_{\textrm{max}}\,.
\end{equation}
For example, in a process where the two outgoing mesons are $\rho$-mesons, $M^2_{12\text{crit}}=\SI{2.7}{GeV^2}$, when one is a $\rho$-meson, and one is a pion, $M^2_{12\text{crit}}=\SI{2.12}{GeV^2}$ and when both mesons are pions,  $M^2_{12\text{crit}}=\SI{1.54}{GeV^2}$.

\begin{figure}[t!]
    \centering
    \includegraphics[width=0.48\linewidth]{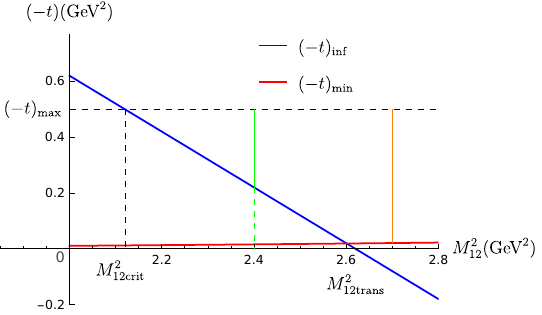}
    \hfill
    \includegraphics[width=0.48\linewidth]{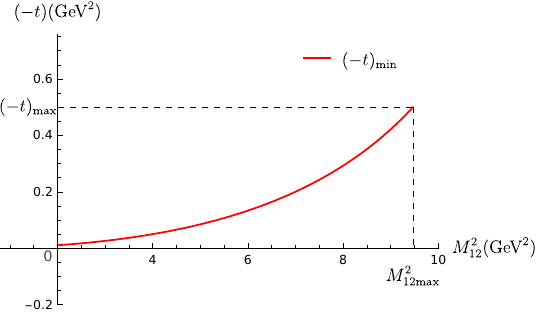}
    \caption{\textbf{Left:} Graphs of $(-t)_{\text{inf}}$ (in blue) and $(-t)_{\text{min}}$ (in red) for the process $\pi^+\rho^0_L$ at $S_{\gamma N}=\SI{20}{GeV^2}$. For $M_{12}^2\leq M_{12\text{trans}}^2$, the phase space is a triangle in the $(-t,-u')$ plane, as can be seen on the left panel of \FIG\ref{fig:oldphasespace}. The integration over $(-t)$ is between $(-t)_{\mathrm{inf}}$ and $(-t)_{\mathrm{max}}$, indicated by the green line. For $M_{12}^2\geq M_{12\text{trans}}^2$, the phase space becomes a trapezoid in the $(-t,-u')$ plane, as can be seen on the right panel of \FIG\ref{fig:oldphasespace}. The range of integration for $(-t)$ is now from $(-t)_{\mathrm{min}}$ to $(-t)_{\mathrm{max}}$, as indicated by the orange line. \textbf{Right:} Graph of $(-t)_{\mathrm{min}}$ for the same process and $S_{\gamma N}= 20 \GeV^2$, but for a wider range of $M_{12}^2$. The function $(-t)_{\mathrm{min}}$ increases with $M_{12}^2$ until it becomes equal to $(-t)_{\mathrm{max}}$ at $M_{12}^2 = M_{12\mathrm{max}}^2$.}
    \label{fig:mt-vs-M12squared}
\end{figure}

 The phase space also becomes empty when $(-t)_{\textrm{min}}=(-t)_{\textrm{max}}$. This condition gives an upper bound on $M_{12}^2$ equal to
 \begin{equation}
 \label{eq:M12Maxsquaredformula}
     M_{12\text{max}}^2=(S_{\gamma N}-M^2)\frac{-(1+2\bar{M}^2)(-\bar{t})_{\textrm{max}}+\sqrt{\displaystyle (-\bar{t})_{\textrm{max}}((-\bar{t})_{\textrm{max}}+4\bar{M}^2)}}{2\bar{M}^2}.
 \end{equation}

\FIG\ref{fig:mt-vs-M12squared} shows the variation of $(-t)_{\mathrm{min}}$ and $(-t)_{\mathrm{inf}}$ as a function of $M_{12}^2$. Note that $(-t)_{\textrm{min}}$ is an increasing function of $M_{12}^2$.

 Finally, the minimal value of $S_{\gamma N}$ is fixed by the condition $M_{12\text{max}}^2=M_{12\text{crit}}^2$, and it depends on the masses of the mesons. $S_{\gamma N\text{max}}$ is obtained when the fraction of momentum carried by the photon with respect to its source is equal to $1$. 

 \subsection{Implications of meson exchange symmetry on the kinematics}

 At low energies, the description of the kinematics in \SEC\ref{sec:exact-kinematics-phase-space} may lead to phenomenological inconsistencies. Indeed, not only $\Delta_\perp$ but also the hadron masses have been neglected in the calculation of the hard part of the amplitude. This is the limit that one works in when employing  collinear factorisation. 
 An important consequence of this approximation is that the unintegrated amplitude is related to its counterpart when the two outgoing mesons are exchanged, by the transformation $p_\perp\to-p_\perp$, $\alpha\to\bar{\alpha}$ and $v\leftrightarrow z$, see \EQ\eqref{eq:approxkinematics} and \FIG\ref{fig:illustration}. After integration over $x$, $v$ and $z$, the corresponding transformations are $p_\perp\to-p_\perp$ and $\alpha\to\bar{\alpha}$, the latter being the same as $-u'\to M_{12}^2-(-u')$, according to \EQ\eqref{eq:approxmandelstam}. 
 
 This symmetry is easily understood by considering \FIG\ref{fig:schemadimeson}. Indeed, exchanging both mesons amounts to $t'\leftrightarrow u'$. From \EQ\eqref{eq:mandelstamrelation}, this symmetry can be expressed as $(-u')\to M_{12}^2-(-u')+(-t)-m_{M_1}^2-m_{M_2}^2$, which collapses to $-u'\to M_{12}^2-(-u')$ when using the approximated kinematics.

 The physical interval for the variable $u'$ must be invariant under this transformation in the exact kinematics. Indeed,
 \begin{align}
      [(-u')_{\textrm{min}},(-u')_{\textrm{max}}(-t)]&\to
      [(-t')_{\textrm{min}},(-t')_{\textrm{max}}(-t)]\nonumber\\
      &= [(-u')_{\textrm{min}},\,M_{12}^2-(-u')_{\textrm{min}}+(-t)-m_{M_1}^2-m_{M_2}^2]\nonumber\\
      &= [(-u')_{\textrm{min}},\,M_{12}^2-(-t')_{\textrm{min}}+(-t)-m_{M_1}^2-m_{M_2}^2]\nonumber\\
      &=  [(-u')_{\textrm{min}},\,(-u')_{\textrm{max}}(-t)]\,,
      \label{eq:uprimeinterval}
 \end{align}
 where we have used
 \begin{align}
      (-t')_{\textrm{max}}(-t)=M^2_{12}-(-u')_{\mathrm{min}}+(-t)-m_{M_1}^2-m_{M_2}^2\,,
 \end{align}
the expression for $(-u')_{\textrm{max}}(-t)$ in    \EQ\eqref{eq:uprimemaxformula}, and  $(-u')_{\textrm{min}}=(-t')_{\textrm{min}}$. So, the physical interval in $(-u')$ is invariant under the exchange $u'\leftrightarrow t'$. Note that the same symmetry would be also observed in the approximated kinematics when $-t$ and the meson masses are neglected. In what follows, it is convenient to introduce the function $B(-t)$ given by
\begin{align}
    B(-t) &= m_{M_1}^2 + m_{M_{2}}^2 - (-t)\,, \qquad
    B_{\mathrm{max}} \equiv B((-t)_{\mathrm{max}})\,.
\end{align}

The issue is that the amplitude itself is not symmetric under $(-u')\to M_{12}^2-(-u')-B(-t)$, but rather, it is symmetric  under $-u'\to M^2_{12}-(-u')$, because $-t$ and the meson masses have been neglected in the calculation of the hard part. Hence, by using the exact kinematics at $\Delta_\perp=0$ for the phase space, we lose the meson exchange symmetry, and this becomes worse as $ M_{12}^2$ becomes closer to  $|B(-t)|$.

    To demonstrate this problem explicitly, consider the photoproduction of a $\rho^0_L\rho_T^0$ pair  with $M_{12}^2=\SI{3}{GeV^2}$. Then, the upper bound of $-u'$ at $(-t) = (-t)_{\mathrm{max}}$, denoted by $(-u')_{\textrm{max}\textrm{Max}}$, is 
    \begin{equation}
    (-u')_{\textrm{max}\textrm{Max}}=(-t)_{\textrm{max}}-2 m_{\rho}^2-(-t')_{\textrm{min}}+M_{12}^2=\SI{1.3}{GeV^2}\,.  
    \end{equation}

\begin{figure}[t!]
        \centering
        \includegraphics[width=0.48\linewidth]{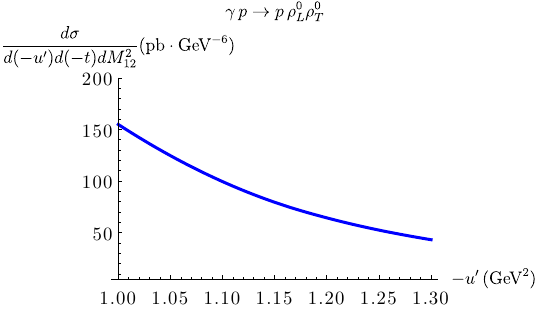}
        \includegraphics[width=0.48\linewidth]{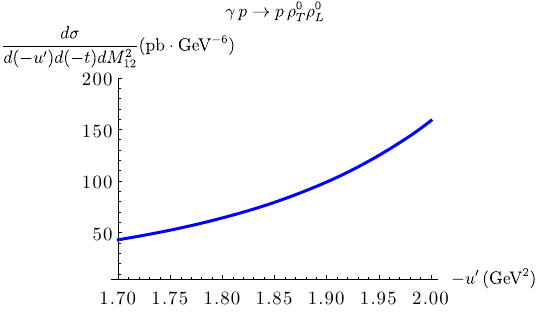}\\[0.2cm]
        (a)\hspace{7.2cm}(b)\\[0.4cm]
        \includegraphics[width=0.48\linewidth]{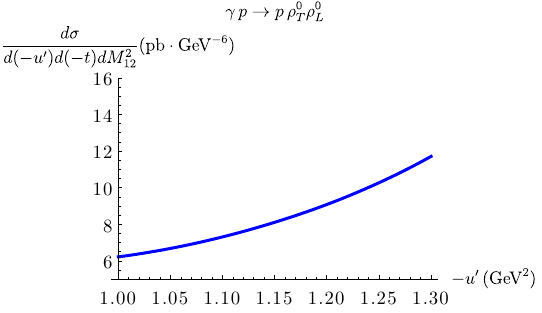}
        \includegraphics[width=0.48\linewidth]{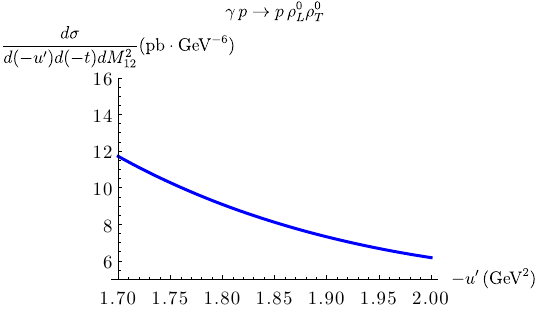}\\[0.2cm]
        \hspace{.16cm}(c)\hspace{7.2cm}(d)
        \caption{Differential cross sections of $\rho^0_L\rho^0_T$ (a) and $\rho^0_T\rho^0_L$ (c) in the exact kinematics at $\Delta_\perp= 0$, showing only the physically allowed range in $-u'$, using $M_{12}^2 = 3 \GeV^2$, $S_{\gamma N} = 20 \GeV^2 $ and $-t=(-t)_{\text{min}}$. Since the two differential cross sections have different orders of magnitude, it is clear that they will yield different cross sections after integration over $(-u')$. To illustrate the symmetry of the amplitude, the corresponding differential cross section of $\rho^0_T\rho^0_L$ (b) and $\rho^0_L\rho^0_T$ (d) are shown for $-u'$ between 1.7 and $\SI{2}{GeV^2}$, which is excluded in the exact kinematics. Each correspond to reflections of the plots on the left in the line $-u'=M_{12}^2/2 = 1.5 \GeV^2$.}
        \label{fig:pbkinerho0Lrho0T}
    \end{figure}
    
    Consider the fully differential cross section of $\rho_L^0 \rho^0_T$ and its counterpart $\rho_T^0 \rho^0_L$ after the exchange of the outgoing mesons, which are shown on the left panels of \FIG\ref{fig:pbkinerho0Lrho0T}. Since $M_{12}^2$ and $-t$ are insensitive to the exchange of the two mesons, we expect $\frac{d\sigma}{d(-t)dM_{12}^2}$ to be the same for both processes. From the plots, it is clear that this is not the case. On the other hand, we know that the amplitudes of each process are related by the transformation $-u'\to M_{12}^2-(-u')$, which is equivalent to taking the reflection with respect to $-u'={M_{12}^2}/{2}=\SI{1.5}{GeV^2}$. This symmetry is indeed what is observed by comparing the left and right panels of \FIG\ref{fig:pbkinerho0Lrho0T}. This is further illustrated in \FIG\ref{fig:rho0Trho0Landsym}.

\begin{figure}[t!]
        \centering
        \includegraphics[width=0.48\linewidth]{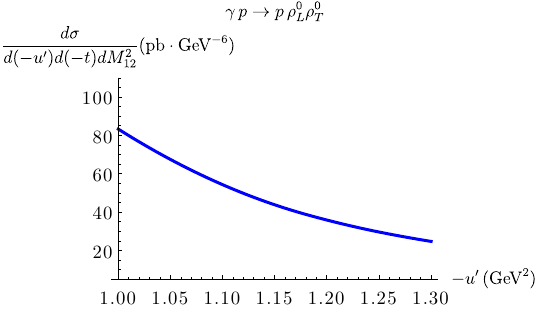}
        \includegraphics[width=0.48\linewidth]{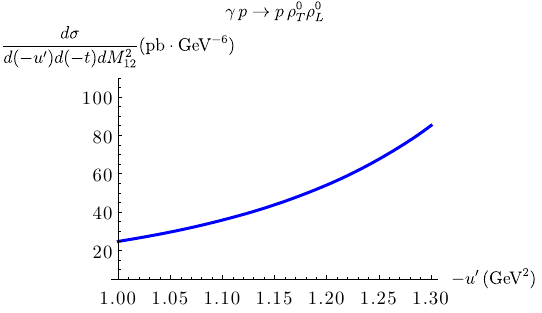}
        \caption{Differential cross section of $\rho^0_L\rho^0_T$ (left) and $\rho^0_T\rho^0_L$ (right) at $S_{\gamma N}=\SI{20}{GeV^2}$, $M_{12}^2=\SI{3}{GeV^2}$ and $(-t)=(-t)_{\text{min}}$. The symmetry in $-u'\to M_{12}^2-(-u')+(-t)_{\textrm{min}}-m_{M_1}^2-m_{M_2}^2$ has been restored on the left (right) through ``brute-force'' by averaging the blue (red) curve and the red (blue) curve after its reflection in the line $-u' = (M_{12}^2 -B(-t_{\text{min}})/2$ in \FIG\ref{fig:rho0Trho0Landsym}}
        \label{fig:averagerho0Lrho0T}
    \end{figure}

\begin{figure}[t!]
          \centering
          \includegraphics[width=0.9\linewidth]{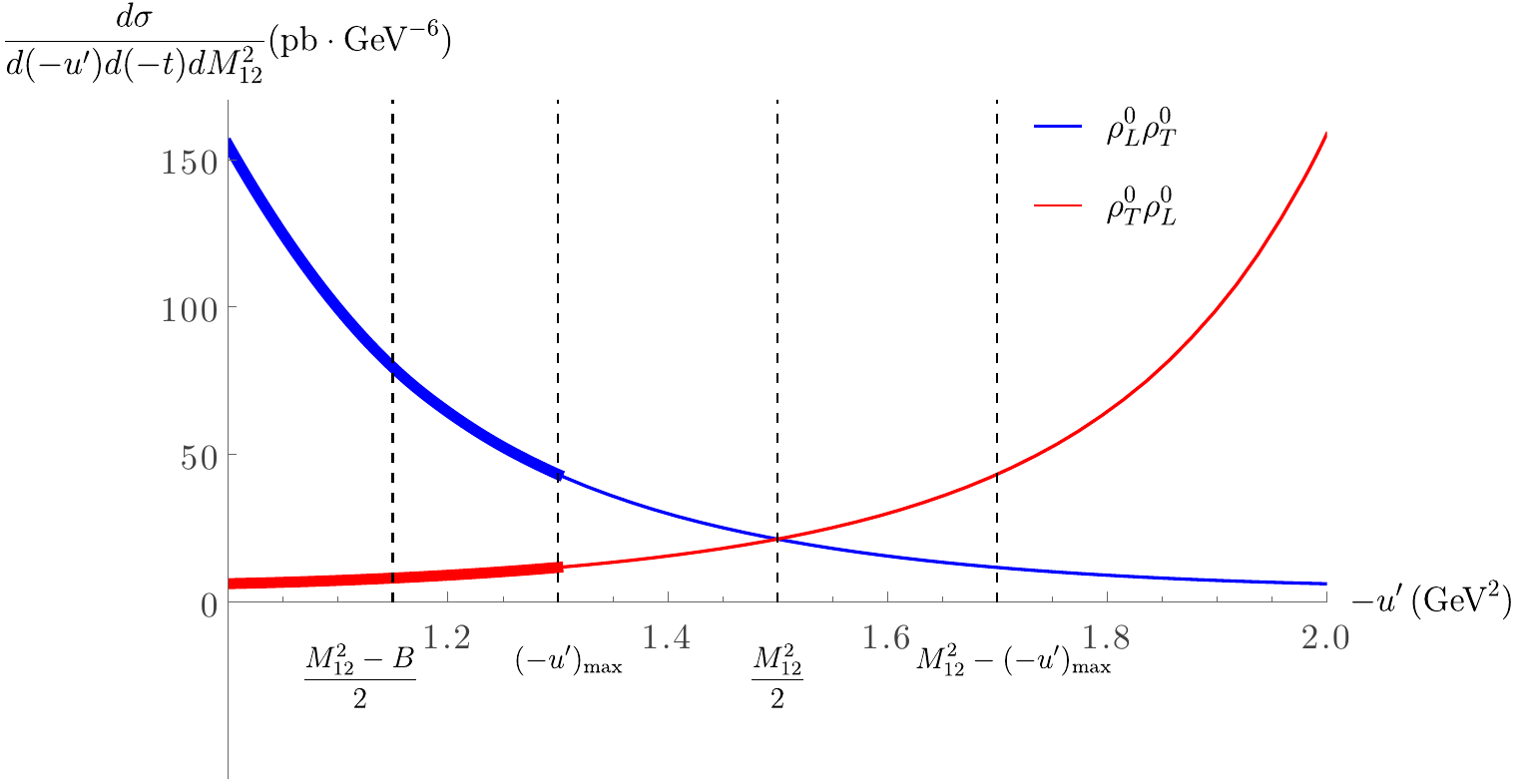}
          \caption{Plots of the differential cross sections of $\rho^0_L\rho^0_T$ (in blue) and $\rho^0_T\rho^0_L$ (in red) for $M_{12}^2=\SI{3}{GeV^2}$, $S_{\gamma N}=\SI{20}{GeV^2}$ and $-t=(-t)_\text{min}$. With these values, $(-u')_{\text{max}}=\SI{1.3}{GeV^2}$ (indicated by a dashed line) and $B=m_{M_1}^2+m_{M_2}^2+(-t)_{\text{max}}=\SI{0.7}{GeV^2}$. The bold part of the curves corresponds to the kinematically allowed region in $(-u')$ in the exact kinematics at $\Delta_\perp=0$. The red and blue curves are symmetric in the line $-u'=\frac{M_{12}^2}{2} = 1.5 \GeV^2$. A possible way to restore the symmetry is to average the bold parts on the left, after performing the reflection of one of them about the $-u'=\frac{M_{12}^2-B}{2}$ axis, resulting in \FIG\ref{fig:averagerho0Lrho0T}. Note that the blue  and red bold parts of the curves correspond to (a) and (c) in \FIG\ref{fig:pbkinerho0Lrho0T} respectively, while the blue and red parts of the curves on the right of $(-u') = M_{12}^2 - (-u')_{\mathrm{max}}$ correspond to (b) and (d) in \FIG\ref{fig:pbkinerho0Lrho0T} respectively.}
          \label{fig:rho0Trho0Landsym}
      \end{figure}

   The main issue is that the transformation on $-u'$ on the amplitude (where the approximated kinematics was used) due to the exchange of the two outgoing mesons is \textit{different} from that on the phase space (where the exact kinematics at $\Delta_{\perp} = 0 $ was used).  One solution to restore the symmetry would be to take the average of the differential cross section of $\rho^0_L\rho^0_T$ in (a) of \FIG\ref{fig:pbkinerho0Lrho0T} and its counterpart after the exchange of the two outgoing mesons and \textit{after} applying the transformation $(-u')\to M_{12}^2-(-u')+(-t)-m_{M_1}^2-m_{M_2}^2$ based on the exact kinematics (this corresponds to reflecting (c) of \FIG\ref{fig:pbkinerho0Lrho0T} in the line $-u' = 1.15 \GeV^2$). This leads to the plot in the left panel of \FIG\ref{fig:averagerho0Lrho0T}. The corresponding procedure can also be performed for $\rho^0_T \rho^0_L$, and this leads to the right panel of \FIG\ref{fig:averagerho0Lrho0T}. In \SEC\ref{sec:phase-space-approx-kinematics}, an alternative solution is discussed, whereby the phase space itself is modified such that it possesses the symmetries consistent with the hard part of the process.
     
    \subsection{Phase space in the approximated kinematics}

    \label{sec:phase-space-approx-kinematics}

\begin{figure}[t!]
    \centering
    \includegraphics[width=0.6\linewidth]{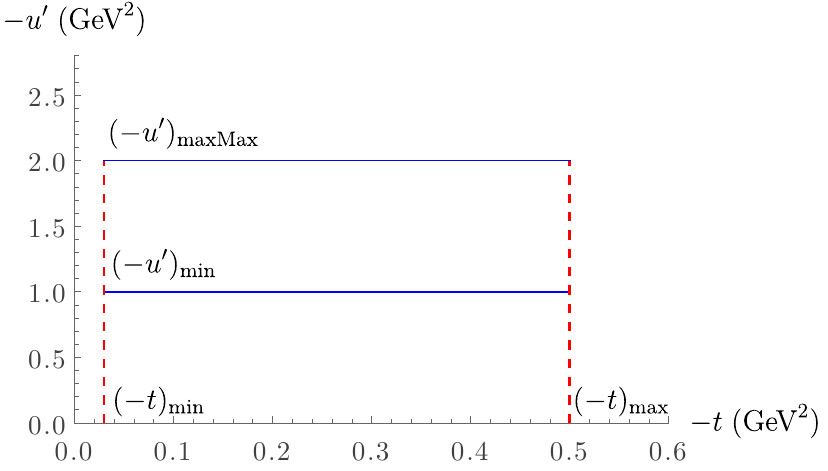}
    \caption{Phase space in the $(-t,-u')$ plane in the approximated kinematics, with $M_{12}^2=\SI{3}{GeV^2}$ and $S_{\gamma N}=\SI{20}{GeV^2}$. Since the hadron masses and the Mandelstam variable $t$ are neglected in the determination of the allowed phase space, the upper bound for $-u'$ is fixed to $M_{12}^2-(-u')_{\text{min}}$. The physically allowed phase space is thus a rectangle in the $(-t,-u')$ plane.}
    \label{fig:newphasespace}
\end{figure}

    As seen above, the lack of symmetry of exchanging the two outgoing mesons is due to  the additional terms $(-t)-m_{M_1}^2-m_{M_2}^2$ in \EQ\eqref{eq:mandelstamrelation} in the treatment of the allowed phase space for $-u'$. Another solution to recover the symmetry, other than averaging the cross section, is to neglect $t$ and the meson masses when determining the phase space boundaries for $-u'$. 
    Thus, \EQ\eqref{eq:uprimemaxformula} becomes 
    \begin{equation}
        (-u')_{\textrm{max}}=M_{12}^2-(-t')_{\textrm{min}}.
    \end{equation}
Thus, the physical region for $-u'$ in the  $(-t,-u')$ plane now lies between two horizontal lines: $-u'=(-u')_{\textrm{min}}$, and $(-u')_{\text{max}}=M_{12}^2-(-t')_{\textrm{min}}$. There is no dependence of $(-u')_{\textrm{max}}$ on $-t$ anymore, so $-t$ varies between $(-t)_{\textrm{min}}$ and $(-t)_{\textrm{max}}$, and the physical region is a rectangle, as shown in \FIG\ref{fig:newphasespace}.
The next things to address are the bounds on the invariant mass $M_{12}^2$. The maximum value of $M_{12}^2$ is unchanged, since it is determined by the condition $(-t)_{\textrm{min}}=(-t)_{\textrm{max}}$. The lower bound, however, is no longer given by \EQ\eqref{eq:Mcritformula}, which was obtained by $(-t)_\mathrm{inf} = (-t)_{\mathrm{max}}$, since $(-t)_\mathrm{inf}$ does not exist here,\footnote{In the exact kinematics, $(-t)_{\mathrm{inf}}$ corresponds to the solution of $(-u')_{\mathrm{max}}(-t) = (-u')_{\mathrm{min}}$.} which is due to the fact that $(-u')_{\mathrm{max}}$ is not dependent on $-t$.

To find the lower bound on $M_{12}^2$, we use the condition $(-u')_{\textrm{max}}=(-u')_{\textrm{min}}$, which occurs when 
\begin{equation}
\label{eq:newMcritformula}
    M_{12\text{crit}}^2=(-u')_{\textrm{min}}+(-t')_{\textrm{min}}=\SI{2}{GeV^2}.
\end{equation}
The bounds for $S_{\gamma N}$ are obtained in a similar way, in particular, the lower bound $S_{\gamma N \mathrm{min}}$ is obtained by equating $M_{12\mathrm{crit}}^2$ in \EQ\eqref{eq:newMcritformula} with $M_{12\mathrm{max}}^2$ in \EQ\eqref{eq:M12Maxsquaredformula}.

The solution presented in this subsection is the one we use to obtain the results in \SEC\ref{sec:results}.

\subsection{Sampling in the $(\alpha,\xi)$ plane}

As discussed in \SEC\ref{sec:factorisation-of-s}, a crucial feature of the average squared amplitude is that the parameter $s$ completely factorises from it. We can always write 
\begin{equation}
\label{eq:sfactorisation}
    |\overline{\mathcal{M}}|^2=\frac{f_1^2 f_2^2}{s^3}g(\alpha,\xi)\,,
\end{equation}
where $g$ is some dimensionless positive function of $\alpha$ and $\xi$. Given the structure of the averaged squared amplitude, rather than sampling the phase space over the 3 dimensionful kinematical variables $S_{\gamma N}$, $M_{12}^2$ and $(-u')$, it makes more sense to instead sample over only 2 dimensionless variables $\alpha$ and $\xi$. In that case, each phase space point corresponds to a \textit{family} of $S_{\gamma N}$, whereby for a chosen value of $S_{\gamma N}$, one can reconstruct the other relevant kinematical variables through 
\begin{align}
\label{eq:sgm12uprime}
    \left(S_{\gamma N},\,
    M_{12}^2,\,
    -u'\right)
=
    \left(S_{\gamma N},\,
    \frac{2\xi}{1+\xi}(S_{\gamma N}-M^2),\,
    \frac{2\xi\alpha}{1+\xi}(S_{\gamma N}-M^2)\right)\,,
\end{align}
where we have used the analogue of \EQ\eqref{eq:approxmandelstam} when $M \neq 0$.
Sampling the phase space in this way is very efficient, and it is in fact the idea behind the rescaling argument described in \SEC5.5 in \cite{Boussarie:2016qop}, where one is able to recycle the phase space points $ \left(S_{\gamma N},\,
    M_{12}^2,\,
    -u'\right)$ for a given $S_{\gamma N}$ for \textit{other} values of $S'_{\gamma N}<S_{\gamma N}$. Nevertheless, explicitly using this sampling procedure gives us a better control on the implementation of \textit{importance sampling}, which becomes very important at high energies.

We want to distribute the integration points in the $(\alpha, \xi)$ plane, in such a way that we use a minimal number of integration points while covering the entire kinematical domain where collinear factorisation is expected to hold. The boundaries of $\alpha$ and $\xi$ should be determined such that one is able to calculate all relevant kinematical points for the range of $S_{\gamma N}$ that we are interested in. Note that $\xi$, given by
\begin{equation}
    \label{eq:xiformula}
    \xi=\frac{M_{12}^2}{2(S_{\gamma N}- M^2)-M_{12}^2}\,,
    \end{equation}
is an increasing function of $M_{12}^2$ and a decreasing function of $S_{\gamma N}$. Hence, for a given beam energy $E_\text{beam}$ in the target rest frame, we determine $S_{\gamma N\text{max}}$, the value for which the momentum fraction of the source carried by the photon equals $1$,
\begin{equation}
  S_{\gamma N\text{max}}=2E_\text{beam}M +M^2\,.
\end{equation}
From this, the minimal value of $\xi$ is determined through
\begin{equation}
    \xi_{\textrm{min}}=\frac{M_{12\text{crit}}^2}{2(S_{\gamma N\text{max}}-M^2)-M_{12\text{crit}}^2}\,,
\end{equation}
since $M^2_{12\text{crit}}$ is the minimum value of $M^2_{12}$, see \EQ\eqref{eq:newMcritformula}.

The determination of $\xi_{\mathrm{max}}$ is slightly more complicated. Na\"ively, one could think that it is given by substituting $M_{12\mathrm{max}}^2$ and $S_{\gamma N \mathrm{min}}$ into \EQ\eqref{eq:xiformula}. However, this argument overlooks the fact that $M_{12\mathrm{max}}^2$ is itself a function of $S_{\gamma N}$, which implies that this configuration may not actually be the maximal value of $\xi$.

The correct way to approach this problem is to simply substitute $M_{12\mathrm{max}}^2$ (since $\xi$ is an increasing function of $M_{12}^2$) into \EQ\eqref{eq:xiformula}, and then maximising over $S_{\gamma N}$. We start by rewriting the expression for $M_{12\mathrm{max}}^2$ in \EQ\eqref{eq:M12Maxsquaredformula} as 
\begin{align}
\label{eq:M12max-kappa}
        M_{12\text{max}}^2=\kappa(S_{\gamma N}-M^2)-(-t)_{\textrm{max}}\,,
\end{align}
where
\begin{align}
    \kappa = \frac{\sqrt{(-t)_{\textrm{max}}((-t)_{\textrm{max}}+4M^2)}-(-t)_{\textrm{max}}}{2M^2}\,.
\end{align}
Taking $(-t)_{\mathrm{max}} = 0.5 \GeV^2$ and $M = 0.938 \GeV$ gives $\kappa \approx 0.52$. Substituting $M_{12\text{max}}^2$ from \EQ\eqref{eq:M12max-kappa} into \EQ\eqref{eq:xiformula}, and maximising over $S_{\gamma N}$ gives
\begin{equation}
\label{eq:max-xi}
    \xi_{\textrm{max}}=\max_{S_{\gamma N}}\left[\frac{\kappa(S_{\gamma N}-M^2)-(-t)_{\textrm{max}}}{(2-\kappa )(S_{\gamma N}-M^2)+(-t)_{\textrm{max}}}\right] = \frac{\kappa(S_{\gamma N\mathrm{max}}-M^2)-(-t)_{\textrm{max}}}{(2-\kappa )(S_{\gamma N\mathrm{max}}-M^2)+(-t)_{\textrm{max}}}\,,
\end{equation}
which can be deduced from the fact that the argument of the maximum function after the first equality is an \textit{increasing} function of $S_{\gamma N}$.

\begin{figure}[t!]
    \centering
    \includegraphics[width=0.48\linewidth]{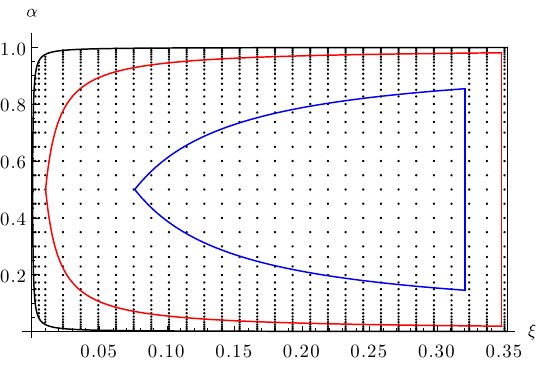}
    \caption{The distribution of phase space points in the $(\alpha,\xi)$ plane is shown. The boundaries in blue, red and black correspond to $S_{\gamma N \mathrm{max}} =$ 15, 100 and 2000 $\GeV^2$ respectively. For each colour, the curved lines correspond to $\alpha_{\mathrm{min}}$ and $\alpha_{\mathrm{max}}$ in \EQ\eqref{eq:alpha-min-max}, while the vertical line corresponds to $\xi_{\mathrm{max}}$ in \EQ\eqref{eq:max-xi}. The black dots represent the phase space that are calculated for $S_{\gamma N \mathrm{max}} = 2000 \GeV^2$, taking into account the importance sampling for small $\xi$, $\alpha$ and $\bar \alpha$. These points are fully enclosed by the black lines.}
    \label{fig:phase-space-distribution-of-points}
\end{figure}

From \EQ\eqref{eq:uprimemaxformula}, $(-u')$ lies in $[(-u')_{\text{min}}, \,M_{12}^2-(-u')_{\text{min}}]$. Since $\alpha=\frac{-u'}{M_{12}^2}$ (see \EQ\eqref{eq:approxmandelstam}), the values of $\alpha$ have to be in $[\frac{(-u')_{\text{min}}}{M_{12}^2},1-\frac{(-u')_{\text{min}}}{M_{12}^2}]$. Inverting \EQ\eqref{eq:xiformula}, $M_{12}^2$ reads 
\begin{equation}
\label{eq:m12formula}
    M_{12}^2(S_{\gamma N},\xi)=\frac{2\xi}{1+\xi}(S_{\gamma N}-M^2)\,.
\end{equation}
Hence, for a given value of $\xi$, the minimum/maximum values of $\alpha$ are obtained by taking the largest possible $M_{12}^2$ compatible with that value of $\xi$. From \EQ\eqref{eq:m12formula}, it is easy to see that this occurs at $S_{\gamma N} = S_{\gamma N \mathrm{max}}$. Consequently, the interval for $\alpha$ may be taken as
\begin{align}
\label{eq:alpha-min-max}
    [\alpha_{\mathrm{min}},\alpha_{\mathrm{max}}]= \left[\frac{(-u')_{\textrm{min}}}{M_{12}^2(S_{\gamma N\text{max}},\xi)},1-\frac{(-u')_{\textrm{min}}}{M_{12}^2(S_{\gamma N\text{max}},\xi)}\right]\,.
\end{align}
The boundaries of the phase space in the $(\alpha,\xi)$ plane for different values of $S_{\gamma N \mathrm{max}}$ are shown in \FIG\ref{fig:phase-space-distribution-of-points}.

It is important to emphasise that the boundaries of $\alpha$ and $\xi$ are determined simply on the basis of covering the \textit{full} phase space. It may be that in practice, the value of the cross section is extremely suppressed in certain regions (e.g.~large $\xi$ at large $S_{\gamma N}$). This could be exploited in a numerical analysis.

\subsection{Constraints based on avoiding resonances}

\label{subsec:constraints-resonances}

 Meson-nucleon resonances (e.g.~$\Delta$ baryon resonance) are expected to be avoided in the collinear factorisation kinematics for large $(-t')$ and $(-u')$ and small $(-t)$. Nevertheless, it is worthwhile to verify whether the values of the kinematical cuts employed in \EQ\eqref{eq:cut-values} are sufficient to ensure that this is true \textit{numerically}. For instance, we could determine whether imposing the cuts
 \begin{equation}
 \label{eq:meson-nucleon-resonances}
     M_{M_1N'}^2\geq\SI{2}{GeV^2}\;,\;\;M_{M_2N'}^2\geq\SI{2}{GeV^2}\,,
 \end{equation}
 leads to further restrictions on the phase space derived in previous sections.

In the approximation where $\Delta_t=m_{M_1}=m_{M_2}=0$,
one can show, from \EQ\eqref{eq:M2Nformula2}, \EQ\eqref{eq:M1Nformula2}, \EQ\eqref{eq:tauformula}  and  \EQ\eqref{eq:xiexact} of \APP\ref{app:further-kinematics}, that
 \begin{align}
 \label{eq:M2Nprimed}
     M_{M_{2} N'}^2&=(M^2-S_{\gamma N}+(-t)_{\mathrm{min}}+M_{12}^2)\frac{u'M_{12}^2}{(M_{12}^2+(-t)_{\mathrm{min}})^2}+M^2+\frac{M^2(u'+M_{12}^2+(-t)_{\mathrm{min}})}{S_{\gamma N}-M^2-M_{12}^2-(-t)_{\mathrm{min}}}\,,\\
     \label{eq:M1Nprimed}
     M_{M_{1} N'}^2&=S_{\gamma N}-M_{12}^2-(M^2-S_{\gamma N}+(-t)_{\mathrm{min}}+M_{12}^2)\frac{u'M_{12}^2}{(M_{12}^2+(-t)_{\mathrm{min}})^2}\nonumber \\
     & -\frac{M^2(u'+M_{12}^2+(-t)_{\mathrm{min}})}{S_{\gamma N}-M^2-M_{12}^2-(-t)_{\mathrm{min}}}\,,
 \end{align}
 where $(-t)_{\text{min}}$ is given in \EQ\eqref{eq:tformulaapprox}.
 
 Using the approximated form in \EQs\eqref{eq:M2Nprimed} and \eqref{eq:M1Nprimed}, it can be verified that the cuts in \EQ\eqref{eq:meson-nucleon-resonances} are always satisfied in the phase space defined by the kinematical cuts specified in \EQ\eqref{eq:cut-values}.
\subsection{Modelling the $t$-dependence}

The amplitude depends on $t$ only through the GPDs, which are computed in \APP\ref{app:GPD-modelling}. Their $t$-dependence is assumed to factorise from the GPDs completely, and taken to have the functional form of the dipole form factor \cite{Dunning:1966xiq,Perdrisat:2006hj}
\begin{equation}
\label{eq:t-dependence}
    F_H(t) = \frac{C^2}{(t-C)^2}\,.
\end{equation}
The single differential cross section then reads
\begin{equation}
\label{eq:integration-over-t-mup}
    \frac{d\sigma}{dM_{12}^2}=\int_{(-t)_{\text{min}}}^{(-t)_{\text{max}}}\;d(-t)\;\int_{(-u')_{\text{min}}}^{(-u')_{\text{max}}}\;d(-u')\,\frac{F_H^2(t)}{F_H^2(t_\text{min})}\times\frac{d\sigma}{d(-t)d(-u')dM_{12}^2}\bigg{|}_{-t=(-t)_{\text{min}}}.
\end{equation}
All the plots for the fully differential cross section in this paper are shown at $(-t)=(-t)_{\text{min}}$.

\section{Results}

\label{sec:results}

\begin{figure}[t!]
\includegraphics[width=.49\linewidth]{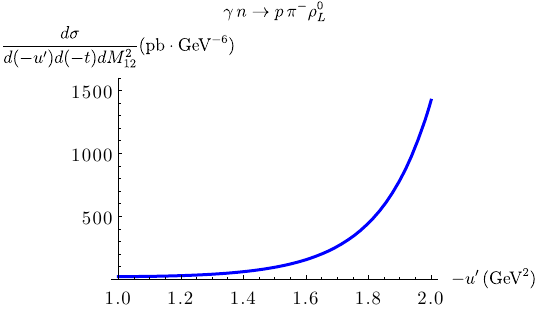}  \includegraphics[width=.49\linewidth]{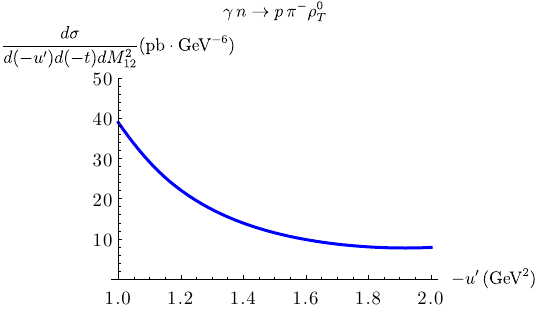}
\\[0.5cm]
\includegraphics[width=.49\linewidth]{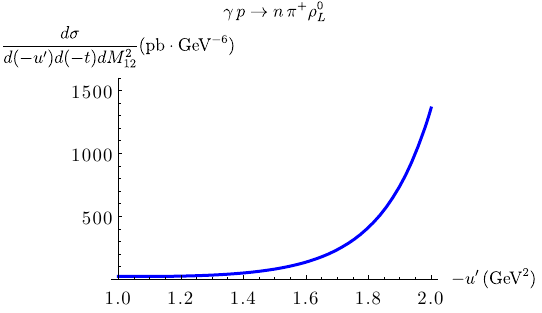}
\includegraphics[width=.49\linewidth]{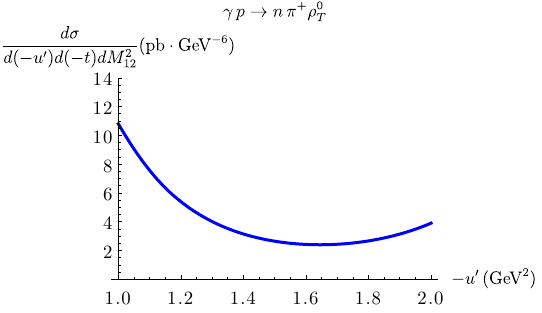}
\caption{Fully differential cross section as a function of $-u'$ for the photoproduction of $\pi^-\rho^0_L$ (top-left), $\pi^-\rho^0_T$ (top-right), $\pi^+\rho^0_L$ (bottom-left) and $\pi^+\rho^0_T$ (bottom-right) at $S_{\gamma N}= \SI{20}{GeV^2}$, $M^2_{12}=\SI{3}{GeV^2}$ and $-t=(-t)_\text{min}$.}
\label{fig:differential-cross-section1}
  \end{figure}

The results obtained in this paper can be used to study exclusive di-meson photoproduction across a wide range of photon-nucleon centre of mass energies $S_{\gamma N}$. In principle, the kinematics in both fixed-target experiments such as those at JLab and COMPASS/AMBER, as well as in collider experiments such as LHC and RHIC in ultra-peripheral collisions (UPCs) and the future EIC, which allows the study of the small $\xi$ limit of GPDs, can be covered.

As a proof of concept, in the present article, we show results for the fully differential cross section as a function of $(-u')$, using values for $S_{\gamma N}$ and $M_{12}^2$  to match the typical set-up at JLab, with $S_{\gamma N } = 20 \GeV^2$ and $M_{12}^2 = 3 \GeV^2$, the latter of which falls in the physical range for $M_{12}^2$, see \EQs\eqref{eq:newMcritformula} and \eqref{eq:M12max-kappa}. The results for four processes,\footnote{We note that in \cite{ElBeiyad:2010pji}, the corresponding plot for the differential cross section for $\gamma p \to n \pi^+\rho^0_T$ in \FIG 6 differs from the one we present in the bottom-right of \FIG\ref{fig:differential-cross-section1}. First, the units for the vertical axis in \cite{ElBeiyad:2010pji} should be $\mathrm{pb\cdot GeV^{-6}}$ instead of $\mathrm{nb\cdot GeV^{-6}}$. Second, the integration is organised differently there and unfortunately, the treatment of the ERBL region of the integration was not done properly (indeed our results match in the DGLAP region).} namely $\pi^- \rho^0_L$, $\pi^- \rho^0_T$, $\pi^+ \rho^0_L$ and $\pi^+ \rho^0_T$, are shown in \FIG\ref{fig:differential-cross-section1}.  While the cross sections of the exclusive photoproduction of a photon-meson pair \cite{Boussarie:2016qop,Duplancic:2018bum,Duplancic:2022ffo,Duplancic:2023kwe} did not exceed a few tens of $\hbox{pb}\cdot \hbox{GeV}^{-6}$, here, the values can reach the order of a thousand, for processes with longitudinal $\rho^0_L$  in the final state. We also note the fact that the cross section is very similar (but not exactly the same) for $\pi^- \rho^0_L$ and $\pi^+ \rho^0_L$.

 \begin{figure}[t!]
\includegraphics[width=.49\linewidth]{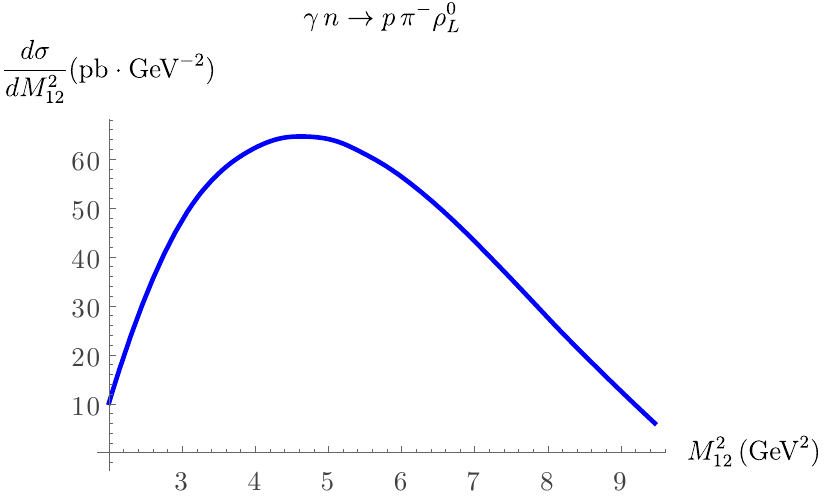}  \includegraphics[width=.49\linewidth]{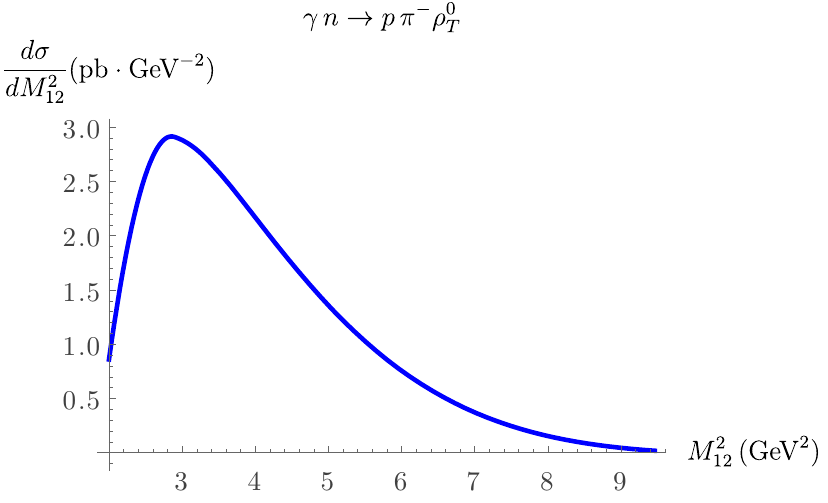}
\\[0.5cm]
\includegraphics[width=.49\linewidth]{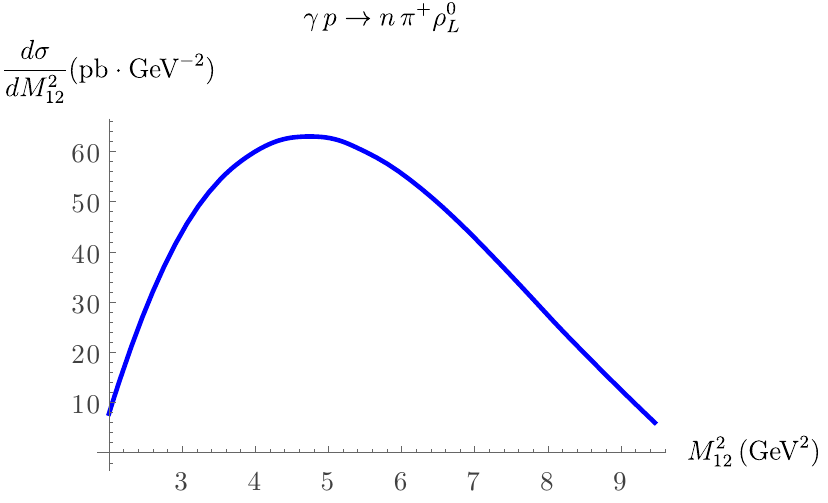}
\includegraphics[width=.49\linewidth]{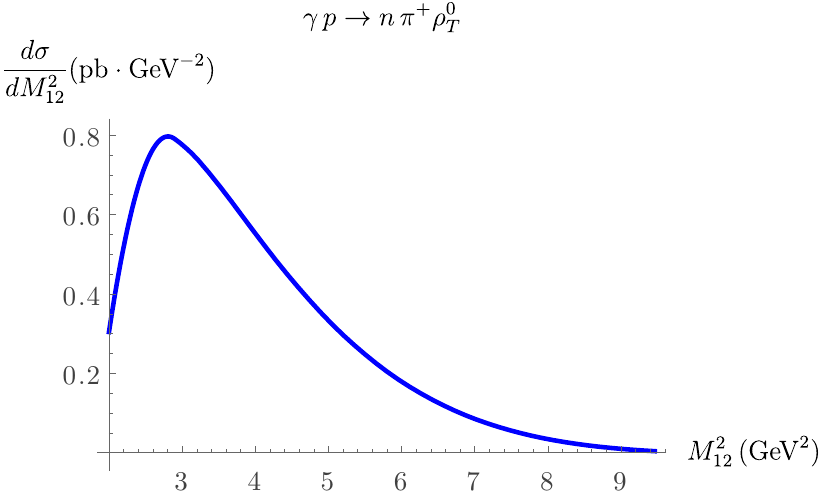}
\caption{Single differential cross sections as a function of $M_{12}^2$ for the photoproduction of $\pi^-\rho^0_L$ (top-left), $\pi^-\rho^0_T$ (top-right), $\pi^+\rho^0_L$ (bottom-left), $\pi^+\rho^0_T$ (bottom-right)  at $S_{\gamma N}=\SI{20}{GeV^2}$.}
\label{fig:singlediff}
  \end{figure}

Performing the integration over $(-u')$ and $(-t)$, as described in \EQ\eqref{eq:integration-over-t-mup}, we obtain single differential cross sections in $M_{12}^2$ which are shown in \FIG\ref{fig:singlediff} for the same four processes considered in \FIG\ref{fig:differential-cross-section1}. According to \EQ\eqref{eq:M12max-kappa}, the upper bound of $M_{12}^2$ is $M^2_{12\text{max}}=\SI{9.46}{GeV^2}$.

 \begin{figure}[t!]
\includegraphics[width=.49\linewidth]{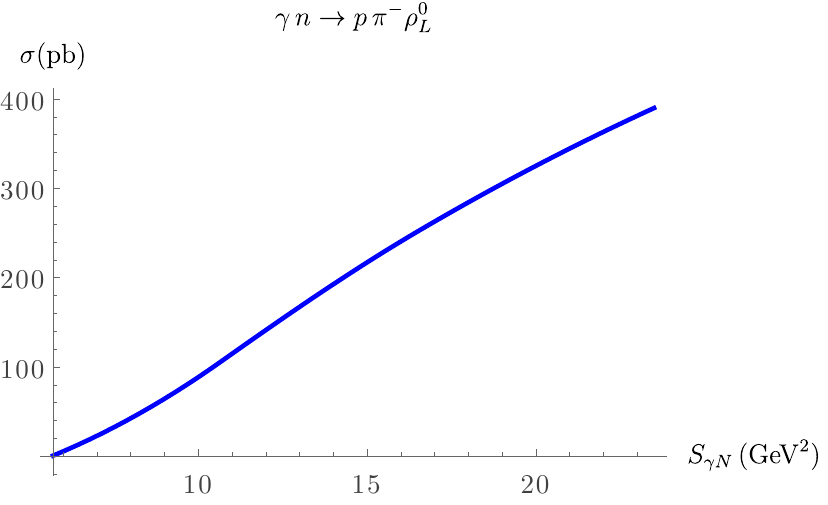}  \includegraphics[width=.49\linewidth]{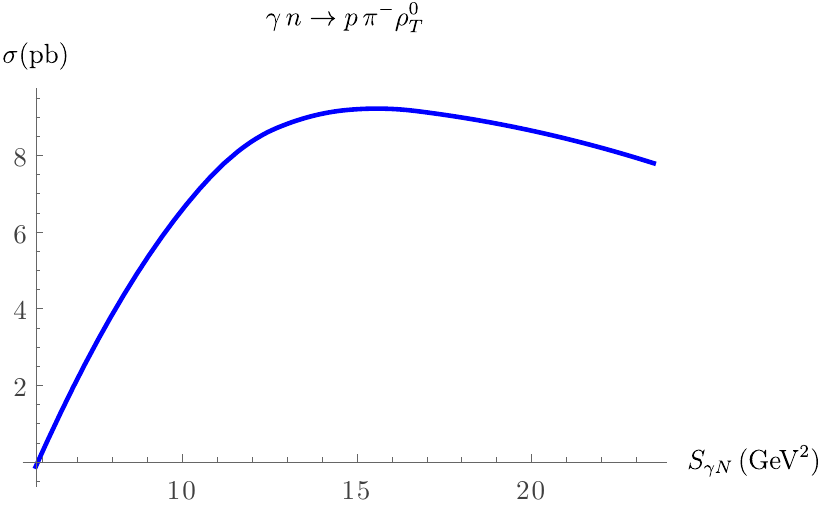}
\\[0.5cm]
\includegraphics[width=.49\linewidth]{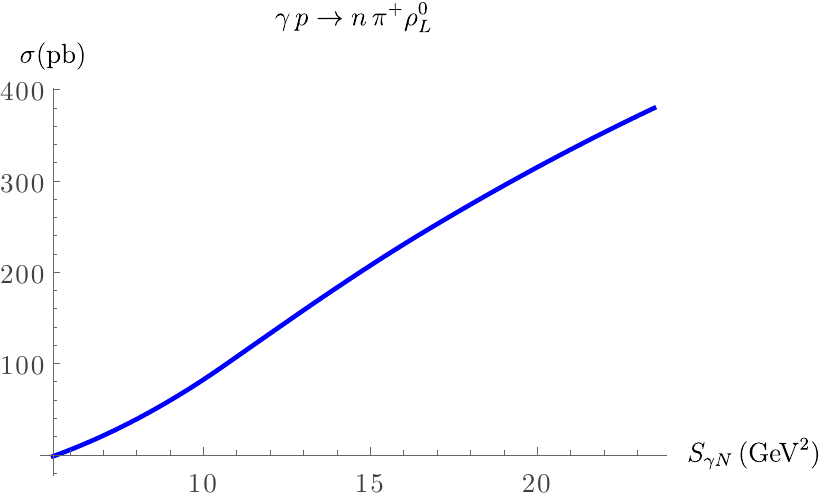}
\includegraphics[width=.49\linewidth]{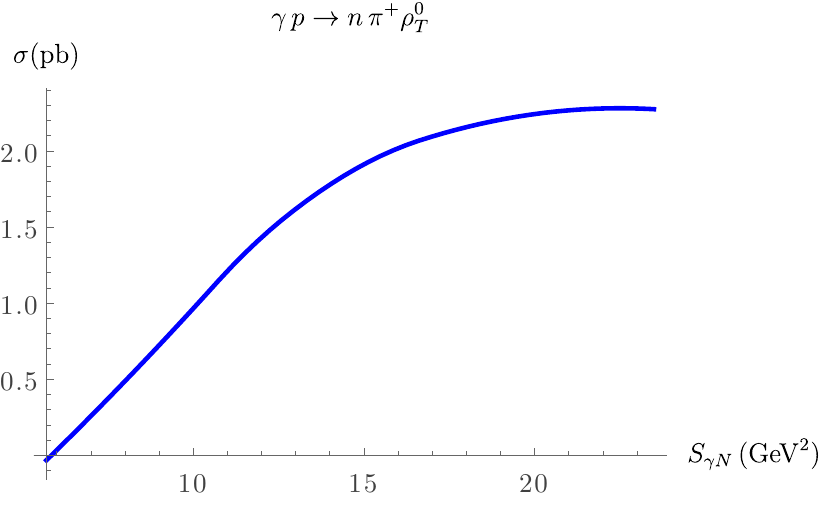}
\caption{Total cross sections as a function of $S_{\gamma N}$ for the photoproduction of $\pi^-\rho^0_L$ (top-left), $\pi^-\rho^0_T$ (top-right), $\pi^+\rho^0_L$ (bottom-left), $\pi^+\rho^0_T$ (bottom-right). The beam energy is chosen to correspond to that of the CLAS12 experiment, with $E_\text{beam}=\SI{12}{GeV}$. }
\label{fig:totcrosssection}
  \end{figure}

Finally, we also show the total cross section as a function of $S_{\gamma N}$ in \FIG\ref{fig:totcrosssection}. From these plots, it is clear that the cross section for some of the processes considered can be as large as hundreds of pb. For this reason, the measurement of these processes is very promising.

\section{Conclusion and Outlook}

\label{sec:conclusions}

In this article, we have employed the collinear factorisation framework to calculate, in a fully automatised way, a whole family of exclusive di-meson photoproduction processes at leading order in $\alpha_s$ and at leading twist. The factorised structure of the amplitude involves a process-dependent but perturbatively calculable hard part, a GPD (associated to the incoming and outgoing nucleons, whose squared momentum difference $t$ is small), and two DAs (associated to the two outgoing mesons, whose squared momentum sum is large). This paper therefore opens the possibility for GPD extraction from such processes, alongside other well-known channels such as DVCS. Furthermore through the exclusive photoproduction of a di-meson pair, depending on the choice of mesons in the final state, one is able to probe chiral-odd (helicity-flip) GPDs at the leading twist. 

This paper discusses in detail the calculation of the amplitude, starting from the construction of quark-level diagrams for the hard part of the process. Particular attention is given to the organisation of the amplitude prior to integration over the momentum fractions entering the GPD and the two DAs in order to achieve numerical stability. We also discuss extensively the treatment of the phase space, and carefully address the issue of the invariance of the cross section with respect to the exchange of the mesons in the final state. Although expected to be kinematically suppressed in the strict collinear factorisation limit, this had been overlooked in a similar calculation for the exclusive photoproduction of a photon-meson pair performed by some of us \cite{Boussarie:2016qop,Duplancic:2018bum,Duplancic:2022ffo,Duplancic:2023kwe}. As a proof of concept, the cross sections for 4 processes out of the 26 possible ones from this set are evaluated for centre-of-mass energies typical of the CLAS12 experiment at JLab. The numbers obtained for some processes are very promising, and are even orders of magnitude higher than those obtained for the exclusive photon-meson photoproduction processes.

This work therefore sets the stage for phenomenological calculations involving the extraction of GPDs from $2 \to 3$ exclusive processes. In particular, the sensitivity of the cross sections on the choice of mesons, the modelling of GPDs and DAs, and on the centre-of-mass energy of the system could be explored further, such as those accessible at COMPASS/AMBER, as well as in collider experiments such as LHC and RHIC in ultra-peripheral collisions (UPCs) and the future EIC. In fact, due to the presence of two effective hard scales ($-u'$ and $M_{12}^2$) in the hard part of the process, such $2 \to 3$ processes provide a better sensitivity to the $x$-dependence of GPDs, which would therefore play a crucial role in their extraction. Furthermore, other observables, e.g.~polarisation asymmetries could be constructed in an attempt to suppress the model dependence of the DA. We intend to address these phenomenological aspects in an upcoming article.

\section*{Acknowledgements}

We thank Valerio Bertone, Jonathan Gaunt, Kornelija Passek-Kumericki, Bernard Pire and Pawel Sznajder for useful discussions. In particular, we would like to express our gratitude to Goran Duplancic for insights on how to improve numerical stability. The work of S.N. was supported by the Science and Technology Facilities Council (STFC) under Grant No. ST/X00077X/1, and by the Royal Society through Grant No. URF/R1/201500. This project has also received funding from the Agence Nationale de la Recherche (ANR) via the grant ANR-20-CE31-0015 (``PrecisOnium'') and via the IDEX Paris-Saclay ``Investissements d’Avenir'' (ANR-11-IDEX-0003-01) through the GLUODYNAMICS project funded by the ``P2IO LabEx (ANR-10-LABX-0038)''. L. S. was supported by the Grant
No. 2024/53/B/ST2/00968 of the National Science Centre in Poland. He also thanks the
support by the "P2I - Graduate School of Physics", in the framework “Investissements d’Avenir” (ANR-11-IDEX-0003-01) managed by the Agence Nationale de la Recherche (ANR), France.
S.N. and L.S. gratefully acknowledge the warm hospitality and financial support of IJCLab, where part of this work was carried out.
S.N. and D.P. also acknowledge the warm hospitality and financial support of Rudjer Bo\v{s}kovi\'{c} Institute in Zagreb.
D.P. further acknowledges the warm hospitality and financial support of National Center for Nuclear Research (NCBJ) in Warsaw.
This project also received funding via the IN2P3 project “QCDFactorisation@NLO”.

\appendix

\section{Further kinematics}

\label{app:further-kinematics}

Here, we present the expressions for all relevant variables at different levels of approximation.

\subsection{Exact kinematics}

\label{app:exact-kinematics}

The aim of this subsection is to express the relevant kinematical variables in the paper in terms of a chosen basis of \textit{independent} kinematical variables, taken to be \{$S_{\gamma N}$, $M_{12}^2$, $u'$,  $t$, $\vec{p}_t\cdot\vec{\Delta}_t$\}, in addition to hadron masses.

Let us first write down a few short useful expressions. We define 
\begin{align}
 \tau=\frac{2\xi}{1+\xi}\,\iff \xi=\frac{\tau}{2-\tau}.
\label{eq:tauformula}
\end{align}
From \EQs\eqref{eq:Sgformula} and \eqref{eq:tformula}, we get 
\begin{align}
\label{eq:Deltat}
    \Delta_t^2&=-(1-\tau)\,t-\tau^2M^2\,,\\
    s&=\frac{S_{\gamma N}-M^2}{1+\xi}\,.
    \label{eq:s-formula}
\end{align}
The minus component of the conservation of the total four-momentum gives
\begin{align}
    1-\alpha_{M_1}-\alpha_{M_2}&=\frac{2\xi M^2}{s(1-\xi^2)}+\frac{\Delta_t^2}{s(1-\xi)}\nonumber\\
    &=\tau\bar{M}^2-\bar{t}\,,
    \label{eq:conservation-minus-comp}
\end{align}
using the same convention as in \SEC\ref{sec:exact-kinematics-phase-space}, i.e.~the bar indicates a rescaling by $\frac{1}{S_{\gamma N}-M^2}$. This relation allows us to eliminate $\alpha_{M_1}$ in favour of $\alpha_{M_2}$ in our expressions. 

The  $+$  component of the total four-momentum conservation yields
\begin{equation}
 2\xi=\frac{(\vec{p}_t-\frac{\vec{\Delta}_t}{2})^2+m_{M_2}^2}{s\alpha_{M_2}}+\frac{(\vec{p}_t+\frac{\vec{\Delta}_t}{2})^2+m_{M_1}^2}{s\alpha_{M_1}}   \,,
\end{equation}
from which we can infer the following expression for $p_t^2$,
\begin{align}
    p_t^2=\frac{1}{1-\tau\bar{M}^2+\bar{t}}&\left[\tau(S_{\gamma N}-M^2)\alpha_{M_2}(1-\alpha_{M_2}-\tau \bar{M}^2+\bar{t})-\vec{p}_t\cdot\vec{\Delta}_t\,(2\alpha_{M_2}-1+\tau\bar{M}^2-\bar{t})\right.\nonumber\\
    &\left.-(1-\alpha_{M_2}-\tau\bar{M}^2+\bar{t})\,m_{M_2}^2-\alpha_{M_2}\,m_{M_1}^2+\frac{1}{4}\left((1-\tau)t+\tau^2M^2\right)(1-\tau\bar{M}^2+\bar{t})\right]\,.\label{eq:ptsquaredformula1}
\end{align}
The Mandelstam variables read
\begin{align}
&M_{12}^2=2\xi s+t= \tau (S_{\gamma N}-M^2)+t\,,\label{eq:M12formula}\\
&-t'=\frac{(\vec{p}_t-\frac{\vec{\Delta}_t}{2})^2+(1-\alpha_{M_2})m_{M_2}^2}{\alpha_{M_2}},\label{eq:tprimeformula}\\
&-u'=\frac{(\vec{p}_t+\frac{\vec{\Delta}_t}{2})^2+(1-\alpha_{M_1})m_{M_1}^2}{\alpha_{M_1}}.\label{eq:uprimeformula}
\end{align}
From \EQ\eqref{eq:M12formula} and \EQ\eqref{eq:tauformula} we can express
\begin{equation}
    \xi=\frac{M_{12}^2-t}{2S_{\gamma N}-2M^2-M_{12}^2+t}.
    \label{eq:xiexact}
\end{equation}
Using \EQ\eqref{eq:tprimeformula} and \EQ\eqref{eq:mandelstamrelation}, one can write $\alpha_{M_2}$ as a function of $p_t^2$
\begin{equation}
    \alpha_{M_2}=\frac{p_t^2-\frac{1}{4}\left((1-\tau)\,t+\tau^2 M^2\right)-\vec{p}_t\cdot\vec{\Delta}_t+m_{M_2}^2}{u'+M_{12}^2-t-m_{M_1}^2}.
    \label{eq:al2formula1}
\end{equation}
Combining \EQ\eqref{eq:ptsquaredformula1} and \EQ\eqref{eq:al2formula1}, one can express $\alpha_{M_2}$ as a function of $\vec{p}_t\cdot\vec{\Delta}_t$, $u'$, $M_{12}^2$, $t$ and $S_{\gamma N}$ only:
\begin{align}
\alpha_{M_2}=\frac{1}{M_{12}^2-t}\left[m_{M_1}^2(\bar{t}-\tau \bar{M}^2)+m_{M_2}^2-2\vec{p}_t\cdot\vec{\Delta}_t-u'(\bar{t}-\tau\bar{M}^2+1)\right]\,,
\label{eq:al2exact}
\end{align}
where $\tau$ through \EQ\eqref{eq:tauformula} is related to $\xi$ which is given by \EQ\eqref{eq:xiexact}.

Moreover, from \EQ\eqref{eq:conservation-minus-comp}, we have
\begin{equation}
    \alpha_{M_1}=1-\alpha_{M_2}-\tau\bar{M}^2+\bar{t}.
    \label{eq:implicit-alpha1}
\end{equation}Replacing \EQ\eqref{eq:al2exact} in \EQ\eqref{eq:ptsquaredformula1}, one gets: 
\begin{align}
    p_t^2& =\frac{1}{M_{12}^2-t}\left[\left(\vec{p}_t\cdot\vec{\Delta}_t-m_{M_1}^2(\bar{t}-\tau\bar{M}^2)\right){(m_{M_1}^2-M_{12}^2+t)}-m_{M_1}^2m_{M_2}^2\right.\nonumber\\
    &+\vec{p}_t\cdot\vec{\Delta}_t(m_{M_1}^2-2u')+u'\left(m_{M_1}^2+m_{M_2}^2+2m_{M_1}^2(\bar{t}-\tau\bar{M}^2)-(\bar{t}-\tau \bar{M}^2+1)(M_{12}^2-t+u')\right)\nonumber\\
    &\left.+\frac{1}{4}(M_{12}^2-t)(\tau^2 M^2+(1-\tau)t)\right]\,.
    \label{eq:ptexact}
\end{align}
Let us now consider the invariant mass of the meson $M_2$ and the outgoing nucleon $N'$, given by
\begin{equation}
\label{eq:M2Nformula1}
    M_{M_2N'}^2=s(1-\xi)\alpha_{M_2}+M^2+\Delta_t^2-2\,\vec{p}_t\cdot\vec{\Delta}_t+m_{M_2}^2+\frac{(M^2+\Delta_t^2)\left((\vec{p}_t-\frac{\vec{\Delta}_t}{2})^2+m_{M_2}^2\right)}{\alpha_{M_2} s(1-\xi)}.
\end{equation}
It follows from \EQs\eqref{eq:tprimeformula} and \eqref{eq:mandelstamrelation} that 
\begin{equation}
    \frac{(\vec{p}_t-\frac{\vec{\Delta}_t}{2})^2+m_{M_2}^2}{\alpha_{M_2}}=m_{M_2}^2-t'=u'+M_{12}^2-t-m_{M_1}^2.
\end{equation}
Thus, using \EQ\eqref{eq:M2Nformula1}, we get
\begin{align}
    M_{M_2N'}^2&=(1-\tau)\alpha_{M_2}(S_{\gamma N}-M^2)+(1-\tau^2)M^2-(1-\tau)\,t-2\,\vec{p}_t\cdot\vec{\Delta}_t+m_{M_2}^2\nonumber\\
    &\quad+\left[(1+\tau)\bar{M}^2-\bar{t}\right]\left[{u}'+{M}_{12}^2-{t}-{m}_{M_1}^2\right]\,,
    \label{eq:M2Nprimedsq}
\end{align}
while $M_{M_1N'}^2$ is given by $M_{M_1N'}^2=M^2+m_{M_1}^2+m_{M_2}^2+S_{\gamma N}-M_{12}^2-M_{M_2N'}^2$, that is 
\begin{equation}
    M_{M_1N'}^2=(1-\tau)(1-\alpha_{M_2})(S_{\gamma N}-M^2)+(1-\tau)M^2+2\,\vec{p}_t\cdot\vec{\Delta}_t+m_{M_1}^2+(m_{M_1}^2-u')\left[(1+\tau)\bar{M}^2-\bar{t}\right].
    \label{eq:M1Nprimedsq}
\end{equation}

To summarise, in this appendix, we have expressed:
\begin{itemize}
    \item $\xi$ in terms of $\SgN$, $M_{12}^2$ and $t$ in \EQ\eqref{eq:xiexact},
    \item $\tau$ in terms of $\xi$ in \EQ\eqref{eq:tauformula},
    \item $\Delta_t^2$ in terms of $t$ and $\tau$ in \EQ\eqref{eq:Deltat},
    \item $s$ in terms of $\SgN$ and $\xi$ in \EQ\eqref{eq:s-formula},
    \item $\alpha_{M_2}$ in terms of $\SgN$, $M_{12}^2$, $u'$, $t$ and $\vec{p}_t \cdot \vec{\Delta}_t$ in \EQ\eqref{eq:al2exact},
    \item $\alpha_{M_1}$ in terms of $\SgN$, $t$, $\tau$ and $\alpha_{M_2}$ in \EQ\eqref{eq:implicit-alpha1},
    \item $\vec{p}_t^2$ in terms of $\SgN$, $M_{12}^2$, $u'$, $t$, $\vec{p}_t \cdot \vec{\Delta}_t$ and $\tau$ in \EQ\eqref{eq:ptexact},
    \item $M_{M_2 N'}^2$ in terms of $\SgN$, $M_{12}^2$, $u'$, $t$, $\vec{p}_t \cdot \vec{\Delta}_t$, $\tau$ and $\alpha_{M_2}$ in \EQ\eqref{eq:M2Nprimedsq},
    \item $M_{M_1 N'}^2$ in terms of $\SgN$, $u'$, $t$, $\vec{p}_t \cdot \vec{\Delta}_t$, $\tau$ and $\alpha_{M_2}$ in \EQ\eqref{eq:M1Nprimedsq}.
\end{itemize}
Finally, we note that the Mandelstam relation in \EQ\eqref{eq:mandelstamrelation} enables $t'$ to be written in terms of $M_{12}^2$, $u'$ and $t$. In this way, we have written all of the relevant kinematical

\subsection{Kinematics at $\Delta_\perp=0$}

\label{app:kinematics-Deltaperp-zero}

Taking the limit $\vec{\Delta}_t=0$, $t$ becomes
\begin{equation}
    t=-\frac{\tau^2}{1-\tau}M^2=-\frac{4\xi^2}{1-\xi^2}M^2\,.
    \label{eq:tapproxformula}
\end{equation}
In \SEC\ref{sec:phase-space}, $(-t)_{\mathrm{min}}$ is defined to be equal to $-t$ in \EQ\eqref{eq:tapproxformula}.
Combined with \EQ\eqref{eq:xiexact}, we can write $t$ as a function of  $M_{12}^2$ and $S_{\gamma N}$ \cite{Boussarie:2016qop}:
\begin{equation}
\label{eq:tformulaapprox}
    t=-\frac{1-\bar{M}_{12}^2(1+2\bar{M}^2)-\sqrt{1+\bar{M}_{12}^2(\bar{M}_{12}^2-2-4\bar{M}^2)}}{2(1+\bar{M}^2)}(S_{\gamma N}-M^2)\,.
\end{equation}
In this limit, \EQs\eqref{eq:al2exact} and \eqref{eq:ptexact} become: 
\begin{align}
    \alpha_{M_2}&=\frac{1}{(1-\tau)M_{12}^2+\tau^2 M^2}\left[(1-\tau)m_{M_2}^2-\tau m_{M_1}^2\bar{M}^2+u'(\tau-1+\tau\bar{M}^2)\right]\,,\\
    p_t^2&=\frac{1}{(1-\tau)M_{12}^2+\tau^2M^2}\left[u'\Bigg((\tau\bar{M}^2+\tau-1)\left(\frac{(1-\tau)M_{12}^2+\tau^2M^2}{1-\tau}+u'\right)-2m_{M_1}^2\tau\bar{M}^2\right.\nonumber\\
&\quad\left.+(1-\tau)(m_{M_1}^2+m_{M_2}^2)\Bigg)
-(1-\tau)m_{M_1}^2m_{M_2}^2+m_{M_1}^2\tau\bar{M}^2\left(m_{M_1}^2-\frac{(1-\tau)M_{12}^2+\tau^2M^2}{1-\tau}\right)\right]\,,
\end{align}
where $\xi$ (present in the expression for $\tau$, see \EQ\eqref{eq:tauformula}) is obtained by replacing \EQ\eqref{eq:tformulaapprox} in \EQ\eqref{eq:xiexact},
\begin{equation}
    \xi=\frac{-1+\sqrt{1+\bar{M}_{12}^2(\bar{M}_{12}^2-2-4\bar{M}^2)}}{\bar{M}_{12}^2-2-4\bar{M}^2}\,.
\end{equation}
We also have 
\begin{equation}
    \alpha_{M_1}=1-\alpha_{M_2}-\frac{\tau\bar{M}^2}{1-\tau}.
    \label{eq:al1approx}
\end{equation}
The invariant masses of each outgoing meson and the outgoing nucleon are now given by
\begin{align}
        M_{M_2N'}^2&=(1-\tau)\alpha_{M_2}(S_{\gamma N}-M^2)+M^2+m_{M_2}^2+\frac{\bar{M}^2}{1-\tau}\left[{u}'+\frac{(1-\tau){M}_{12}^2+\tau^2M^2}{1-\tau}-{m}_{M_1}^2\right]\,,\\
    M_{M_1N'}^2&=(1-\tau)(1-\alpha_{M_2})(S_{\gamma N}-M^2)+(1-\tau)M^2+m_{M_1}^2+\frac{\bar{M}^2}{1-\tau}(m_{M_1}^2-u')\,.
\end{align}

It is worth pointing out that the approximated kinematics discussed in this subsection is much simpler to the exact kinematics in \APP\ref{app:exact-kinematics}. Indeed, the independent variable $\vec{p}_t \cdot \vec{\Delta}_t$ vanishes and drops out, and $t$ becomes constant. Therefore, everything is expressible in terms of three kinematical variables, namely $\SgN$, $M_{12}^2$ and $u'$.

\subsection{Kinematics at $\Delta_\perp=m_{M_1}=m_{M_2}=0$}

In this limit, the expressions for $\alpha_{M_2}$, $p_t^2$, $M^2_{M_2 N'}$ and $M^2_{M_1 N'}$ respectively become
\begin{align}
    &\alpha_{M_2}=\frac{1}{(1-\tau)M_{12}^2+\tau^2M^2}u'(\tau(1+\bar{M}^2)-1)\,,\\
    &p_t^2=\frac{1}{(1-\tau)M_{12}^2+\tau^2M^2}u'(\tau(1+\bar{M}^2)-1)\left(u'+\frac{(1-\tau)M_{12}^2+\tau^2M^2}{1-\tau}\right)\,,\\
    &M_{M_2N'}^2=(1-\tau)\alpha_{M_2}(S_{\gamma N}-M^2)+M^2+\frac{\bar{M}^2}{1-\tau}\left(u'+\frac{(1-\tau)M_{12}^2+\tau^2M^2}{1-\tau}\right)\,\label{eq:M2Nformula2},\\
    &M_{M_1N'}^2=(1-\tau)(1-\alpha_{M_2})(S_{\gamma N}-M^2)+(1-\tau)M^2-u'\frac{\bar{M}^2}{1-\tau}\label{eq:M1Nformula2}.
\end{align}
Within the approximation here, $\xi$ and $\alpha_{M_1}$ have the same expressions as in \EQ\eqref{eq:xiformula} and \EQ\eqref{eq:al1approx}.
\subsection{Approximated kinematics in collinear factorisation framework}
Here, we set $\Delta_\perp=m_{M_1}=m_{M_2}=M=0$. Therefore, one has
\begin{align}
s&=\frac{\SgN}{1+\xi}\,,\\
\xi&=\frac{M_{12}^2}{2 S_{\gamma N}-M_{12}^ 2}\,,\\
   \alpha_{M_2}&=\frac{-u'}{M_{12}^2}\,,\\
   \alpha_{M_1}&=1-\alpha_{M_2}\,,\\
   p_t^2&=\frac{-u'}{M_{12}^2}(M_{12}^2-(-u'))\,,\\
   M_{M_2N'}^2&=(S_{\gamma N}-M_{12}^2)\frac{-u'}{M_{12}^2}\,,\\
   M_{M_1N'}^2&=(S_{\gamma N}-M_{12}^2)\left(1-\frac{-u'}{M_{12}^2}\right)\,.
\end{align}

\section{GPD modelling}

\label{app:GPD-modelling}

\subsection{Double distribution parametrisation}
In this study, we model the GPD via a \textit{double distribution}, which follows \cite{Radyushkin:1997ki}:
\begin{equation}
\label{eq:RDDA}
    {H}^q(x,\xi,t)=\int_{\{|\beta|+|\alpha|\leq 1\}}d\beta\,d\alpha\,\delta(\beta+\xi\alpha-x)F^q(\beta,\alpha,t)\,,
\end{equation}
where $F^q$ is constructed using a profile function $\Pi(\beta,\alpha)=\frac{3}{4}\frac{(1-\beta)^2-\alpha^2}{(1-\beta)^3}$ and PDFs through \cite{Goeke:2001tz,Belitsky:2001ns,Radyushkin:1998es,Radyushkin:1998bz,Musatov:1999xp}
\begin{align}
    &F^q(\beta,\alpha,t)=\left(\Pi(\beta,\alpha) {q}(\beta)\Theta(\beta)-\Pi(-\beta,\alpha) {\bar{q}}(-\beta)\Theta(-\beta)\right)\frac{C^2}{(t-C)^2}\,,\\
&\tilde{F}^q(\beta,\alpha,t)=\left(\Pi(\beta,\alpha) {\Delta q}(\beta)\Theta(\beta)-\Pi(-\beta,\alpha) {\Delta\bar{q}}(-\beta)\Theta(-\beta)\right)\frac{C^2}{(t-C)^2}\,,\\
    &F^q_ T(\beta,\alpha,t)=\left(\Pi(\beta,\alpha) {\delta q}(\beta)\Theta(\beta)-\Pi(-\beta,\alpha) {\delta \bar{q}}(-\beta)\Theta(-\beta)\right)\frac{C^2}{(t-C)^2}\,.
\end{align}
In the above, $q$, $\Delta q$ and $\delta q$ denote respectively the unpolarised, polarised and transversity PDF of flavour $q$. The dependence on $t$, which is assumed to be factorised from the $x$ and $\xi$ dependence of the GPD, is taken to be the standard dipole form factor \cite{Dunning:1966xiq,Perdrisat:2006hj}, see \EQ\eqref{eq:t-dependence}. The double distribution representation has the advantage of fulfilling three main properties of a GPD, namely: it reduces to a PDF in the forward limit, to an elastic form factor upon integration over $x$, and its Mellin moments are polynomials in $\xi$ \cite{Diehl:2003ny}.

The manipulation of PDFs is facilitated by the use of the \textit{Mathematica} package \textit{ManeParse} \cite{ManeParse}. For this  paper, the set of PDFs are fixed to the central values of CT10NNLO (NNPDFpol11), for the unpolarised (polarised) PDFs. The factorisation scale is fixed to $\mu_F^2 = 3 \GeV^2$, such that it corresponds to the typical value of $M_{12}^2$ for which the cross-section is peaked, see \FIG\ref{fig:singlediff}.

For the transversely polarised PDFs, we use the following parametrisation: 
\begin{equation}
    \delta q(x)=\frac{1}{2}\mathcal{N}^T_q(x)[q(x)+\Delta q(x)],
\end{equation}
with 
\begin{equation}
    \mathcal{N}_q ^T(x)=N_q^Tx^\alpha(1-x)^\beta\frac{(\alpha+\beta)^{\alpha+\beta}}{\alpha^\alpha\beta^\beta}\,,
\end{equation}
where $N_u^T=0.46$ and $N_d^T=-1$.
This form of transversely polarised PDF was obtained as a limiting case of the fits of TMDs based on semi-inclusive deep inelastic scattering \cite{Anselmino:2013vqa}.

\subsection{Sampling of the GPD}

\label{app:GPD-sampling}

Each evaluation of the GPD in $x$ and $\xi$ requires an integration of the PDFs as indicated by \EQ\eqref{eq:RDDA}. In order to be efficient in the numerical evaluation, we use an interpolation of the GPD in $x$ at different specified values of the skewness parameter $\xi$ when performing the convolution integral in the amplitude. Using a few evaluations of the GPD may result in an interpolation not perfectly smooth, which in turn may distort the integrand in \EQ\eqref{eq:amplitude}. Conversely, having too many points in the interpolation may increase the integration time. Therefore, we fix the number of points to the minimum necessary so that the numerical derivative of the interpolation is smooth. 

At small $\xi$, the GPD varies rapidly in the region $|x| \lesssim \xi$. As a result, it is necessary to ensure that enough points are sampled in this region to obtain a reliable GPD. Concretely, if $\xi<10^{-3}$, we use 60 points in the DGLAP region ($\xi<|x|<1$) and 200 points in the ERBL region enlarged by a factor $1+\delta=2.2$ (i.e.~$|x|<2.2\,\xi$), see \EQ\eqref{eq:xinterval}. If $\xi>10^{-3}$, we take only 40 points in the DGLAP region and 150 points in the ERBL region which we enlarge by a factor $1+\delta=1.5$ (i.e.~$|x|<1.5\,\xi$). This allows us to perfectly capture the variations of the GPD, which are more pronounced in the ERBL region.

\section{Processes sensitive to gluon GPDs}

\label{app:gluon-GPDs}

Throughout this paper, we do not consider processes which could have an accompanying gluon GPD channel. This is because it was shown in \cite{Nabeebaccus:2023rzr,Nabeebaccus:2024mia} that such processes break collinear factorisation. The origin of this factorisation breaking is a Glauber pinch contribution, which is of leading power. In fact, na\"ively assuming collinear factorisation leads to a \textit{divergent} amplitude. The interpretation of the divergence is that the collinear pinch contribution, which is captured by collinear factorisation, does not correspond to the complete leading power contribution of the amplitude.

\begin{figure}[t!]
    \centering
    \includegraphics[width=0.45\linewidth]{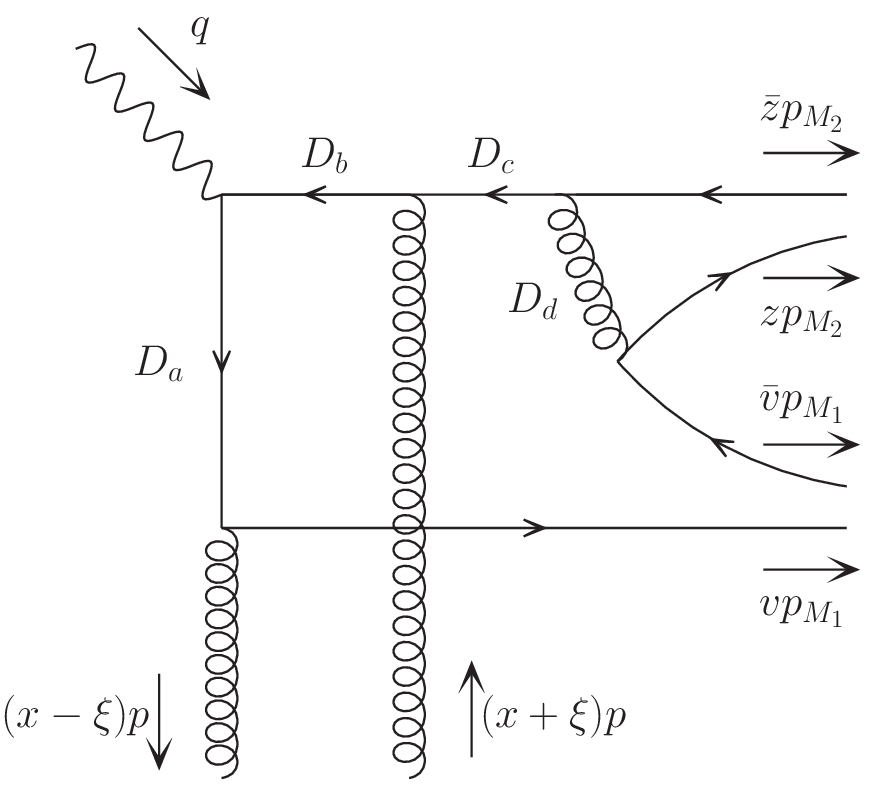}
    \caption{Diagram for di-meson photoproduction, with a  2-gluon exchange with the nucleon sector (gluon GPD contribution) which diverges upon integration over $x$ and $v$.}
    \label{fig:gluon-GPD-di-meson}
\end{figure}

In this Appendix, we show the appearance of  the divergence in such an amplitude for $\pi^0 \rho^0_L$ photoproduction. Note that at leading order, not all diagrams contributing to the amplitude are divergent - Indeed, only the ones compatible with the Glauber pinch contribution identified in \cite{Nabeebaccus:2023rzr,Nabeebaccus:2024mia} are divergent. To this end, we consider the Feynman diagram in \FIG\ref{fig:gluon-GPD-di-meson}. The hard part of the process is given by
\begin{align}
    {\cal M} \propto &\int _{-1}^1 dx\int_0^1 dv \int_0 ^1 dz\,\frac{\Tr\left[\gamma^5 \slashed{p}_{M_1} \gamma_\mu (v \slashed{p}_{M_1}+(x-\xi )\slashed{p})\slashed{\epsilon}_q (v \slashed{p}_{M_1}+(x-\xi)\slashed{p}-\slashed{q}) \gamma_\nu (\bar v \slashed{p}_{M_1}+\slashed{p}_{M_2}) \slashed{p}_{M_2}  \right]}{\left[v(x-\xi) p \cdot p_{M_1}+i\epsilon\right]
    \left[v(x-\xi)p \cdot p_{M_1} - (x-\xi)p\cdot q -v\, q\cdot p_{M_1}+i\epsilon\right]}\nonumber\\[5pt]
    &\qquad \times \frac{\phi_{1}(v)\,\phi_{2}(z)}{
    \left[\bar v \,p_{M_1}\cdot p_{M_2}+i\epsilon\right]
    \left[\bar v  z \,p_{M_1}\cdot p_{M_2} +i\epsilon\right]}\frac{H_g(x)g_{\perp}^{\mu\nu}}{(x+\xi-i\epsilon)(x-\xi+i\epsilon)}\,.
\end{align}

We now study this amplitude in the limit $x \to \xi$ and $ v \to 0$. From \FIG\ref{fig:gluon-GPD-di-meson}, this corresponds to the limit where the $D_a$ propagator is \textit{soft}, the $D_b$ propagator is \textit{collinear} (to the incoming photon), and the $D_c$ and $D_d$ propagators are \textit{hard}. In this limit, taking the DAs to have the simple asymptotic form in \EQ\eqref{eq:asDA}, the amplitude simplifies to
\begin{align}
    {\cal M} 
    &\propto \int_{-1}^1 dx \int_0^1 dv \int_0^1 dz \frac{\bar z\,g^{\mu\nu}_{\perp}H_g(\xi)}{2\xi\,(x-\xi + i\epsilon)}
    \frac{(x-\xi )\,
    \Tr
    \left[
    \gamma^5
    \slashed{p}_{M_1} 
    \gamma^\mu 
    \slashed{p}\,
    \slashed{\epsilon}_q 
    ((x-\xi)\slashed{p}-\slashed{q})
    \gamma^\nu
    \slashed{p}_{M_1}
    \slashed{p}_{M_2} 
    \right]
    }{\left[(x-\xi) p \cdot p_{M_1}+i\epsilon\right]
    \left[ (x-\xi)p\cdot q +v\, q\cdot p_{M_1}-i\epsilon\right]}\nonumber\\
    &\propto \int_{-1}^1 dx \int_0 ^1 dv
    \frac{g^{\mu\nu}_{\perp}H_g(\xi)}{(x-\xi + i\epsilon)^2}
    \frac{(x-\xi )\,
    \Tr
    \left[
    \gamma^5
    \slashed{p}_{M_1} 
    \gamma^\mu 
    \slashed{p}\,
    \slashed{\epsilon}_q 
    ((x-\xi)\slashed{p}-\slashed{q})
    \gamma^\nu
    \slashed{p}_{M_1}
    \slashed{p}_{M_2} 
    \right]
    }{
    \left[ (x-\xi) + A\, v-i\epsilon\right]}\,,
\end{align}
where $A = (q \cdot p_{M_1})/(p \cdot q) \sim {\cal O}(1)$ is positive. 
In the above, we have also defined a coordinate system such that, as before, $p$ is a vector in the $+$ direction, while $p_{M_1}$ (instead of $q$) is a vector in the $-$ direction. Note that we have dropped overall prefactors (including the $z$ integration) in going to the second line. To simplify the trace, we note that
\begin{align}
    \gamma^\mu \gamma^\alpha \gamma^\beta \gamma ^\delta \gamma_\mu = -2 \gamma^\delta \gamma^\beta \gamma^\alpha\,.
\end{align}
Using the cyclicity of the trace, and the fact that $p_{M_1}$ is in the $-$ direction,
\begin{align}
    g^{\mu\nu}_{\perp} 
    \gamma^\nu
    \slashed{p}_{M_1}
    \slashed{p}_{M_2} 
    \slashed{p}_{M_1} 
    \gamma^\mu \propto 
        \slashed{p}_{M_1}
            \slashed{p}\,
                \slashed{p}_{M_1}\,.
\end{align}
This leads to 
\begin{align}
     {\cal M} 
   \propto \int_{-1}^1 dx \int_0^1 dv
    \frac{H_g(\xi)}{(x-\xi + i\epsilon)^2}
    \frac{(x-\xi )\,
    \Tr
    \left[
    \gamma^5
       \slashed{p}\,
    \slashed{\epsilon}_q 
    ((x-\xi)\slashed{p}-\slashed{q})
     \slashed{p}_{M_1}
            \slashed{p}\,
                \slashed{p}_{M_1}
    \right]
    }{
    \left[ (x-\xi) + A\, v-i\epsilon\right]}\,.
\end{align}
Noting that the $(x-\xi) \slashed{p}$ term inside the trace leads to a finite contribution, we are left with
\begin{align}
\label{eq:divergent-diagram}
      {\cal M} 
   &\propto [\mathrm{finite}] + \int_{-1}^1 dx \int_0^1 dv
    \frac{(x-\xi )\,
    }{(x-\xi + i\epsilon)^2
    \left[ (x-\xi) + A\, v-i\epsilon\right]}\nonumber\\
    &\propto 
    [\mathrm{finite}] + i \log(\epsilon)\,,
\end{align}
which diverges when $\epsilon \to 0$, and is purely imaginary. Note that the integral in the second term of the first line of \EQ\eqref{eq:divergent-diagram} is similar to the integral in \EQ(A.3) of \cite{Nabeebaccus:2023rzr}. 

This calculation thus confirms that the amplitude for $\pi^0\rho^0$ photoproduction is divergent if one assumes collinear factorisation, due to the gluon GPD channel. More generally, this is expected to be the case for any process where one can exchange two gluons with the nucleon, such as $\pi^+\pi^-$ photoproduction (which has both $C=+1$ and $C=-1$ contributions). In fact, the exclusive photoproduction of any $C=-1$ state would allow the exchange of two gluons in the $t$-channel, which would lead to collinear factorisation breaking effects.

\section{Absence of an expected tensor structure in the VTT case}

\label{app:absence-VTT-structure}

The VTT case corresponds to having a vector GPD (which corresponds to $\slashed{p}$), and two tensor DAs which represent two transversely polarised $\rho^0_T$ mesons. In this appendix, we show that for specific choices of the gauge fixing for the polarisation vectors $\epsilon_q$, $\epsilon_{M_1}$ and $\epsilon_{M_2}$, the tensor structure
\begin{align}
\label{eq:relevant-tensor-structure}
    (p_{\perp} \cdot \epsilon_{q\perp})(p_{\perp}\cdot \epsilon_{M_1\perp})(p_{\perp}\cdot \epsilon_{M_2\perp})\,,
\end{align}
cannot appear in any LO diagram for that process, as noted in \SEC\ref{sec:tensor-structures}. This is quite remarkable, since a general decomposition of the full amplitude, on the basis of gauge invariance, is supposed to have four independent coefficients.

We note that at leading twist, the way the tensor DA is implemented, through $(\epsilon_{M_i}^\mu p_{M_i}^\nu - \epsilon_{M_i}^\nu p_{M_i}^\mu)$ (see \EQ\eqref{eq:tensor-DA}), implies that taking $\epsilon_{M_i}^\mu \to p_{M_i}^\mu$ causes \textit{any} diagram that it enters to vanish. Therefore, at least from a calculational point of view, it is possible to interpret this feature as gauge invariance. In fact, the polarisation vector constructed in \EQ\eqref{eq:transpolparam} satisfies $p \cdot \epsilon_{M_i} = 0 $, which can be interpreted as gauge fixing.\footnote{The ``freedom'' in constructing $\epsilon_{M_i}(p_{M_i},T)$ is connected to the freedom in choosing the lightlike vector when constructing $\epsilon_{{M_i}}(p_{M_i},L)$, where we have used $p^\mu$ as that lightlike vector in \EQ\eqref{eq:longpolparam}.}

In the first step, let us identify the topologies from \FIG\ref{fig:topologies} which are relevant here. They are topologies 1, 2 and 4. Otherwise, each Dirac trace contains an odd number of Dirac $\gamma$ matrices. Since the exact positioning of the mesons does not matter for investigating the tensor structures (exchanging the mesons merely corresponds to a relabelling), we can simply focus on topologies 1 and 2, which have 2 and 1 Dirac traces respectively.

Due to the tensor Dirac nature of the DAs, see \EQ\eqref{eq:tensor-DA}, the structures $DA_1$ and $DA_2$ can effectively be represented by $\slashed{\epsilon}_{M_i} \slashed{p}_{M_i}$ since
\begin{align}
    \sigma_{\mu\nu} (\epsilon_{M_i}^\mu p_{M_i}^\nu - \epsilon_{M_i}^\nu p_{M_i}^\mu) = i \slashed{\epsilon}_{M_i} \slashed{p}_{M_i}\,.
\end{align}
Since $p_{M_i} \cdot \epsilon_{M_i} = 0$, $\slashed{\epsilon}_{M_i} \slashed{p}_{M_i} = -\slashed{p}_{M_i}  \slashed{\epsilon}_{M_i} $ can be freely anti-commuted. This implies that any term involving  $\slashed{p}_{M_i}$ next to $\slashed{\epsilon}_{M_i} \slashed{p}_{M_i}$ vanishes, as well as
\begin{align}
\label{eq:two-gamma-contraction}
    \gamma^\mu \slashed{\epsilon}_{M_i} \slashed{p}_{M_i} \gamma_\mu = 0\,.
\end{align}

In what follows, we focus mainly on two gauge choices:
\begin{itemize}
    \item[(i)] $p \cdot \epsilon_{q} = p \cdot \epsilon_{M_{1}} = p \cdot \epsilon_{M_{2}} = 0$, which is the gauge fixing choice used throughout this paper.
    \item[(ii)] $q \cdot \epsilon_{M_{1}} = q \cdot \epsilon_{M_{2}} = 0$. In this case, we keep the gauge fixing choice general for $\epsilon_q$. Whenever we work in this gauge, we eliminate $p^\mu$ using momentum conservation, expressing all momenta in terms of $q^\mu$, $p_{M_1}^\mu$ and $p_{M_2}^\mu$.
\end{itemize}
Note that in both gauge fixing choices, we have a single vector (either $p^\mu$ or $q^\mu$) which when contracted with all three polarisation vectors give zero.
In the first (second) gauge fixing choice, we eliminate $q^\mu$ ($p^\mu$) using momentum conservation, expressing all momenta in terms of $p^\mu$ ($q^\mu$), $p_{M_1}^\mu$ and $p_{M_2}^\mu$. A crucial feature of both gauge fixing choices is that the appearance of a scalar product of $p_\perp$ with any of the three polarisation vectors necessarily comes from the scalar product of either $p_{M_1}$ or $p_{M_2}$ with that polarisation vector. Thus, to show the absence of the tensor structure in \EQ\eqref{eq:relevant-tensor-structure}, it suffices to show that the Dirac traces \textit{cannot} produce a term of the form
\begin{align}
\label{eq:product-3-SPs-with-pMi}
    (p_{M_i} \cdot \epsilon_{q})(p_{M_2} \cdot \epsilon_{M_1})(p_{M_1} \cdot \epsilon_{M_2})\,,
\end{align}
where $i=1$ or 2. It is important to highlight that while such a tensor structure does not appear for specific choices of gauge fixing, the  underlying cause comes from the fact that there is a ``hidden'' linear dependence due to the restrictive nature of diagrams that can be constructed at LO, as we show below. Thus, we find that there exists a further relation that connects the 4 coefficients which one may naively believe to be independent on the basis of gauge invariance of the general decomposition of the complete LO amplitude.

\subsection{Topology 1}

\begin{figure}
    \centering
    \includegraphics[width=0.45\linewidth]{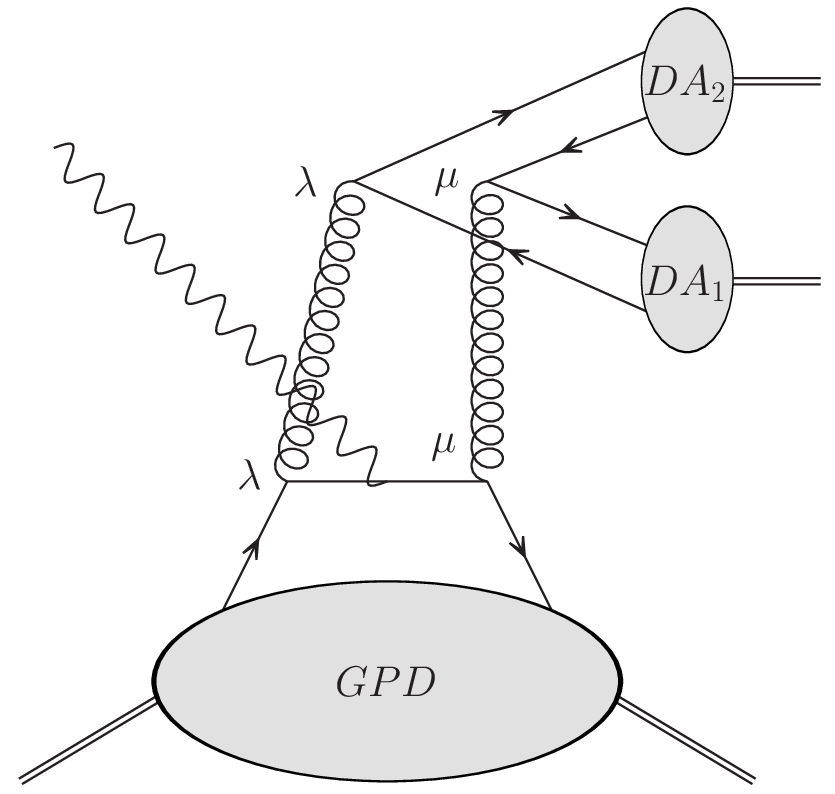}
    \caption{An example of a diagram within topology 1.}
    \label{fig:double-trace-VTT-tensor-structure}
\end{figure}

For topology 1, the photon must necessarily be attached to the quark line connected to the GPD, by virtue of $C$-parity conservation. Furthermore, there should be two gluon lines attached to the quark line connected to the GPD, since only the colour singlet part contributes. An example of such a configuration is shown in \FIG\ref{fig:double-trace-VTT-tensor-structure}.
The numerator contains
\begin{align}
    \Tr\left(\slashed{p}_{M_1} \slashed{\epsilon}_{M_1} \gamma_\mu \slashed{p}_{M_2}\slashed{\epsilon}_{M_2} \gamma_\lambda \right) \times A^{\mu \lambda}\,,
\end{align}
where $A^{\mu\lambda}$ corresponds to the result of the quark trace involving the GPD. Without even writing the exact form of $A^{\mu\lambda}$, we can determine the restrictions on the possible tensor structures from such a class of diagrams:
\begin{itemize}
    \item $A^{\mu\lambda} \propto g^{\mu\lambda}$. This gives zero by virtue of \EQ\eqref{eq:two-gamma-contraction}.
    \item $A^{\mu\lambda} \propto p^\mu p^\lambda$  [$A^{\mu\lambda} \propto q^\mu q^\lambda$] with gauge fixing choice (i) [(ii)].  In this case, after performing the Dirac trace, each momentum $p$ [$q$] is necessarily contracted with $p_{M_1}$ or $p_{M_2}$, since $p \cdot \epsilon_q = p \cdot \epsilon_{M_i} = p^2 = 0$, implying there are no $p_{M_j}$, $j=1,2$ left for contraction with the polarisation vectors.
    \item  $A^{\mu\lambda} \propto \epsilon_q^\mu p^\lambda$  [$A^{\mu\lambda} \propto \epsilon_q^\mu q^\lambda$] with gauge fixing choice (i) [(ii)]. Here, one has a trace involving three polarisation vectors, but only 2 $p_{M_j}$.
\end{itemize}
We note that the appearance of any $p_{M_i}^\mu$ or $p_{M_i}^\lambda$ will immediately cause the trace to vanish, since $\slashed{p}_{M_i}$ will be necessarily sandwiched between $\slashed{p}_{M_1} \slashed{\epsilon}_{M_1} $ and $\slashed{p}_{M_2} \slashed{\epsilon}_{M_2} $. This is why only the three cases considered above are relevant, where $A^{\mu \lambda}$ contains neither $p_{M_i}^\mu$ nor $p_{M_i}^\lambda$.

Since none of the diagrams from topology 1 can ever produce a term of the form in \EQ\eqref{eq:product-3-SPs-with-pMi}, it implies that the tensor structure in \EQ\eqref{eq:relevant-tensor-structure} cannot appear.

\subsection{Topology 2}

Topology 2 involves a single Dirac trace. We start by deriving an interesting feature of any Dirac trace that can appear from such a topology.

The basic Dirac structures that are always present in that single Dirac trace are
\begin{align}
\label{eq:Dirac-structures}
    \slashed{p}_{M_1}\slashed{\epsilon}_{M_1},\;
\slashed{p}_{M_2}\slashed{\epsilon}_{M_2},\;
\slashed{\epsilon}_q,\; 
\slashed{p},\;
\slashed{p}_{q1},\;
\slashed{p}_{q2}\,,
\end{align}
where $p_{q1}$ and $p_{q2}$ represent generic momenta.  One of these momenta is always adjacent to the photon vertex, while the other originates either directly from the triple gluon vertex Feynman rule, or from an extra quark propagator in the case where there are no triple gluon vertex. Note that the list in \EQ\eqref{eq:Dirac-structures} applies to both gauge fixing choices (i) and (ii), where in the latter case, $\slashed{p}$ (which comes from the vector GPD) is rewritten in terms of $\slashed{q}$, $\slashed{p}_{M_1}$ and $\slashed{p}_{M_2}$ by momentum conservation.

Now, the presence of a triple gluon vertex will cause the appearance of $\slashed{p}_{q2}$ in the Dirac trace, as well as the appearance of a pair of contracted $\gamma$ matrix. We observe that one of the elements of the pair will \textit{always} be adjacent to one of the two Dirac structures $\slashed{p}_{M_i} \slashed{\epsilon}_{M_i}$. Furthermore, the number of $\gamma$ matrices between the contracted ones will be at most 4, since there are 8 $\gamma$ matrices (see \EQ\eqref{eq:Dirac-structures}) in addition to the two contracted ones. This will have an important implication later.

The case without a triple gluon vertex has a Dirac trace which contains two quark propagators (taken into account already in \EQ\eqref{eq:Dirac-structures}), and two pairs of contracted $\gamma$ matrices. It is necessary for at least one element of each pair to be adjacent a $\slashed{p}_{M_i} \slashed{\epsilon}_{M_i}$ Dirac structure. The maximum number of $\gamma$ matrices between any pair of contracted $\gamma$ matrices is 5, given by
\begin{align}
    \Tr\left(\gamma^\mu\slashed{p}_{M_1} \slashed{\epsilon}_{M_1}
    \gamma^\lambda
    \slashed{p}_{M_2} \slashed{\epsilon}_{M_2}
    \gamma_\mu \slashed{p}[.]\slashed{p}_{q1} [.]\slashed{p}_{q2}
    \right)\,,
    \label{eq:Dirac-trace-two-contracted-pairs}
\end{align}
where the $[.]$ represents a slot for one of $\slashed{\epsilon}_{q}$ and $\gamma_{\lambda}$. Note that it is impossible to have only  $\slashed{\epsilon}_q$ between $\slashed{p}_{M_1} \slashed{\epsilon}_{M_1}$ and $\slashed{p}_{M_2} \slashed{\epsilon}_{M_2}$, as it corresponds to a disconnected diagram not compatible with collinear factorisation. One can thus deduce that the pair of contracted $\gamma^\lambda$ matrices will be separated by a maximum of 4 $\gamma$ matrices, no matter how $\slashed{\epsilon}_{q}$ and $\gamma_{\lambda}$ are organised in the slots $[.]$. For simplifying such contractions, we use the following identities in 4D
\begin{align}
\label{eq:Dirac-gamma-identity-2}
    \gamma^\mu \slashed{p}_{M_i} \slashed{\epsilon}_{M_i} \gamma_\mu &= 0\,,\\
    \gamma^\mu \slashed{p}_{M_i} \slashed{\epsilon}_{M_i} \gamma^\alpha\gamma_\mu&=-2\gamma^\alpha \slashed{p}_{M_i} \slashed{\epsilon}_{M_i}\,,  \\
    \gamma^\mu \slashed{p}_{M_i} \slashed{\epsilon}_{M_i} \gamma^\alpha \gamma^\beta \gamma_\mu&= 2 \gamma^\alpha \slashed{p}_{M_i} \slashed{\epsilon}_{M_i}  \gamma^\beta + 2 \gamma^\beta \slashed{p}_{M_i} \slashed{\epsilon}_{M_i}  \gamma^\alpha\,.
    \label{eq:Dirac-gamma-identity-4}
\end{align}
Thus, once the Dirac structure $\gamma^\lambda ... \gamma_\lambda$ in \EQ\eqref{eq:Dirac-trace-two-contracted-pairs} is simplified, eliminating $\gamma^\lambda$ and $\gamma_{\lambda}$ from the Dirac chain in the process, the other pair $\gamma^\mu ...\gamma_\mu$ will necessarily be separated by at most 4 $\gamma$ matrices, analogous to the case discussed above for the triple gluon vertex. Of course, the argument is independent of the exact orderings of $\slashed{p}$, $\slashed{p}_{q1}$ and $\slashed{p}_{q2}$ in the Dirac trace in \EQ\eqref{eq:Dirac-trace-two-contracted-pairs}.

Having established that the simplification of contracted $\gamma$ matrices in the single trace involves separations of at most 4 $\gamma$ matrices between them, and that at least one of each pair of contracted $\gamma$ matrices is adjacent a $\slashed{p}_{M_i} \slashed{\epsilon}_{M_i}$ structure, we can determine an important property of the Dirac trace. From the $\gamma$ matrix identities in \EQ\eqref{eq:Dirac-trace-two-contracted-pairs}, we find that the number of $\gamma$ matrices eliminated/added on either side of $\slashed{p}_{M_i} \slashed{\epsilon}_{M_i} $ is \textit{always} even  (including zero), after the simplification of each pair of contracted $\gamma$ matrices. Furthermore, the $\slashed{p}_{M_i} \slashed{\epsilon}_{M_i} $ structure remains intact after the simplification, i.e.~$\slashed{p}_{M_i}  $ does not get separated from $\slashed{\epsilon}_{M_i}$. The starting configuration (before any simplification of the $\gamma$ matrices chain) must have an odd number of $\gamma$ matrices between the  $\slashed{p}_{M_i} \slashed{\epsilon}_{M_i} $. Therefore, the number of $\gamma$ matrices between the  $\slashed{p}_{M_i} \slashed{\epsilon}_{M_i} $ \textit{after} all contracted $\gamma$ matrices have been simplified through \EQs\eqref{eq:Dirac-gamma-identity-2} to \eqref{eq:Dirac-gamma-identity-4} must remain odd.\footnote{This may not be true at NLO. Indeed, the trace structure becomes much more involved in that case. Furthermore, the identity in \EQ\eqref{eq:Dirac-gamma-identity-4} becomes
\begin{align}
     \gamma^\mu \slashed{p}_{M_i} \slashed{\epsilon}_{M_i} \gamma^\alpha \gamma^\beta \gamma_\mu&= 2 \gamma^\alpha \slashed{p}_{M_i} \slashed{\epsilon}_{M_i}  \gamma^\beta + 2 \gamma^\beta \slashed{p}_{M_i} \slashed{\epsilon}_{M_i}  \gamma^\alpha-(2 \varepsilon) \slashed{p}_{M_i} \slashed{\epsilon}_{M_i} \gamma^\alpha \gamma^\beta\,,
\end{align}
in $4-2\varepsilon$ dimensions, where the last term corresponds to removing an \textit{odd} number of Dirac $\gamma$ matrices on either side of $\slashed{p}_{M_i} \slashed{\epsilon}_{M_i}$.
} This dramatically simplifies the subsequent analysis.

Hence, we need to simply consider all possible arrangements of the Dirac structures in \EQ\eqref{eq:Dirac-structures}. They are
\begin{enumerate}
    \item \label{item:case-1} $\Tr\left(\slashed{p}_{M_1} \slashed{\epsilon}_{M_1} \slashed{p}_{q1}\slashed{p}_{M_2} \slashed{\epsilon}_{M_2}\slashed{\epsilon}_{q}\slashed{p}_{q2}\slashed{p}_{q3}\right)$,
    \item \label{item:case-2} $\Tr\left(\slashed{p}_{M_1} \slashed{\epsilon}_{M_1} \slashed{p}_{q1}\slashed{p}_{M_2} \slashed{\epsilon}_{M_2}\slashed{p}_{q2}\slashed{\epsilon}_{q}\slashed{p}_{q3}\right)$,
    \item \label{item:case-3} $\Tr\left(\slashed{p}_{M_1} \slashed{\epsilon}_{M_1} \slashed{p}_{q1}\slashed{p}_{q2}\slashed{p}_{q3}\slashed{p}_{M_2} \slashed{\epsilon}_{M_2}\slashed{\epsilon}_{q}\right)$,
\end{enumerate}
where one of the three $p_{qi} = p$, and we have chosen to write it in this way to keep the analysis both simple and general. Note that the trace $\Tr\left(\slashed{p}_{M_1} \slashed{\epsilon}_{M_1} \slashed{p}_{q1}\slashed{p}_{M_2} \slashed{\epsilon}_{M_2}\slashed{p}_{q2}\slashed{p}_{q3}\slashed{\epsilon}_{q}\right)$ is actually equivalent to case \ref{item:case-1}, by reversing the direction of the trace and relabelling $M_1 \leftrightarrow M_2$ and $p_{q2} \leftrightarrow p_{q3}$.

Recall that the gauge fixing choices we consider in (i) and (ii) at the beginning of this section are such that the lightlike vector $\ell^\mu$, which can be either $p^\mu$ or $q^\mu$, contracted with \textit{all} three polarisation vectors gives zero. After using momentum conservation to write all three momenta $p_{qi}$ in terms of $\ell^\mu$, $p_{M_1}^\mu$ and $p_{M_2}^\mu$, this implies that one can have at most 1 lightlike vector $\slashed{\ell}$ in the trace. This is because such a lightlike vector $\ell$ is necessarily contracted with $p_{M_i}$, which means there are not enough $p_{M_i}$ momenta left to obtain the structure in \EQ\eqref{eq:product-3-SPs-with-pMi}. Thus, the only cases to consider are those with no $\ell^\mu$ (only relevant for the gauge fixing choice (ii)), or one $\ell^\mu$.
The case with no $\ell^\mu$ is simple: In cases \ref{item:case-1} and \ref{item:case-2}, $\slashed{p}_{q1}$ is sandwiched between $\slashed{p}_{M_1} \slashed{\epsilon}_{M_1}$ and $\slashed{p}_{M_2} \slashed{\epsilon}_{M_2}$, which immediately vanishes. In case \ref{item:case-3}, one is forced to choose $p_{q1} = p_{M_2}$ and $p_{q3} = p_{M_1}$, in which case there is no way to choose $p_{q2}$ (without any $\ell$) in order to make the Dirac chain non-vanishing.

The case with one $\ell^\mu$ requires some more work. Here, one can freely anti-commute $\slashed{p}_{M_1}$ and $\slashed{p}_{M_2}$. This is because there are four $\slashed{p}_{M_i}$ in the Dirac trace, hence removing a factor $p_{M_1}\cdot p_{M_2}$ from the trace would leave only two $p_{M_i}$ in the trace, which is insufficient to produce the tensor structure in \EQ\eqref{eq:product-3-SPs-with-pMi}. Starting with case \ref{item:case-1}, we find that the only non-vanishing possibility for the trace is
\begin{align}
    \Tr\left(\slashed{p}_{M_1} \slashed{\epsilon}_{M_1} \slashed{\ell}\slashed{p}_{M_2} \slashed{\epsilon}_{M_2}\slashed{\epsilon}_{q}\slashed{p}_{M_1}\slashed{p}_{M_2}\right) \sim  \Tr\left(\slashed{p}_{M_1} \slashed{\epsilon}_{M_1} \slashed{\ell}\slashed{p}_{M_2} \slashed{\epsilon}_{M_2}\slashed{\epsilon}_{q}\slashed{p}_{M_2}\slashed{p}_{M_1}\right) = 0\,,
\end{align}
where the ``$\sim$'' symbol represents an equality after dropping terms which cannot lead to the tensor structure in \EQ\eqref{eq:product-3-SPs-with-pMi}. Next, with case \ref{item:case-2}, we have
\begin{align}
     \Tr\left(\slashed{p}_{M_1} \slashed{\epsilon}_{M_1} \slashed{\ell}\slashed{p}_{M_2} \slashed{\epsilon}_{M_2}\slashed{p}_{M_1}\slashed{\epsilon}_{q}\slashed{p}_{M_2}\right)
     &=
     2 (\epsilon_q \cdot p_{M_2})
     \Tr\left(\slashed{p}_{M_1} \slashed{\epsilon}_{M_1} \slashed{\ell}\slashed{p}_{M_2} \slashed{\epsilon}_{M_2}\slashed{p}_{M_1}\right)\nonumber\\
     &\qquad
     -
     \Tr\left(\slashed{p}_{M_1} \slashed{\epsilon}_{M_1} \slashed{\ell}\slashed{p}_{M_2} \slashed{\epsilon}_{M_2}\slashed{p}_{M_1}\slashed{p}_{M_2}\slashed{\epsilon}_{q}\right)\nonumber\\
     &\sim0\,,
\end{align}
where the first term on the RHS of the first equality vanishes, while the second term is equivalent to case \ref{item:case-1}, after reversing the direction of the Dirac trace and taking $M_{1} \leftrightarrow M_2$. Finally, with case \ref{item:case-3}, we have two distinct possibilities:
\begin{itemize}
    \item Either $\slashed{\ell}$ is next to a $\slashed{p}_{M_i} \slashed{\epsilon}_{M_i}$ structure, which corresponds to a trace of the type
    \begin{align}
        \Tr\left(\slashed{p}_{M_1} \slashed{\epsilon}_{M_1} \slashed{\ell}\slashed{p}_{M_2}\slashed{p}_{M_1}\slashed{p}_{M_2} \slashed{\epsilon}_{M_2}\slashed{\epsilon}_{q}\right)\sim \Tr\left(\slashed{p}_{M_1} \slashed{\epsilon}_{M_1} \slashed{\ell}\slashed{p}_{M_1}\slashed{p}_{M_2}\slashed{p}_{M_2} \slashed{\epsilon}_{M_2}\slashed{\epsilon}_{q}\right)= 0\,.
    \end{align}
    \item Or $\slashed{\ell}$ is sandwiched between two $p_{qi}$, giving
    \begin{align}
           \Tr\left(\slashed{p}_{M_1} \slashed{\epsilon}_{M_1} \slashed{p}_{M_2}\slashed{\ell}\slashed{p}_{M_1}\slashed{p}_{M_2} \slashed{\epsilon}_{M_2}\slashed{\epsilon}_{q}\right) &= 2(\ell \cdot p_{M_2})\Tr\left(\slashed{p}_{M_1} \slashed{\epsilon}_{M_1} \slashed{p}_{M_1}\slashed{p}_{M_2} \slashed{\epsilon}_{M_2}\slashed{\epsilon}_{q}\right)\nonumber\\
           &\qquad -\Tr\left(\slashed{p}_{M_1} \slashed{\epsilon}_{M_1} \slashed{\ell}\slashed{p}_{M_2}\slashed{p}_{M_1}\slashed{p}_{M_2} \slashed{\epsilon}_{M_2}\slashed{\epsilon}_{q}\right)\nonumber\\ &\sim 0\,,
    \end{align}
    where the first term on the RHS of the first equality vanishes, while the second term is equivalent to the previously considered possibility.
\end{itemize}

Thus, we have established that for the two gauge choices in (i) and (ii), $p \cdot \epsilon_{q} = p \cdot \epsilon_{M_{1}} = p \cdot \epsilon_{M_{2}} = 0$ and $q \cdot \epsilon_{M_{1}} = q \cdot \epsilon_{M_{2}} = 0$ respectively, the tensor structure  $(p_{\perp} \cdot \epsilon_{q\perp})(p_{\perp}\cdot \epsilon_{M_1\perp})(p_{\perp}\cdot \epsilon_{M_2\perp})$ \textit{cannot} appear. We highlight that while we have explicitly shown the derivation for the VTT structure, a similar demonstration should also exist for the ATT structure, where a specific tensor structure never appears for each individual diagram (after using the Schouten identity).

\section{Poles in $x$ are always in the ERBL region}

\label{app:ERBL-poles}

Here, we show that the feature in \TAB\ref{tab:denLO-cases}, namely that all poles in $x$ are in the ERBL region $-\xi \leq x \leq  \xi$ is in fact the consequence of a more general aspect of \textit{physical} LO scattering amplitudes.

First, let us focus on the DGLAP region $|x| > \xi$. In this case, the hard partonic scattering subprocess can be represented as a $2 \to 5$ process, where incoming momenta and outgoing momenta all have positive energies. The two incoming particles are the photon, and a quark (or anti-quark) from the nucleon. Consequently, all scalar products between any two momenta are non-negative. Our aim will be to show that the virtuality of each propagator at LO \textit{cannot} change sign within the \textit{physical} phase space.

Since the argument easily generalises to any number of particles in the final state, we will consider a generic $2 \to n$ process, where the two incoming momenta are $p_a$ and $p_b$, while the outgoing ones are $p_{j}$, $j=1,...,n$. All particles are taken to be massless. By momentum conservation, the momentum $k$ of any internal propagator is fixed, and its virtuality can be generically represented as
\begin{align}
    k^2 = \left(\sum_{i \subset I} p_{i} -\sum_{j \subset F} p_{j}  \right)^2\,,
    \label{eq:construct-virtualityk2}
\end{align}
where $i \subset I$ ($j \subset F$) represents any subset of the incoming (outgoing) momenta. Of course, one omits the case where no momentum or all of the momenta are chosen, since such a virtuality cannot actually occur and is in fact zero.

Now, if either none or both incoming momenta are picked, it is clear that the virtuality has to be non-negative. For the case where both incoming momenta are picked, one simply uses momentum conservation to express $k$ in terms of final-state momenta only.

The interesting case corresponds to when one incoming momentum (taken to be $p_a$ without loss of generality) and a subset of final-state momenta are picked. Let us denote the sum of momenta included (excluded) in this subset to be $\pin$ ($\pout$),
\begin{align}
    \pin = \sum_{j \subset F} p_{j} \,,\qquad \pout = \sum_{j=1}^n p_j - \pin\,.
\end{align}
One can therefore effectively represent such a situation as a $2 \to 2$ scattering, which significantly simplifies the analysis, with the caveat that the final state effective particles may now be massive, i.e.~$\pin^2,\,\pout^2 \geq 0$.

The virtuality of $k$ can thus be written as
\begin{align}
    k^2 = -2 p_a \cdot \pin  + \pin^2\,.
\end{align}
At first glance, this virtuality may appear to have a sign depending on the relative signs of $p_a \cdot \pin$ and $\pin^2$. However, as we will find, for any \textit{physical} phase space points, this virtuality is \textit{always} negative. Working in the centre-of-mass frame of the two incoming particles, we find that
\begin{align}
\label{eq:k2-CM-frame}
    k^2 = -2 E_a (E_\mathrm{inc}-|\vec{p}_{\mathrm{inc}}|\cos \theta)+\pin^2\,,
\end{align}
where $\theta$ is the angle between $\vec{p}_a$ and $\vec{p}_\mathrm{inc}$. It is straight-forward to show that
\begin{align}
    E_a = \frac{\sqrt{s_{ab}}}{2}\,,\quad E_{\mathrm{inc}} = \frac{s_{ab}+\pin^2-\pout^2}{2\sqrt{s_{ab}}}\,,\quad |\vec{p}_{\mathrm{inc}}| = \frac{\lambda^{\frac{1}{2}}(s_{ab},\,\pin^2,\,\pout^2)}{2\sqrt{s_{ab}}}\,,
\end{align}
where $s_{ab} = (p_a+p_b)$ is the square of the centre-of-mass energy, and the function $\lambda$ is the K\"all\'en function, see \EQ\eqref{eq:Kallen-function}. Substituting into the expression for $k^2$ in \EQ\eqref{eq:k2-CM-frame}, we obtain
\begin{align}
    k^2 &= \pin^2 - \frac{1}{2}\left[(s_{ab}+\pin^2-\pout^2) - \cos\theta\sqrt{(s_{ab}-\pin^2-\pout^2)^2-4 \pin^2\, \pout^2} \right]\nonumber\\
    &\leq - \frac{1}{2}\left[(s_{ab}-\pin^2-\pout^2) - \sqrt{(s_{ab}-\pin^2-\pout^2)^2-4 \pin^2\, \pout^2} \right]\nonumber\\
    &\leq 0\,,
    \label{eq:virtualityk2}
\end{align}
where we have used $\cos \theta \leq  1$ in the first inequality. The second inequality follows using the fact that any physical phase space point yields a non-negative argument for the square root, as well as the observation that
\begin{align}
    &s_{ab} = (p_a+p_b)^2 = (\pin+\pout)^2 = \pin^2 +2 \pin \cdot \pout + \pout^2\\ 
    &\implies (s_{ab}-\pin^2-\pout^2) = 2 \pin \cdot \pout \geq 0\,.
\end{align}
Finally, we highlight that the edge case $k^2=0$ can only occur when $\cos \theta =1$, and either $\pin^2 =0 $ or $\pout^2 = 0$, as can be seen from \EQ\eqref{eq:virtualityk2}.

Thus, we have demonstrated that, for a generic $2 \to n $ process with massless legs,
\begin{equation}
\label{eq:k2-cases}
    k^2 \begin{cases}
        \geq 0\,,& \textrm{either none or both incoming momenta included in $k^2$},\\[5pt]
        \leq 0\,, & \textrm{one incoming momentum included in $k^2$},
    \end{cases}
\end{equation}
for any physical point in the phase space. Consequently, there cannot be poles in $x$ in the DGLAP region, as we have found in \TAB\ref{tab:denLO-cases}, since a given propagator falls into only one of the two categories in \EQ\eqref{eq:k2-cases}.

We note that the arguments presented above break down for a generic $3 \to n$ process, where representing it as a $2 \to 2 $ scattering now corresponds to combining two of the incoming particles into an effective one. There are only two distinct cases to consider for the virtuality of a generic propagator in such a case: First, where none or all of the incoming momenta are in $k^2$ in \EQ\eqref{eq:construct-virtualityk2}, in which case $k^2$ is forced to be non-negative, as in the $2 \to n$ case. Second, where one or two of the incoming momenta are picked. By momentum conservation, we can focus only on the case where one of the incoming momenta are picked. Without loss of generality, we consider momentum $p_a$ to be the picked incoming momentum, while the other effective incoming momentum is $p_{b'}$. A crucial distinction from the $2 \to n$ scattering case is that $p_{b'}^2 \geq 0$. As before, we obtain \EQ\eqref{eq:k2-CM-frame} for $k^2$, but now with
\begin{align}
   E_a= \frac{s_{ab'}-p_{b'}^2}{2\sqrt{s_{ab'}}}\,,
\end{align}
in the centre of mass frame of the $ab'$ system. The same expressions for $E_{\mathrm{inc}}$ and $|\vec{p}_{\mathrm{inc}}|$ hold, after the simple replacement $s_{ab} \to s_{ab'}$. This time, we have
\begin{align}
\label{eq:newvirtualityk2}
    k^2 &=  \pin^2 - \frac{s_{ab'}-p_{b'}^2}{2 s_{ab'}}\left[(s_{ab'}+\pin^2-\pout^2) - \cos\theta\sqrt{(s_{ab'}-\pin^2-\pout^2)^2-4 \pin^2\, \pout^2} \right]\nonumber\\
    &=\frac{p_{b'}^2\pin^2}{s_{ab'}}- \frac{s_{ab'}-p_{b'}^2}{2 s_{ab'}}\left[(s_{ab'}-\pin^2-\pout^2) - \cos\theta\sqrt{(s_{ab'}-\pin^2-\pout^2)^2-4 \pin^2\, \pout^2} \right]\,.
\end{align}
It can be deduced by inspection that $k^2$ can be either positive or negative inside the phase space, depending on the exact values of kinematical variables, as long as $p_{b'}^2>0$. To see this clearly, consider a situation where one fixes $\pin^2$ to some positive value. We then treat $p_{b'}^2$ as a phase space parameter that we can freely vary from $0$ to $s_{ab'}$. Close to the threshold, i.e.~when $p_{b'}^2\lesssim s_{ab'}$, the first term after the second equality in \EQ\eqref{eq:newvirtualityk2} is $\frac{p_{b'}^2\pin^2}{s_{ab'}} \sim \pin^2 $. However, the second term is negative as it corresponds to a very small negative number $- \frac{s_{ab'}-p_{b'}^2}{2 s_{ab'}}$ multiplied by a positive number (in the square brackets). It is thus clear that one can always choose $p_{b'}^2$ to be close enough to $s_{ab'}$, such that $k^2>0$. On the other hand, when $p_{b'}^2 \to 0$, the expression for $k^2$ collapses to the first line of \EQ\eqref{eq:virtualityk2}, and in that case $k^2 \leq 0$. This unambiguously shows that the sign of $k^2$ is not fixed in massless $3 \to n$ scattering.

Thus, one concludes that propagator denominators can change sign as a function of $x$, $v$ and $z$ when in the ERBL region, defined by $|x|<\xi$.

\section{Symmetries}

\label{app:symmetries}

In this appendix, we study three types of symmetries:  charge conjugation symmetry (at the diagram level),  isospin symmetry and meson exchange symmetry.  These relations provide a good check of the numerical results.

\subsection{Charge conjugation symmetry}

The hard part of any diagram (after Fierz projection) can be mapped onto a different one of the same process through charge conjugation, up to a sign. To illustrate this symmetry, let us consider a diagram of topology 2 in \FIG\ref{fig:topologies}, and let us denote by $C$ the charge conjugation matrix. The numerator of the amplitude (up to a prefactor) reads
\begin{equation}
\label{eq:trace}
\mathrm{Tr}\left(GPD\,x_3DA_1x_2DA_2x_1\right)=\mathrm{Tr}\left(C\,GPD\,C^{-1}C\,\tilde x_3\,C^{-1}C\,DA_1\,C^{-1}C\,\tilde x_2\,C^{-1}C\,DA_2\,C^{-1}C\,\tilde x_1\,C^{-1}\right)\,,
\end{equation}
where $x_1$, $x_2$ and $x_3$ correspond to Dirac matrices coming from propagators, quark-gluon vertices and/or a quark-photon vertex.  The tilde on the $x_i$ on the RHS correspond to rewriting $x_i$, by inserting $C^{-1}C$ in between \textit{all} $\gamma$ matrices in $x_i$, i.e.
\begin{align}
    x_i = \gamma^\mu\gamma^\nu \cdots \gamma^\alpha=\tilde x_i = \gamma^\mu \,C^{-1}C\, \gamma^\nu \,C^{-1}C \cdots \gamma^\alpha\,.
\end{align}
We recall that 
\begin{equation}
\label{eq:Cproperties}
    C\;\gamma_\mu\;C^{-1}=-\, \gamma_\mu^t,\qquad C\;\sigma_{\mu\nu}\;C^{-1}=-\,\sigma_{\mu\nu}^t ,\qquad C\;(\gamma_5\gamma_\mu)\;C^{-1}=\,(\gamma_5\gamma_\mu)^t .
\end{equation}
The number of $\gamma$ matrices different from $\gamma^5$ must be even for the trace to be non-zero. Moreover, we know that the transversity matrices $\sigma_{\mu\nu}$ must appear in a pair, for the same reason. Therefore, the minus signs coming from the two first equalities in \EQ\eqref{eq:Cproperties} compensate in \EQ\eqref{eq:trace}. However, if there is an odd number of $\gamma^5$ in the trace, then the minus signs will not compensate according to the last equality of \EQ\eqref{eq:Cproperties}, since any $\gamma^\mu$ matrix associated with a $\gamma^5$ (which only comes from $GPD$/$DA$) does \textit{not} produce a minus sign. Thus, the resulting sign of the trace is $(-1)^{\sharp\gamma^5}$. By taking the transpose of the argument of the trace in \EQ\eqref{eq:trace}, which leaves the trace invariant, we finally get 
\begin{equation}
\label{eq:transposedtrace}
\mathrm{Tr}\left(GPD\;x_3DA_1x_2DA_2x_1\right)=(-1)^{\sharp\gamma^5}\mathrm{Tr}\left(GPD \;\overleftarrow{x_1}\; DA_2 \;\overleftarrow{x_2}\; DA_1 \;\overleftarrow{x_3}\right)\,,
\end{equation}
where $\overleftarrow{x_i}$ means that the order of the $\gamma$ matrices inside $x_i$ is reversed.

\begin{figure}[t!]
\centering
\includegraphics[width=0.3\linewidth]{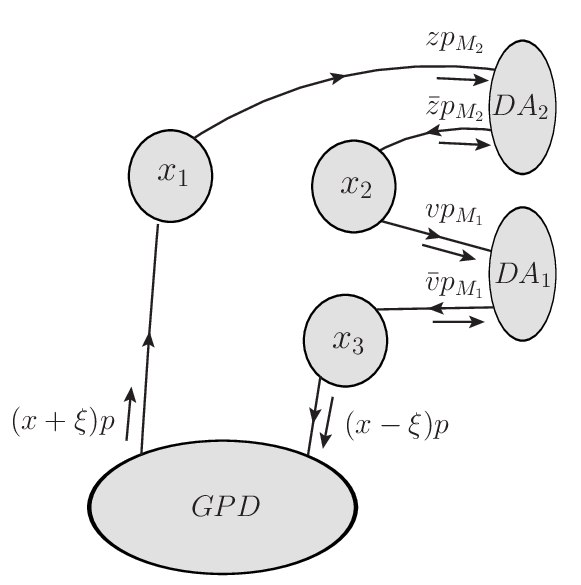}\hspace{2cm}
\includegraphics[width=0.3\linewidth]{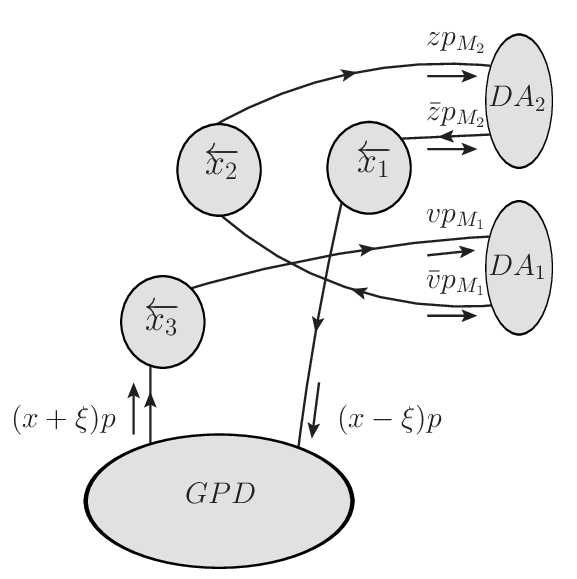}\\
\hspace{-.5cm}\includegraphics[width=0.37\linewidth]{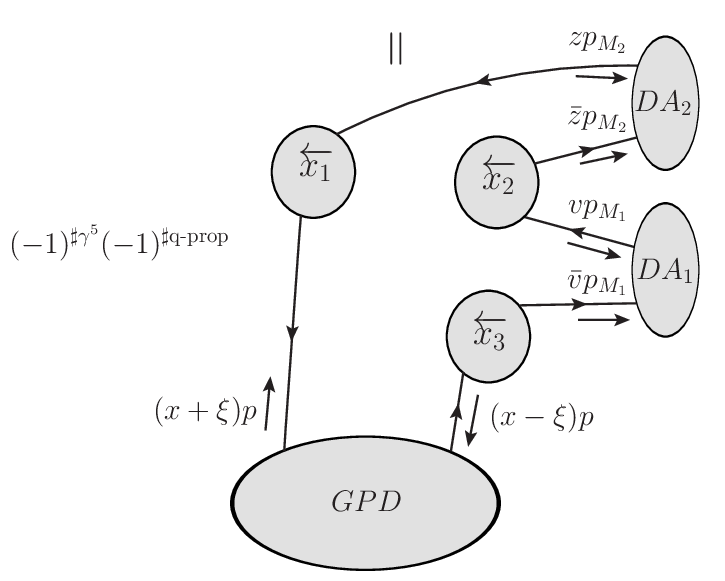}\hspace{2cm}
\includegraphics[width=0.35\linewidth]{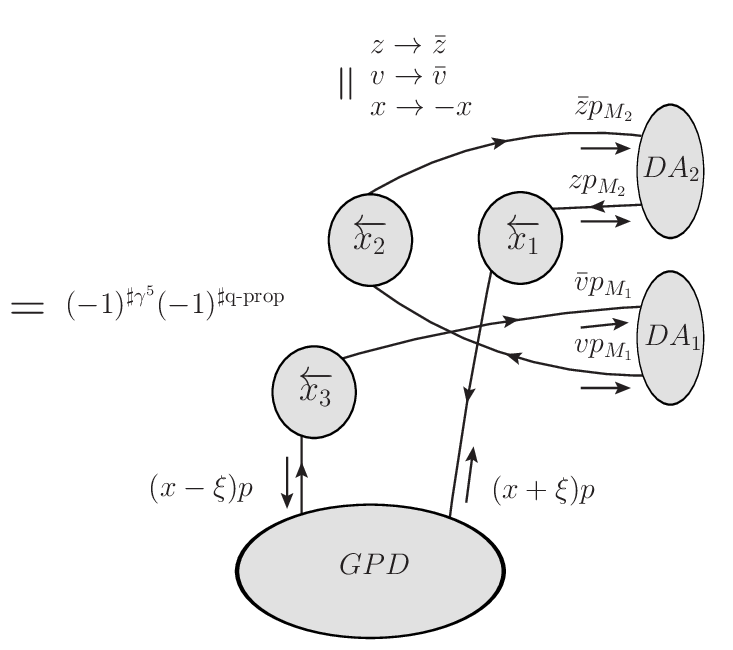}
    \caption{Charge conjugation of a diagram of topology 2 (top left). After a series of manipulations, such a diagram is related to one in topology 4 (top right), see \FIG\ref{fig:topologies}. }
    \label{fig:conjsym}
\end{figure}

Consider two diagrams which we label as A and B, whose Dirac traces are given by the LHS and RHS (without $(-1)^{\sharp\gamma^5}$) of \EQ\eqref{eq:transposedtrace} respectively. They are represented in diagrammatric form in the top panel of \FIG\ref{fig:conjsym}. Let us now determine the transformations required to transform diagram A onto B. First, we note that diagram A, with the direction of the quark lines reversed, corresponds to reversing the order of Dirac matrices, which means that the direction of momentum is against the spinor flow. Each quark propagator picks up a negative sign, and hence the first equality relating the two diagrams on the left hand side of \FIG\ref{fig:conjsym} is accompanied by $(-1)^{\sharp\gamma^5}(-1)^{\sharp\text{q-prop}}$, where $\sharp\text{q-prop}$ is the number of quark propagators. The second equality simply redraws the same diagram. Finally, performing the transformation $x \to -x$, $z \to \bar z$, and $v \to \bar v$ maps the result onto diagram B, as indicated by the third equality in \FIG\ref{fig:conjsym}. Inside the 3-dimensional convolution integral, using the notation of \EQ\eqref{eq:amplitude}, this relation implies that 
\begin{align}
    &\int_{-1}^{1}dx \int_{0}^1dz\int_0^{1}dv\;T_ {HA}(x,v,z)H(x)\phi_1(v)\phi_2(z)+\int_{-1}^{1}dx \int_{0}^1dz\int_0^{1}dv\;T_{HB}(x,v,z)H(x)\phi_1(v)\phi_2(z)\nonumber\\=&\int_{-1}^{1}dx \int_{0}^1dz\int_0^{1}dv\;T_ {HA}(x,v,z)H(x)\phi_1(v)\phi_2(z)\nonumber\\
    &\hspace{3cm}+(-1)^{\sharp \gamma^5+\sharp \text{q-prop}}\int_{-1}^{1}dx \int_{0}^1dz\int_0^{1}dv\;T_{HA}(-x,\bar{v},\bar{z})H(x)\phi_1(v)\phi_2(z)\nonumber\\
    =&\int_{-1}^{1}dx \int_{0}^1dz\int_0^{1}dv\;T_ {HA}(x,v,z)(H(x)+(-1)^{\sharp \gamma^5+\sharp \text{q-prop}}H(-x))\phi_1(v)\phi_2(z)\,,
    \label{eq:integral-charge-conjugation}
\end{align}
where we have used the symmetry property of the DAs,  $\phi_{1}(v) = \phi_{1}(\bar v)$ and $\phi_{2}(z) = \phi_{2}(\bar z)$.  Thus, the relation in \EQ\eqref{eq:transposedtrace} obtained from charge conjugation enables the reduction of the number of diagrams to integrate by 2, provided the appropriate modification on the GPD $H(x)$ is made, as is done in the last line of \EQ\eqref{eq:integral-charge-conjugation}.

In the above explicit example, we have shown that through conjugation symmetry, diagrams of topology 2 can be related to those of topology 4, both of which involve a single Dirac trace. On the other hand, charge conjugation symmetry connects diagrams within the \textit{same} topology for topologies 1, 3 and 5, all of which involve two Dirac traces.

So far, we have assumed that all the quark lines have the same flavour. We stress however that the charge conjugation symmetry discussed above is exact for the hard part of the diagram when the electric charge is factorised, modulo a minus sign given by $(-1)^{\sharp\gamma^5+\sharp\text{q-prop}}$. In the general case, one needs to examine on a case-by-case basis the proper transformations on the charges and GPDs that enter the full amplitude. For concreteness, let us discuss the example of a generic diagram from topology 2 (top left diagram of \FIG\ref{fig:conjsym}). We distinguish between three types of processes:
\begin{itemize}
    \item First, when the two mesons are neutral, in which case all fermion lines have the same flavour, which can be either $u$ or $d$. Therefore, the application of charge conjugation will not affect the flavours of the quark lines in the process. 
    \item Second, when one of the mesons is neutral and the other is charged. This case necessarily involves a transition GPD. Note also that the flavour of each quark line is fixed once a specific process is chosen. For example, when $M_1 = \pi^0$ and $M_2 = \pi^+$, the quark lines associated with $x_1$, $x_2$ and $x_3$ have flavours $u$, $d$ and $d$, respectively. In this case, strict application of charge conjugation will change the process to $M_2 = \pi^-$. Thus, to obtain the corresponding diagram for the same original process, it is necessary to swap the flavours of the quark-antiquark lines entering each charged outgoing states. For this particular example, this means that the quark lines associated to $\overleftarrow{x_1}$, $\overleftarrow{x_2}$ and $\overleftarrow{x_3}$ should reassigned to be $d$, $u$ and $u$, respectively in order to recover the original process with $M_1 = \pi^0$ and $M_2 = \pi^+$. Note that after reassigning the quark flavours, the same transition GPD as in the original diagram appears. However, the transformed diagram picks up a negative sign due to the fact that the neutral wavefunction has a relative minus sign between the $|\bar u u\rangle$ and $|\bar d d\rangle$ states.
    \item Third, when the two mesons have opposite charges. As in the second case, the flavours of all three quark lines are fixed when a process is specified. For example, when $M_1=\pi^+$ and $M_2 = \pi^-$, the quark lines associated with $x_1$, $x_2$ and $x_3$ have flavours $d$, $u$ and $d$, respectively. Applying charge conjugation alone will change the process by swapping the charges of the mesons. Therefore, it is once again necessary to adjust the flavours of the quark line to map  diagrams within the same process. For the example given here, this means that $\overleftarrow{x_1}$, $\overleftarrow{x_2}$ and $\overleftarrow{x_3}$ should be associated with flavours $u$, $d$ and $u$, respectively. However, in that process, the initial $d$ flavour GPD has changed to a $u$ flavour one.
\end{itemize}

\begin{figure}[t!]
    \centering
    \includegraphics[scale=0.5]{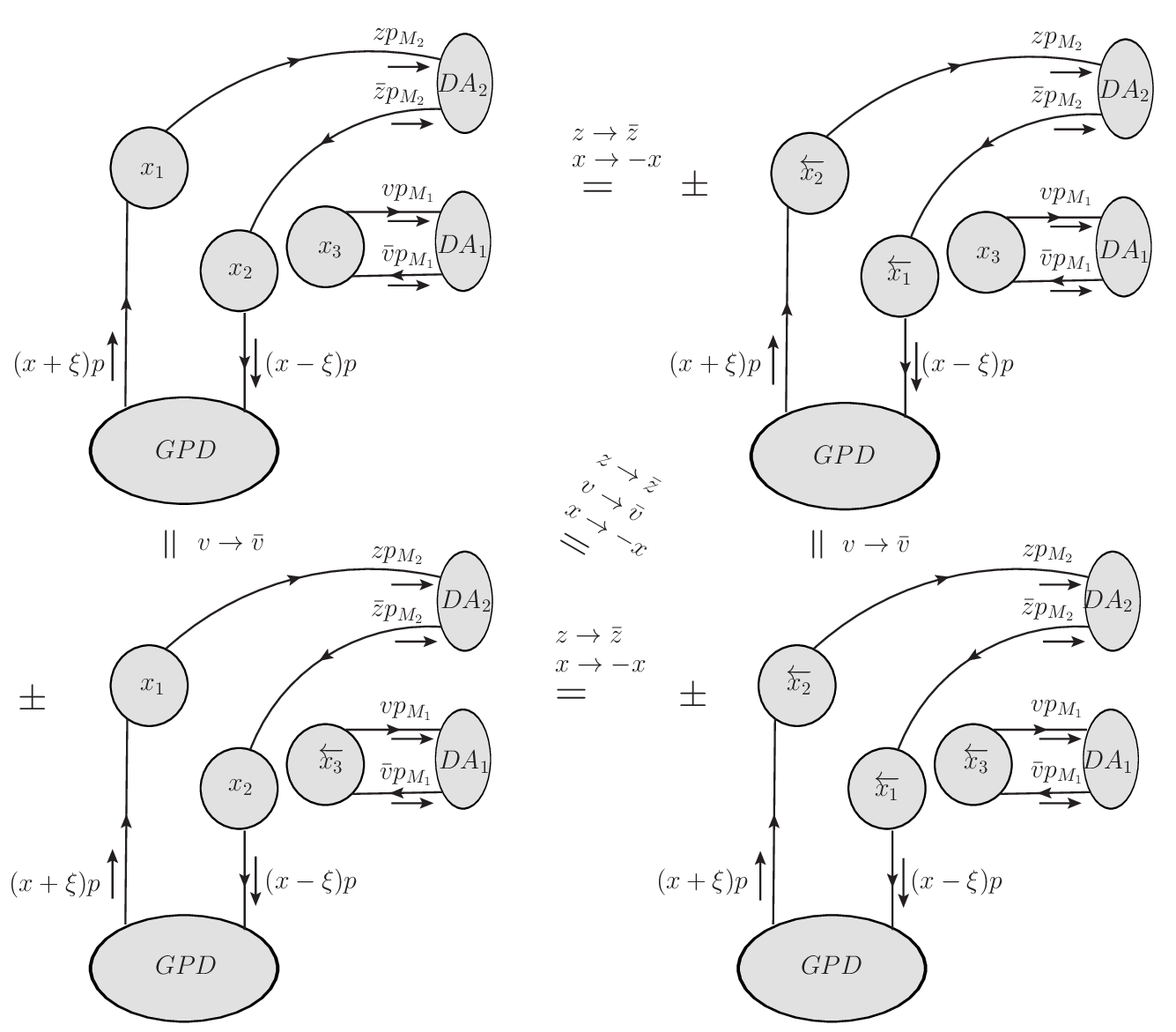}
    \caption{A diagram of topology 5 can be mapped on three different diagrams by charge conjugation. }
    \label{fig:conjsym2}
\end{figure}

In the case of topologies 1, 3 and 5, which consist of two traces, we can choose to apply the charge conjugation operation on only one of them. This results in two additional relations between diagrams which reduces further the number of diagrams to integrate in the amplitude, as indicated in \FIG\ref{fig:conjsym2}. In some cases, these symmetries can even lead to a complete cancellation of the topologies involved. For instance, consider the topology 5 with $M_1 = \pi^0$. By colour conservation, two gluons must be attached to the fermion line corresponding to $\pi^0$. Due to the flavour antisymmetry of the pion wavefunction,  diagrams where the photon is not connected to the fermion line of the $\pi^0$ meson cancel. Hence, the trace associated to the $\pi^0$ meson contains two propagators and one $\gamma^5$ matrix coming from the pion DA, such that upon reversing the direction of the quark line ($(-1)^{\sharp \gamma^5+\sharp \text{q-prop}} = -1$) and adjusting the momentum fraction of the pion DA ($v \to \bar v$), one obtains a minus sign relative to the original diagram. Therefore, pairs of diagrams cancel in this way, and the whole set of diagrams corresponding to that topology adds up to zero.

This symmetry has an important consequence for topology 1, namely, that only the process $\pi^0\rho^0_L$ has a non-vanishing contribution from that topology when the incoming photon is attached to the fermion lines connected to the DAs. This is because such diagrams cancel each other due to the symmetry $v\to\bar{v}$ and $z\to\bar{z}$, when the two  DAs either both contain a $\gamma^5$ or neither does, since there is an odd number of quark propagators in the fermion loop that connects the two DAs (and so, one has $(-1)^{\sharp \gamma^5+\sharp \text{q-prop}} = -1$).
This is consistent with the observation that gluon GPD contributions, which correspond to a modified version of topology 1 where the GPD  does not have the quark lines and the blob attached to it as in \FIG\ref{fig:topologies}, all cancel for $\pi^0\pi^0$ and $\rho^0\rho^0$ production. These diagrams are precisely the problematic ones that are involved in collinear factorisation breaking effects \cite{Nabeebaccus:2023rzr,Nabeebaccus:2024mia}.
This means that the amplitude is integrable for the processes $\pi^0\pi^0$ and $\rho^0\rho^0$, but it is divergent for $\pi^0\rho^0$ due the gluon GPD contribution,\footnote{The divergence actually occurs at leading order for the gluon GPD contribution. The corresponding leading order calculation for the quark GPD contribution for $\pi^0\rho^0$ indicates that it is finite. However, it is expected to diverge at NLO, based on the arguments in \cite{Nabeebaccus:2023rzr,Nabeebaccus:2024mia}.} see \APP\ref{app:gluon-GPDs}. The fact that the processes $\pi^0\pi^0$ and $\rho^0\rho^0$ do not have a contribution from topology 1 when the photon is connected to the fermion lines of the DAs can of course be understood on general grounds of charge parity conservation in QED/QCD. Indeed, the incoming photon has $C=-1$, while both the pair of mesons and the pair of gluons exchanged in the $t$-channel have $C=+1$.

\subsection{Isospin symmetry}
Isospin symmetry is assumed to model the neutron GPD from the proton PDFs. This symmetry implies that
\begin{equation}
    u^p=d^n,\;\;d^p=u^n\;\;\text{and}\;\; \bar{u}^p=\bar{d}^n,\;\;\bar{d}^p=\bar{u}^n\,,
\end{equation}
where $u^p$, $d^p$ ($u^n$, $d^n$) denote the PDFs of flavour $u$ and $d$ in the proton (neutron).

The consequence is that the transition GPDs from a proton to a neutron, denoted by $H^{ud}$, and from a neutron to a proton ($H^{du}$) are equal: 
\begin{equation}
\label{eq:transGPDsym}
H^{ud}=H_p^u-H_p^d=H^{du},
\end{equation}
where the subscript $p$ indicates that the GPD is that of a proton. This isospin symmetry also leads to 
\begin{equation}
    H_n^u=H_p^d\quad\text{and}\quad H_n^d=H_p^u\,.
    \label{eq:GPDsym}
\end{equation}
In other words, probing a $u$
 quark inside a proton is, from the point of view of QCD, the same as probing a $d$ quark inside a neutron. 
\EQ\eqref{eq:transGPDsym} implies that the matrix element for a proton that the active parton is a $u$ quark and that the absorbed one is a $d$ quark, is equal to the one for a neutron, that the active parton is a $d$ quark, and that the absorbed one is a $u$ quark.

Exchanging the flavours $u$ and $d$ amounts to inverting the charges of the mesons and to transforming the proton (neutron) into a neutron (proton). The isospin symmetry, due to the properties of the nucleon GPD under the $u\leftrightarrow d$ transformation, relates the coefficient of the tensor structures of a process to the one of its symmetric partner. 

Indeed, the GPD, whether it is a transition GPD or one with a well-defined flavour, is invariant under the isospin transformation, according to \EQ\eqref{eq:transGPDsym} and \EQ\eqref{eq:GPDsym}. Consequently, the only effect of the isospin transformation is to transform the charge of the fermionic line probed by the photon. Once this charge has been factored out, we expect that the coefficients associated to the charge $Q_u$ (resp. $Q_d$) in front of the tensor structures of a specific process must be equal to the one associated to the charge $Q_d$ (resp. $Q_u$) of the symmetric process. This is indeed what we observe for the pairs of processes: $\rho^+_T\rho^0_T\leftrightarrow\rho^-_T\rho^0_T$, $\rho^+_L\rho^0_T\leftrightarrow\rho^-_L\rho^0_T$, $\rho^+_L\rho^0_L\leftrightarrow\rho^-_L\rho^0_L$, $\rho^+_T\rho^0_L\leftrightarrow\rho^-_T\rho^0_L$, $\pi^+\rho^0_T\leftrightarrow\pi^-\rho^0_T$, $\pi^+\rho^0_L\leftrightarrow\pi^-\rho^0_L$, $\pi^0\rho^+_T\leftrightarrow\pi^0\rho^-_T$, $\pi^0\rho^+_L\leftrightarrow\pi^0\rho^-_L$, $\pi^+\pi^0\leftrightarrow\pi^-\pi^0$, $p\pi^0\pi^0\leftrightarrow n\pi^0\pi^0$, $p\rho^0_L\rho^0_L\leftrightarrow n\rho^0_L\rho^0_L$, $p\rho^0_L\rho^0_T\leftrightarrow n\rho^0_L\rho^0_T$ and $p\rho^0_T\rho^0_T\leftrightarrow n\rho^0_T\rho^0_T$, where we explicitly specify the target ($p$ or $n$) when both final state mesons are neutral in order to avoid ambiguity. This is true also for $p\pi^+\pi^-\leftrightarrow n\pi^-\pi^+$ and $p\rho^+_L\rho^-_L\leftrightarrow n\rho^-_L\rho^+_L$. However, the latter two processes involve collinear factorisation breaking effects due to the possibility of exchanging two gluons with the nucleon target, see \APP\ref{app:gluon-GPDs}.

\subsection{Meson exchange symmetry}

\begin{figure}[t!]
\centering
\begin{minipage}{0.45\textwidth}
    \includegraphics[width=\linewidth]{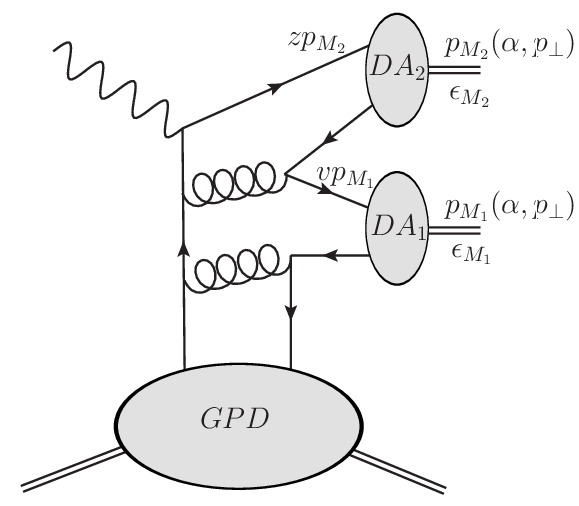}
\end{minipage}
\hfill
\Huge{$=$}
\hfill
\begin{minipage}{0.45\textwidth}
    \includegraphics[width=\linewidth]{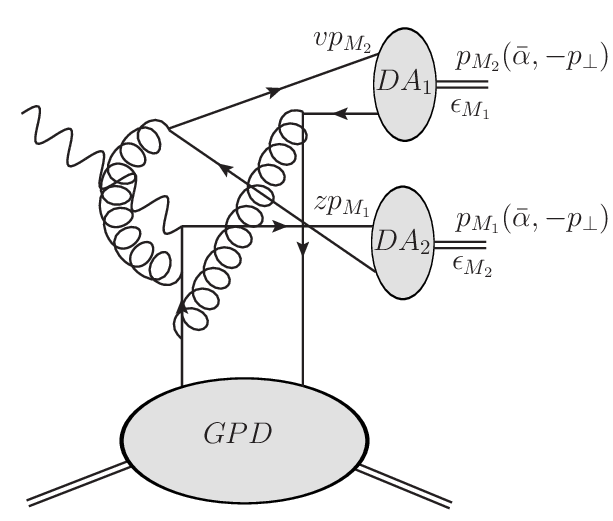}
\end{minipage}
\caption{A diagram can be transformed into a diagram of the symmetric process where the mesons $M_1$ and $M_2$ have been exchanged. For the right-hand-side diagram to correspond to the symmetric process, the polarisation vectors have to be exchanged (in the case of polarised vector mesons). The variables $v$ and $z$ are dummy variables since they are integrated, so we can further exchange $v$ and $z$ in the right-hand-side diagram. The momenta appearing in the left-hand-side have to be understood as  $p_{M_1}(\alpha, p_\perp)$ and $p_{M_2}(\alpha, p_\perp)$ (see \EQ\eqref{eq:approxkinematics}), whereas, on the right-hand-side, they have to be understood as $p_{M_1}(\bar{\alpha}, -p_\perp)$ and $p_{M_2}(\bar{\alpha}, -p_\perp)$ (i.e.~\EQ\eqref{eq:approxkinematics} with $\alpha \to \bar \alpha$ and $p_{\perp} \to -p_{\perp}$). }
\label{fig:exchangesym}
\end{figure}

According to \EQs\eqref{eq:pM1} and \eqref{eq:pM2}, exchanging the two mesons in final state amounts to the transformation
\begin{equation}
    \alpha \rightarrow \bar{\alpha}\equiv1-\alpha\,,
    \quad
    p_{\perp}\rightarrow-p_{\perp}\,,
    \quad
    \epsilon_{M_1}\leftrightarrow\epsilon_{M_2}\,.
\end{equation}
Each diagram of the process $\gamma N\to N'M_1M_2$ can be mapped on a diagram of the symmetric process $\gamma N\to N'M_2M_1$. If both mesons are identical, and identically polarised, this gives an interesting symmetry between the diagrams. Let us denote by $ \{C_i(\alpha)\}$ ($ \{C'_i(\alpha)\}$) the coefficients in front of the various tensor structures $\{T_i\}$ ($\{T'_i\}$) which enter in the amplitude of the process $\gamma N\rightarrow N'M_1M_2$ ($\gamma N\rightarrow N'M_2M_1$), so that
\begin{equation}
    i\mathcal{M}(\gamma N\rightarrow N'M_1M_2)=\sum_i C_i(\alpha)\; T_i\,.
\end{equation}
Because of the above-mentioned one-to-one correspondence between diagrams of both processes, we have
\begin{equation}
\label{eq:coefrelation}
    \sum_i C_i(\alpha)\;T_i=\sum_i C'_i(\bar{\alpha})\;T'_i\bigg{|}_{p_\perp\rightarrow-p_\perp,\,\epsilon_{M_1}\leftrightarrow\epsilon_{M_2}}\,.
\end{equation}
Since the tensors are always either linear or cubic in $p_\perp$ (see \TAB\ref{tab:tensorbasis}), the transformation $p_\perp\rightarrow-p_\perp$ amounts to a minus sign in the above formula.
If we define
\begin{equation}
\phi(T'_i)= T'_i\bigg{|}_{\epsilon_{M_1}\leftrightarrow\epsilon_{M_2}}\,,
\end{equation}
 then \EQ\eqref{eq:coefrelation} becomes \begin{equation}
\label{eq:coefrelation3}\sum_iC_i(\alpha)T_i=-\sum_iC'_{i}(\bar{\alpha})\,\phi(T_i')\,.
\end{equation}
This relation holds not only for the coefficient in front of the tensor structures, but also for the coefficients in front of each electric charge. Indeed, in the transformation shown in \FIG\ref{fig:exchangesym}, focusing on two specific diagrams, the  flavour of the quark line to which the photon is attached remains unchanged.

If the mesons are identical, then this symmetry has a further implication. Instead of \EQ\eqref{eq:coefrelation}, one has the stronger constraint that
\begin{align}
\label{eq:identical-mesons-symmetry}
     \sum_i C_i(\alpha)\;T_i=\sum_i C_i(\bar{\alpha})\;T_i\bigg{|}_{p_\perp\rightarrow-p_\perp,\,\epsilon_{M_1}\leftrightarrow\epsilon_{M_2}}= -\sum_i C_i(\bar{\alpha})\;\phi(T_i) \,.
\end{align}
In other words, this relation implies that the labelling of the outgoing mesons as $M_1 M_2$ or $M_2 M_1$ leaves the amplitude invariant.

\begin{figure}[t!]
\includegraphics[width=0.5\linewidth]{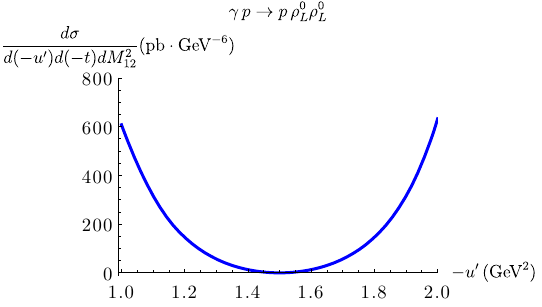}
\includegraphics[width=0.5\linewidth]{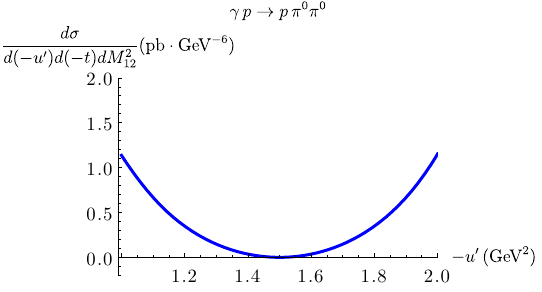}
\caption{Differential cross section at $S_{\gamma N}=\SI{20}{GeV^2}$, $M_{12}^2=\SI{3}{GeV^2}$ and $-t = (-t)_{\mathrm{min}}$  for the photoproduction of two $\rho^0_L$ mesons (left) and two $\pi^0$ mesons (right). The symmetry $-u'\rightarrow M_{12}^2-(-u')$ and the cancellation at $-u'=\frac{M_{12}^2}{2}$ are clearly visible.}
\label{fig:cancellationcrosssection}
\end{figure}

Now suppose that both mesons are either $\pi^0$ or $\rho^0_L$.  Since there are no polarisation vector associated to the mesons in the amplitude, the set of tensor structures are identical,  $\phi(T_i)=T_i$. From \EQ\eqref{eq:identical-mesons-symmetry}, we obtain
\begin{align}
    \label{eq:coefrelation2}
    &C^q_i(\alpha)=-C^q_{i}(\bar{\alpha})\,,
\end{align}
where the superscript $q$ implies that the relation holds for the coefficient in front of each charge $Q_q$. \EQ\eqref{eq:coefrelation2}
 then implies that
\begin{equation}
   \label{eq:coefcancellation}
    C^q_i\left(\frac{1}{2}\right)=0\,.
\end{equation}
\EQs\eqref{eq:coefrelation2} to \eqref{eq:coefcancellation} imply that the differential cross section is symmetric with respect to $-u'=\frac{M_{12}^2}{2}$ where it vanishes exactly. \FIG\ref{fig:cancellationcrosssection} illustrates these two features.

Next, we examine the case of $\rho^0_T\rho^0_T$
photoproduction. 
In this case, the tensor structures contain polarisation vectors from the transversely polarised mesons. These tensor structures, according to \TAB\ref{tab:tensorbasis}, are 
\begin{align}
    &\text{TA}_{11}=(\epsilon_{M_1 \perp}\cdot p_\perp)(\epsilon_{M_2\perp}\cdot\epsilon_{q\perp})\,,&\text{TB}_{11}&=(\epsilon_{M_2 \perp}\cdot p_\perp)(\epsilon_{M_1\perp}\cdot\epsilon_{q\perp})\,,&\text{TC}_{11}&=(\epsilon_{M_2 \perp}\cdot\epsilon_{M_1\perp})(p_\perp\cdot\epsilon_{q\perp}),\\
    &\text{TA}_{12}=(\epsilon_{M_1\perp}\cdot p_\perp)\epsilon^{\epsilon_{M_2 \perp}\,\epsilon_{q\perp}p\,q}\,
    ,&\text{TB}_{12}&=(\epsilon_{M_2 \perp}\cdot\epsilon_{q\perp})\epsilon^{\epsilon_{M_1\perp} p\,p_\perp q}\,
    ,&\text{TC}_{12}&=(\epsilon_{M_2 \perp}\cdot p_\perp)\epsilon^{\epsilon_{M_1\perp}\epsilon_{q\perp}p\,q}.
\end{align}

\begin{figure}[t!]
    \centering
    \includegraphics[width=0.5\linewidth]{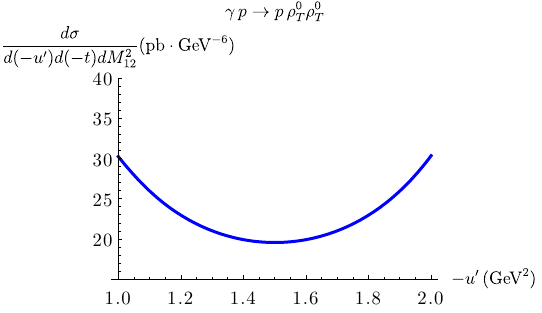}
    \caption{Differential cross section of the photoproduction of $\rho^0_T\rho^0_T$ at $S_{\gamma N}=\SI{20}{GeV^2}$, $M_{12}^2=\SI{3}{GeV^2}$ and $-t = (-t)_{\mathrm{min}}$. Once again, the symmetry with respect to the $-u'=\frac{M_{12}^2}{2}$ axis is visible.}
    \label{fig:transverse-rho0T-symmetry}
\end{figure}

Under the transformation $\epsilon_{M_1}\leftrightarrow\epsilon_{M_2}$, these tensor structures are transformed in the following way:
\begin{align}
    \phi(\text{TA}_{11})&= \text{TB}_{11}\,,\\ \phi(\text{TC}_{11})&= \text{TC}_{11}\,,\\ \phi(\text{TA}_{12})&=\text{TC}_{12}\,,\\
\phi(\text{TB}_{12})&= -\text{TA}_{12}+\text{TB}_{12}+\text{TC}_{12}\,,
\end{align}
where the Schouten identity was used to obtain the last equation.
Therefore, \EQ\eqref{eq:identical-mesons-symmetry} leads to the following relations:
\begin{align}
&C_{A11}^q(\alpha) = -C_{B11}(\bar \alpha )\,,\\
& C_{C11}^q(\alpha) = -C_{C11}(\bar \alpha)\,,\\
&C_{A12}^q(\alpha)=-C_{B12}^q(\alpha)-C_{C12}^q(\bar{\alpha})\,,\\
&C_{B12}^q(\alpha)=-C_{B12}^q(\bar{\alpha})\,.
\end{align}
Interestingly, the amplitude here does not vanish when $\alpha = \bar \alpha  $. Nevertheless, the cross section is symmetric in $\alpha \to \bar \alpha$, as can be seen in \FIG\ref{fig:transverse-rho0T-symmetry}.

These relations provide an interesting sanity check for the results after integration of the amplitude.

\section{Derivation of the cross-section formula}

\label{app:crosssectionproof}

In this Appendix, we derive the differential cross section formula in \EQ\eqref{eq:crosssectionformula}.
As we will show, two angles have already been integrated out in this formula because the cross section is independent of them within the framework of collinear factorisation (due to the neglect of $\Delta_{\perp}$). Consequently, the cross section is not ``fully differential'' in the strict sense.

\subsection{The $2\to 3$ phase space}

To compute the phase space measure of the three final state particle \cite{Byckling:1971vca},  we separate the $2\to 3$ process in two processes, first a $2 \to 2$ process, from which a final state particle subsequently decays into two. Both phase space measures are well-known. 
Let us denote for simplicity the photoproduction process by 
\begin{equation}
\label{eq:2-to-3-process}
    b+a\rightarrow1+2+3\,,
\end{equation}
where $a\equiv N$, $b\equiv\gamma$, $1\equiv M_1$, $2\equiv M_2$ and $3\equiv N'$.
The Mandelstam variables are
\begin{align*}
    &(p_a-p_3)^2=t\,,\qquad
    (p_1+p_2)^2=M_{12}^2\,,\qquad
    (p_b-p_1)^2=u'\,.
\end{align*}
We generically define the Lorentz-invariant phase space factor for a $1 \to n$ or $2 \to n$ process as\footnote{We use the notation of \cite{Byckling:1971vca}, and thus the $(2\pi)$ factors are not included in the phase space, and are instead moved to the flux factor.} 
\begin{align}
      R_n=\int \left(\prod_{i=1}^n \frac{d\vec{p}_i}{2E_i}\right)\delta^4\left(P-\sum_{i=1}^n p_n\right)\,,
\end{align}
where $E_i = \sqrt{\vec{p}_i^{\,2} + m_i^2}$, and $P$ is the total momentum in the initial state. 
For the process in \EQ\eqref{eq:2-to-3-process}, the Lorentz-invariant phase space measure for this process is 
\begin{equation}
    R_3=\int \frac{d\vec{p}_1}{2E_1}\frac{d\vec{p}_2}{2E_2}
\frac{d\vec{p}_3}{2E_3}\delta^4(p_a+p_b-p_1-p_2-p_3)\,.
\end{equation}

\begin{figure}[t!]
    \centering
    \includegraphics[width=0.8\linewidth]{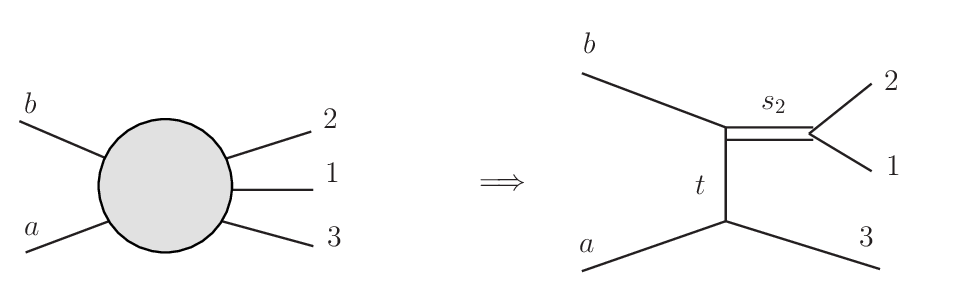}
    \caption{Separation of the $2\to3$ process in two subprocesses, using a fictitious particle of mass $M_{12}^2$. }
    \label{fig:2to3}
\end{figure}
In accordance with \FIG\ref{fig:2to3}, we introduce a fictitious particle of mass $M_{12}$ which decays into 1 and 2. We denote its four-momentum by $p_{12}$. Using
\begin{equation}
    \int_{0}^\infty dM_{12}^2\int \frac{d \vec{p}_{12}}{2E_{12}}\delta^4(p_{12}-p_1-p_2)=1\,,
    \end{equation}
    we have 
\begin{align}
    R_3&=\int_0^\infty dM_{12}^2 \left(\int \frac{d\vec{p}_1}{2E_1}\frac{d\vec{p}_{12}}{2E_{12}}\delta^4(p_a+p_b-p_3-p_{12})\right)\left(\int \frac{d\vec{p}_1}{2E_1}\frac{d\vec{p}_2}{2E_2}\delta^4(p_{12}-p_1-p_2)\right)\nonumber\\
    &=\int_0^\infty  dM_{12}^2 \;R_2(S_{\gamma N};m_3^2,M_{12}^2) R_2(M_{12}^2;m_1^2,m_2^2)\,.\label{eq:R3formula}
\end{align}
To express the two particle phase space measure $R_2$, we introduce the K\"all\'en function
\begin{equation}
\label{eq:Kallen-function}
    \lambda(x,y,z)=(x-y-z)^2-4yz.
\end{equation}
One can show that
\begin{equation}
\label{eq:R1to2}
R_2(M_{12}^2;m_1^2,m_2^2)=\frac{\sqrt{\lambda(M_{12}^2,m_1^2,m_2^2)}}{8\,M_{12}^2}\int d\Omega_1^{R12}\;\Theta\left(M_{12}-m_1-m_2\right)\,.
\end{equation}
The above $\Theta$ function originates from imposing positive energies for particles 1 and 2, as well as $\lambda(M_{12}^2,m_1^2,m_2^2)>0$. Furthermore, $\Omega^{R12}_1$ is the solid angle that specifies the vector $\vec{p}_1$ in the centre of mass frame of particles 1 and 2, denoted by $R12$. 

On the other hand, we can write
\begin{align}
\label{eq:R2to2}
    R_2(S_{\gamma N};m_3^2,M_{12}^2)&=\int \frac{d\vec{p}_{12}}{2E_{12}}\frac{d\vec{p}_3}{2E_3}\delta^4(p_a+p_b-p_3-p_{12})\nonumber\\
    &=\int \frac{d\vec{p}_3}{2E_3}d^4p_{12}\;\delta(p_{12}^2-M_{12}^2)\;\Theta(p_{12}^0)\;\delta^4(p_a+p_b-p_3-p_{12})\nonumber\\
    &=\int \frac{d\vec{p}_3}{2E_3}\;\delta((p_a+p_b-p_3)^2-M_{12}^2)\;\Theta(p_a^0+p_b^0-p_3^0)\;\nonumber\\
    &=\int \frac{d\vec{p}_3}{2E_3}\;dt\;\delta((p_a+p_b-p_3)^2-M_{12}^2)\;\Theta(p_a^0+p_b^0-p_3^0)\;\delta(t-(p_a-p_3)^2).
\end{align}
We now go to spherical coordinates in the center of mass frame $\vec{p}_a+\vec{p}_b=0$, 
\begin{equation}
    \frac{d\vec{p}_3}{2E_3}=\frac{|\vec{p}_3^{\,*}|^{2}d|\vec{p}_3^{\,*}|d\Omega_3^*}{2E_3^*}=\frac{1}{2}|\vec{p}_3^{\,*}|dE_3^* \,d\phi\, d\cos\theta_{a3}^*\,,
\end{equation}
where $|\vec{p}_3^{\,*}|=\sqrt{E_3^{* 2}-m_3^2}$, and $\theta_{a3}^*$ and $\phi$ are, respectively, the polar and azimuthal angles of $\vec{p}_3$ with respect to the axis defined by $\vec{p}_a$ in the centre of mass frame. \EQ\eqref{eq:R2to2} becomes
\begin{align}
    R_2(S_{\gamma N};m_3^2,M_{12}^2)=&\int _0^{2\pi}d\phi\int_{m_3}^{\infty}\,\,dE_3^*\int_{-1}^1d\cos\theta_{a3} ^*\;
    \int dt\;\frac{|\vec{p}_3^{\,*}|}{2}\bigg[\delta(S_{\gamma N}-2E_3^* \sqrt{S_{\gamma N}}+m_3^2-M_{12}^2)\nonumber\\
    &\qquad\times \Theta(\sqrt{S_{\gamma N}}-E_3^*)\;
    \delta(t-m_a^2-m_3^2+2E_a^*E_3^*-2|\vec{p}_a^{\,*}||\vec{p}_3^{\,*}|\,\cos\theta_{a3}^*)\bigg]\,.
\end{align}
 The integration over $E_3^*$ using the first $\delta$ function yields
\begin{align}
 R_2(S_{\gamma N};m_3^2,M_{12}^2)&= \frac{|\vec{p}_3^{\,*}|}{4\sqrt{S_{\gamma N}}}\int_0^{2\pi} d\phi\int_{-1}^1 d\cos\theta_{a3}^*\int dt\Bigg[\Theta\left(\sqrt{S_{\gamma N}}-E_3^*\right)\nonumber\\
 &\qquad\times
\Theta\left(E_3^*-m_3\right)
 \delta(t-m_a^2-m_3^2+2E_a^*E_3^*-2|\vec{p}_a^{\,*}||\vec{p}_3^{\,*}|\,\cos\theta_{a3}^*)\Bigg]\,,
 \label{eq:R2-intermediate}
\end{align}
where $E_3^*$ now should be understood as
\begin{align}
\label{eq:E3starformula}
E_3^*=\frac{S_{\gamma N}+m_3^2-M_{12}^2}{2\sqrt{S_{\gamma N}}}\,.
\end{align}
The integration over $\cos\theta_{a3}^*$ can be done using the remaining $\delta$ function, resulting in 
\begin{align}
R_2(S_{\gamma N};m_3^2,M_{12}^2)=&\frac{1}{8|\vec{p}_a^{\,*}| \sqrt{S_{\gamma N}}}\int_0^{2\pi} d\phi\int dt\Bigg[\Theta
\left(\sqrt{S_{\gamma N}}-E_3^*\right)
\Theta
 \left(E_3^*-m_3\right)
\nonumber\\
&\times\Theta\left(-1\leq\frac{t-(m_a^2+m_3^2)+2E_a^*E_3^*}{2|\vec{p}_a^{\,*}||\vec{p}_3^{\,*}|}\leq 1\right)\Bigg]\,.
\end{align}
Using \EQ\eqref{eq:E3starformula} and 
\begin{equation}
|\vec{p}_3^{\,*}|=\frac{\sqrt{\lambda(S_{\gamma N},m_3^2,M_{12}^2)}}{2\sqrt{S_{\gamma N}}}\,,\qquad    |\vec{p}_a^{\,*}|=\frac{\sqrt{\lambda(S_{\gamma N},m_a^2,m_b^2)}}{2\sqrt{S_{\gamma N}}}\,,\qquad  E^*_a=\frac{S_{\gamma N}+m_a^2-m_b^2}{2\sqrt{S_{\gamma N}}},
\end{equation}
we finally get
\begin{align}
         &  R_2(S_{\gamma N};m_3^2,M_{12}^2)=\frac{1}{4\sqrt{\lambda(S_{\gamma N},m_a^2,m_b^2)}}\int d\phi\int dt\Bigg[
         \Theta\left(\frac{S_{\gamma N}-M_{12}^2+m_3^2}{2\sqrt{S_{\gamma N}}}-m_3\right)
           \nonumber\\
           \label{eq:R2to22}
    &\times 
    \Theta\left(\frac{M_{12}^2-m_3^2+S_{\gamma N}}{2\sqrt{S_{\gamma N}}}\right)
      \Theta\left(-1\leq\frac{2S_{\gamma N}(t-m_a^2-m_3^2)+(S_{\gamma N}+m_3^2-M_{12}^2)(S_{\gamma N}+m_a^2-m_b^2)}{\sqrt{\lambda(S_{\gamma N},m_a^2,m_b^2)\lambda(S_{\gamma N},m_3^2,M_{12}^2)}}\leq 1\right)\Bigg].
\end{align}

Let us now define the kinematical function $G$, which is related to the Gram determinant,
\begin{align}
\label{eq:gramdet}
G(S_{\gamma N},t,M_{12}^2,m_a^2,m_b^2,m_3^2)&=-4\left|\begin{array}{ccc}
          p_a^2 & p_a\cdot p_b & p_a\cdot p_3 \\
          p_a\cdot p_b & p_b^2 & p_b\cdot p_3\\
         p_a\cdot p_3 & p_b\cdot p_3 & p_3^2
    \end{array}\right|\nonumber
    \\[5pt]
    &=\frac{\lambda(S_{\gamma N},m_3^2,M_{12}^2)\lambda(S_{\gamma N},m_a^2,m_b^2)}{-4S_{\gamma N}}\;\sin^2\theta_{a3}^*.
\end{align}
It can be shown that, in terms of Lorentz invariants, this function is explicitly given by 
\begin{align}
G(x,y,z,u,v,w)=x^2y+xy^2+z^2u+zu^2+v^2w+vw^2+xzw+xuv+yzw+yuw\nonumber\\-xy(z+u+v+w)-zu(x+y+v+w)-vw(x+y+z+u) \,.
\end{align}
The third $\Theta$ function in \EQ\eqref{eq:R2to22}, which imposes that $|\cos\theta_{a3}^*|\leq1$, is equivalent to\\ $G(S_{\gamma N },t,M_{12}^2,m_a^2,m_b^2,m_3^2)\leq0$. Therefore, one can replace the third $\Theta$ function by\\ $\Theta(-G(S_{\gamma N},t,M_{12}^2, m_a^2,m_b^2,m_3^2))$, leading to
\begin{align}
           R_2(S_{\gamma N};m_3^2,M_{12}^2)&=\frac{1}{4\sqrt{\lambda(S_{\gamma N},m_a^2,m_b^2)}}\int d\phi\int dt\Bigg[\Theta\left(\frac{M_{12}^2-m_3^2+S_{\gamma N}}{2\sqrt{S_{\gamma N}}}\right)
           \nonumber\\
     &\qquad\times \Theta\left(\frac{S_{\gamma N}-M_{12}^2+m_3^2}{2\sqrt{S_{\gamma N}}}-m_3\right)
     \Theta\left(-G(S_{\gamma N},t,M_{12}^2,m_a^2,m_b^2,m_3^2)\right)\Bigg]\,.
     \label{eq:R2to23}
\end{align}
Combining \EQs\eqref{eq:R3formula}, \eqref{eq:R1to2} and \eqref{eq:R2to23}, we get
\begin{align}
    R_3=&\frac{1}{32\sqrt{\lambda(S_{\gamma N},m_a^2,m_b^2)}}\int dM_{12}^2\int d\phi\int dt\int d\Omega_1^{R12}
    \Bigg[\frac{\sqrt{\lambda(M_{12}^2,m_1^2,m_2^2)}}{M_{12}^2}
    \Theta\left(-G\right)
    \nonumber\\
    &\!\!\! \times\Theta\left(M_{12}-m_1-m_2\right)
    \Theta\left(\frac{M_{12}^2-m_3^2+S_{\gamma N}}{2\sqrt{S_{\gamma N}}}\right)
    \Theta\left(\frac{S_{\gamma N}-M_{12}^2+m_3^2}{2\sqrt{S_{\gamma N}}}-m_3\right)\Bigg]\,,
     \label{eq:R3}
\end{align}
where the arguments of $G$ have been suppressed for conciseness.

\subsection{Integration over the angles $\phi$ and $\phi_b$}

\begin{figure}[t!]
    \centering
    \includegraphics[width=0.6\linewidth]{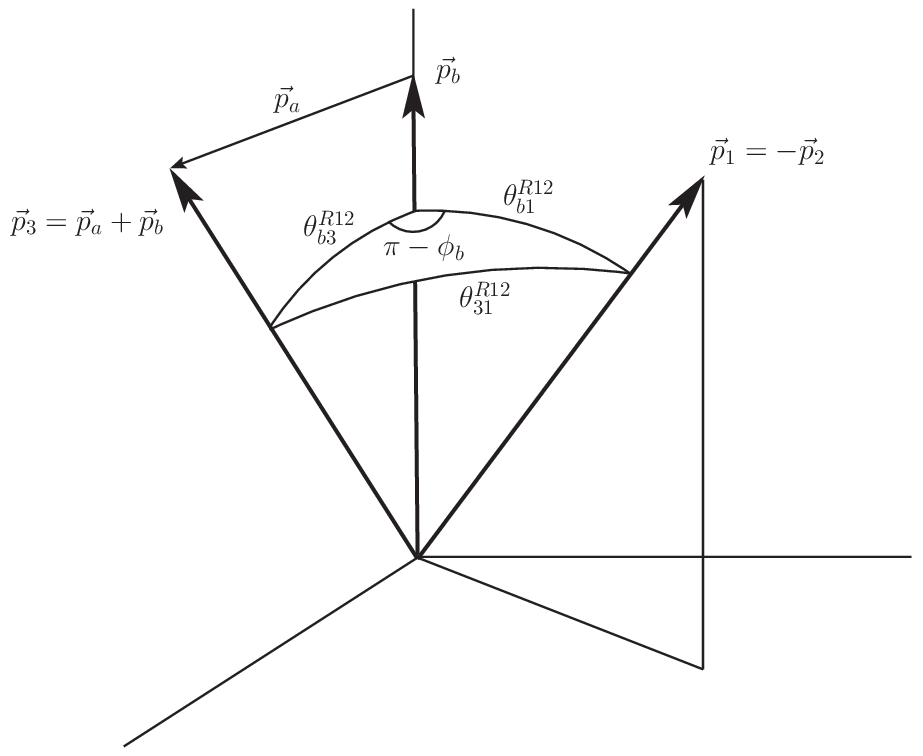}
    \caption{In the Jackson frame defined by $\vec{p}_1+\vec{p}_2=0$ and the $z$-axis taken to be in the direction of $\vec{p}_b$, one has $d\Omega_1^{R12}=d\cos\theta_{b1}^{R12}\,d\phi_b$.}
    \label{fig:jacksonframe}
\end{figure}

To simplify further, we need to specify the direction of the $z$-axis for the angular integration over $d\Omega_1^{R12}$ which is the solid angle that specifies the orientation of $\vec{p}_1$. In the so-called \textit{Jackson frame}, where $\vec{p}_1+\vec{p}_2=0$, and $\vec{p}_b$ sets the $z$-axis, one has $d\Omega_1^{R12}=d\cos\theta_{b1}^{R12}\,d\phi_b$, where the exact definitions of the angles are shown in \FIG\ref{fig:jacksonframe}.

To obtain the differential cross section as a function of $u'$, we start by writing
\begin{equation}
    u'=(p_b-p_1)^2=m_b^2+m_1^2-2E_b^{R12}E_1^{R12}+2P_b^{R12}P_1^{R12}\cos\,\theta_{b1}^{R12}\,,
\end{equation}
where the superscript $R12$ implies that the kinematical variables are measured in the rest-frame of particles 1 and 2, and $P_i$ denotes the magnitude of the 3-momentum of particle $i$.
A short calculation shows that
\begin{align}
  E_b^{R12}&=\frac{m_b^2+M_{12}^2-t}{2M_{12}}
  \,,&
  E_1^{R12}&=\frac{M_{12}^2+m_1^2-m_2^2}{2M_{12}}\,,
  \\  P_b^{R12}&=\frac{\sqrt{\lambda(M_{12}^2,m_b^2,t)}}{2M_{12}}
  \,,& P_1^{R12}&=\frac{\sqrt{\lambda(M_{12}^2,m_1^2,m_2^2)}}{2M_{12}}\,.
\end{align}
Thus, one obtains
\begin{equation}
    u'=m_b^2+m_1^2-\frac{(m_b^2+M_{12}^2-t)(M_{12}^2+m_1^2-m_2^2)}{2M^2_{12}}
    +\frac{\sqrt{\lambda(M_{12}^2,m_b^2,t)\,\lambda(M_{12}^2,m_1^2,m_2^2)}}{2M^2_{12}}\;\cos\,\theta_{b1}^{R12}\,,
\end{equation}
which allows us to change variables from $\cos\,\theta_{b1}^{R12}$ to $u'$ in \EQ\eqref{eq:R3}, giving
\begin{align}
    & R_3=\frac{1}{16\sqrt{\lambda(S_{\gamma N},m_a^2,m_b^2)}}\int dM_{12}^2\int d\phi\int dt\,\int du'\int d\phi_b\frac{1}{\sqrt{\lambda(M_{12}^2,m_b^2,t)}}\Theta\left(M_{12}-m_1-m_2\right)\nonumber
    \\&   \times
    \Theta\left(\frac{M_{12}^2-m_3^2+S_{\gamma N}}{2\sqrt{S_{\gamma N}}}\right)\Theta\left(\frac{S_{\gamma N}-M_{12}^2+m_3^2}{2\sqrt{S_{\gamma N}}}-m_3\right)
     \Theta\left(-G(S_{\gamma N},t,M_{12}^2,m_a^2,m_b^2,m_3^2)\right).
\end{align}
If there is no preferred direction in the transverse plane of the CM frame, the amplitude must be independent of $\phi$. The integration over $\phi$ is thus trivial. In the collinear kinematics, working in the rest frame of particles 1 and 2, the momenta $\vec{p}_3$, $\vec{p}_b$ and $\vec{p}_a$ are aligned, so once again, there is no preferred direction in the transverse plane. The amplitude does not depend on $\phi_b$ in this approximation, and the integration is also trivial.

After performing the two trivial integrals over $\phi$ and $\phi_b$, the three-body phase space becomes
\begin{align}
\label{eq:finalR3}
    & R_3=\frac{(2\pi)^2}{16(S_{\gamma N}-M^2)}\int dM_{12}^2\int dt\,\int du'\frac{1}{(M_{12}^2-t)}\;\Theta\left(M_{12}-m_1-m_2\right)\nonumber\\&\Theta\left(\frac{M_{12}^2-M^2+S_{\gamma N}}{2\sqrt{S_{\gamma N}}}\right)\Theta\left(\frac{S_{\gamma N}-M_{12}^2+M^2}{2\sqrt{S_{\gamma N}}}-M\right)
     \Theta\left(-G(S_{\gamma N},t,M_{12}^2,M^2,0,M^2)\right).
\end{align}
It can be checked that all $\Theta$ functions in \EQ\eqref{eq:finalR3} are satisfied in the phase space defined by the kinematics within collinear factorisation, see \SEC\ref{sec:phase-space}.

The flux factor is $F=2(2\pi)^{5}\sqrt{\lambda(S_{\gamma N},M^2,0)}$, so the differential cross-section finally reads
\begin{equation}
    \frac{d\sigma}{dM_{12}^2\,d(-t)\,d(-u')}=\frac{|\overline{\mathcal{M}}|^2}{32(2\pi)^3S_{\gamma N}(M_{12}^2-t)(S_{\gamma N}-M^2)}\approx \frac{|\overline{\mathcal{M}}|^2}{32(2\pi)^3M_{12}^2S_{\gamma N}^2}\,,
\end{equation}
in the generalised Bjorken limit. This concludes the derivation of \EQ\eqref{eq:crosssectionformula}.

\bibliographystyle{utphys}

\bibliography{masterrefs.bib}

\end{document}